\documentclass[fleqn,12pt]{wlscirep}
\usepackage[utf8]{inputenc}
\usepackage[T1]{fontenc}
\usepackage{xcolor,color,soul}
\usepackage{subcaption}
\usepackage{amsmath}
\usepackage{gensymb}
\usepackage{caption}
\usepackage{amssymb}  
\usepackage{amsfonts}
\def\kpar{k_\parallel}

\usepackage{xr}
\makeatletter
\newcommand*{\addFileDependency}[1]{
  \typeout{(#1)}
  \@addtofilelist{#1}
  \IfFileExists{#1}{}{\typeout{No file #1.}}
}
\makeatother

\newcommand*{\myexternaldocument}[1]{%
    \externaldocument{#1}%
    \addFileDependency{#1.tex}%
    \addFileDependency{#1.aux}%
}
\myexternaldocument{supporting-info}

\usepackage{hyperref}
\hypersetup{
    colorlinks,
    linkcolor={blue!90!black},
    citecolor={magenta!80!black},
    urlcolor={blue!90!black}
}
\title{Photonics in Flatland: Challenges and Opportunities for Nanophotonics with 2D Semiconductors}

\author[4]{Azimi, Ali}
\author[5]{Barrier, Julien}
\author[6]{Barreda, Angela}
\author[1]{Bauer, Thomas}
\author[7]{Bouzari, Farzaneh}
\author[17]{Brokkelkamp, Abel}
\author[8]{Buatier de Mongeot, Francesco}
\author[12]{Parsons, Timothy}
\author[28]{Christianen, Peter}
\author[17,*]{Conesa-Boj, Sonia}
\author[2,3,*]{Curto, Alberto G.}
\author[9]{Das, Suprova}
\author[1]{Dias, Bernardo}
\author[10]{Epstein, Itai}
\author[9]{Fedorova, Zlata}
\author[5,29]{García de Abajo, F. Javier}
\author[27]{Goykhman, Ilya}
\author[11]{Greten, Lara}
\author[1]{Grönqvist, Johanna}
\author[1]{Guarneri, Ludovica}
\author[2,3]{Guo, Yujie}
\author[1]{Hoekstra, Tom}
\author[12]{Hu, Xuerong}
\author[9]{Laudert, Benjamin}
\author[26]{Lynch, Jason}
\author[11]{Meyer, Sabrina}
\author[13]{Munkhbat, Battulga}
\author[14]{Neshev, Dragomir}
\author[1,15]{Ogienko, Masha}
\author[16]{Papadopoulos, Sotirios}
\author[2,3]{Parappurath, Aparna}
\author[17]{Sangers, Jeroen}
\author[18]{Soubelet, Pedro}
\author[17]{Soukaras, Chris}
\author[9]{Soavi, Giancarlo}
\author[9,*]{Staude, Isabelle}
\author[19,*]{Sun, Zhipei}
\author[20,21]{Tielrooij, Klaas-Jan}
\author[19]{Uddin, MD Gius}
\author[9]{Ustinov, Alexey}
\author[1,*]{van de Groep, Jorik}
\author[1]{van Wezel, Jasper}
\author[22]{Vermeulen, Nathalie}
\author[23]{Wang, Hai}
\author[12]{Wang, Yadong}
\author[13]{Xiao, Sanshui}
\author[2,3]{You, Bingying}
\author[24]{Zambrana-Puyalto, Xavier}

\affil[1]{Institute of Physics, University of Amsterdam, 1098 XH Amsterdam, The Netherlands}
\affil[2]{Photonics Research Group, Ghent University-imec, 9052 Ghent, Belgium}
\affil[3]{Center for Nano- and Biophotonics, Ghent University, 9052 Ghent, Belgium}
\affil[4]{Experimental Physics and Functional Materials, Brandenburg University of Technology, 03046 Cottbus, Germany}
\affil[5]{ICFO - The Institute of Photonic Sciences, 08860 Castelldefels, Barcelona, Spain}
\affil[6]{Department of Electronic Engineering, University Carlos III of Madrid, 28911 Leganés, Spain}
\affil[7]{Department of Electronics and Nanoengineering, Aalto University, 2150 Espoo, Finland}
\affil[8]{Dipartimento di Fisica, Università di Genova, 16146 Genova, Italy}
\affil[9]{Institute of Solid-State Physics, Abbe Center of Photonics, Friedrich Schiller University Jena, 07743 Jena, Germany}
\affil[10]{School of Electrical Engineering, Faculty of Engineering, Tel Aviv University, 6997801 Tel Aviv, Israel}
\affil[11]{Institut für Physik und Astronomie, Technische Universit\"{a}t Berlin, 10623 Berlin, Germany}
\affil[12]{School of Mathematical and Physical Sciences, The University of Sheffield, S3 7RH Sheffield, United Kingdom}
\affil[13]{Department of Electrical and Photonics Engineering, Technical University of Denmark, 2800 Kongens Lyngby, Denmark}
\affil[14]{ARC Centre of Excellence for Transformative Meta-Optical Systems (TMOS), Research School of Physics, Australian National University, 2601 Acton, Australia}
\affil[15]{NWO Institute AMOLF, 1098 XG Amsterdam, The Netherlands}
\affil[16]{Institut de Physique et Chimie des Matériaux de Strasbourg, Université de Strasbourg, CNRS, 67000 Strasbourg, France}
\affil[17]{Quantum Nanoscience, Delft University of Technology, 2628 CJ Delft, The Netherlands}
\affil[18]{Walter Schottky Institut, TUM School of Natural Sciences, Technische Universität München, 85748 Garching, Germany}
\affil[19]{QTF Centre of Excellence, Department of Electronics and Nanoengineering, Aalto University, 2150 Espoo, Finland}
\affil[20]{Applied Physics and Science Education, Eindhoven University of Technology, 5612 AZ Eindhoven, The Netherlands}
\affil[21]{CSIC \& BIST, Catalan Institute of Nanoscience and Nanotechnology (ICN2), 08193 Bellaterra (Barcelona), Spain}
\affil[22]{Brussels Photonics, Department of Applied Physics and Photonics, Vrije Universiteit Brussel, 1050 Brussel, Belgium}
\affil[23]{Nanophotonics, Debye Institute for Nanomaterials Science, Utrecht University, 3584 CC Utrecht, The Netherlands}
\affil[24]{Department of Physics, Technical University of Denmark, Fysikvej, DK-2800 Kongens Lyngby, Denmark}
\affil[25]{Technion, Israel Institute of Technology, 3200003 Haify, Israel}
\affil[26]{Electrical and Systems Engineering, University of Pennsylvania, Philadelphia, Pennsylvania 19104, USA}
\affil[27]{Institute of Applied Physics, The Faculty of Science, The Hebrew University of Jerusalem, Jerusalem 91904, Israel
}
\affil[28]{High Field Magnet Laboratory (HFML—EMFL), Radboud University, 6525 ED Nijmegen, the Netherlands}
\affil[29]{ICREA-Institucio Catalana de Recerca i Estudis Avançats, Passeig Lluis Companys 23, 08010 Barcelona, Spain}
\affil[*]{Corresponding authors: j.vandegroep@uva.nl; Alberto.Curto@UGent.be; s.conesaboj@tudelft.nl; zhipei.sun@aalto.fi; isabelle.staude@uni-jena.de}

\keywords{Keywords}

\begin{abstract}
Two-dimensional (2D) semiconductors are emerging as a versatile platform for nanophotonics, offering unprecedented tunability in optical properties through exciton resonance engineering, van der Waals heterostructuring, and external field control.
These materials enable active optical modulation, single-photon emission, quantum photonics, and valleytronic functionalities, paving the way for next-generation optoelectronic and quantum photonic devices.
However, key challenges remain in achieving large-area integration, maintaining excitonic coherence, and optimizing amplitude-phase modulation for efficient light manipulation. 
Advances in fabrication, strain engineering, and computational modelling will be crucial to overcoming these limitations. 
This perspective highlights recent progress in 2D semiconductor-based nanophotonics, emphasizing opportunities for scalable integration into photonics.
\end{abstract}

\begin{document}

\flushbottom
\maketitle
\newcommand{\todo}{\hl{\textbf{TODO}}}

\thispagestyle{empty}

\section*{Introduction}
Two-dimensional (2D) semiconductors, including transition metal dichalcogenides (TMDs, such as MoS$_2$ and WSe$_2$)~\cite{radisavljevic2011single,srivastava2015optically} and emerging layered systems~\cite{novoselov2005two} (e.g. black phosphorus, InSe, and GaSe), are redefining the landscape of nanophotonics.
These atomically thin crystals offer unprecedented control over excitonic resonances, enabling dynamic modulation of optical properties via electrostatic gating~\cite{Sun_NP_16,Dai_AN_20}, mechanical strain~\cite{Liang_NL_17}, dielectric environment engineering~\cite{Du_NRP_21}, and the creation of tailored van der Waals (vdW) heterostructures~\cite{wang2021stacking,Guo_AM_23}.
Their exceptional optical characteristics, large exciton binding energies, high oscillator strengths, and intrinsic valley-dependent optical selection rules~\cite{xiao2012coupled,Mak2012PL,cao2012valley}, position them as ideal platforms for the realization of active optical modulators~\cite{Wang_AP_21}, single-photon emitters~\cite{he2015single,iff2021purcell,zhao2021site}, and integrated quantum photonic devices~\cite{shields2007semiconductor,wang2020integrated,maring2024versatile}.

Yet, despite rapid experimental advances, transitioning these remarkable physical attributes into scalable and reliable technologies remains a considerable challenge.
Achieving large-area integration demands wafer-scale uniformity and precise fabrication control~\cite{Liao_NC_20}, whereas preserving excitonic coherence across functional devices requires meticulous interface engineering and robust material encapsulation strategies~\cite{Hu_NC_17,Vincent_LAM_23}.
Furthermore, fully harnessing the distinctive valleytronic and quantum photonic features inherent to 2D semiconductors depends critically on deeper theoretical insights into many-body excitonic phenomena and robust methods for device-scale predictive modelling~\cite{Du_NM_24,Lin_NC_24}.

This Perspective is structured to reflect the progression from fundamental insights toward technological realization.  
We begin by discussing key mechanisms underpinning excitonic tunability, such as electrical, mechanical, and optical control, and highlight their implications for designing active nanophotonic devices~\cite{Dai_AN_20,Sun_NP_16,Akkanen_am_2022}.
We then examine vdW heterostructure engineering, emphasizing opportunities presented by atomically precise stacking and interface quality for enhancing optical performance~\cite{wang2021stacking,Guo_AM_23,Du_S_23}.
Building upon these foundations, we explore the emerging field of valleytronics, underscoring the potential of valley-dependent optical phenomena for novel quantum-state manipulation~\cite{Mak2012PL,cao2012valley,Yang2015Kerr}.
Our discussion subsequently addresses quantum photonics, illustrating how 2D semiconductors uniquely facilitate single-photon generation, nonlinear photon-pair sources, and quantum sensing technologies~\cite{shields2007semiconductor,srivastava2015optically,he2015single,iff2021purcell,zhao2021site}.
A dedicated theoretical section follows, underscoring the necessity for advanced quantum-mechanical modelling capable of accurately describing exciton dynamics and predicting macroscopic optical device performance~\cite{Du_NM_24,Lin_NC_24,Herrmann2023SHG}.
Finally, we address the pressing challenges of scalability and reproducibility, identifying strategies for wafer-scale growth, integration, and metrology that are essential for translating laboratory advances into practical photonic systems~\cite{Liao_NC_20,Hu_S_23,Uddin_NC_24}.
By systematically linking the physics of atomically thin semiconductors with challenges and opportunities in device engineering and integration, we aim to provide a cohesive roadmap for the development of scalable and tunable nanophotonic and quantum devices.

\begin{figure}[h!]
\centering
\includegraphics[width=0.85\textwidth]{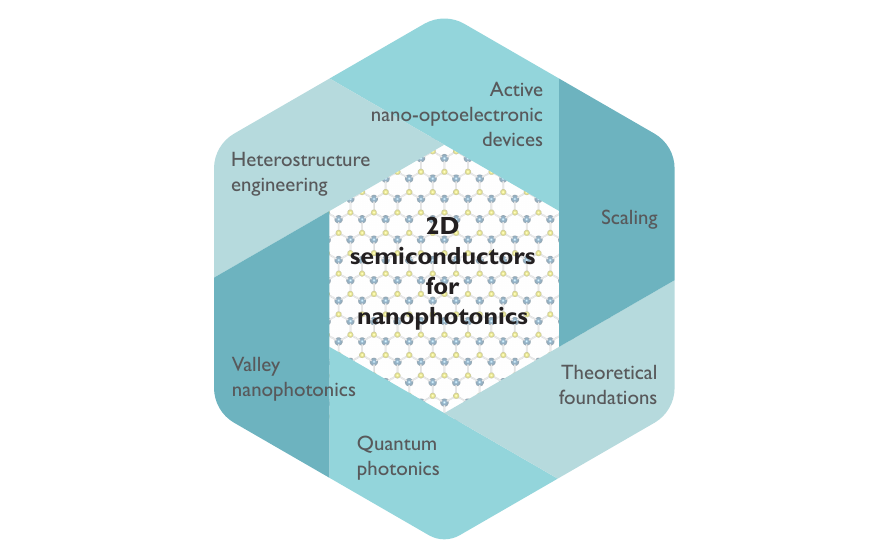}
    \caption[]{Challenges and opportunities for 2D semiconductors in nanophotonics.}
\label{TOC}
\end{figure}

\section*{Active 2D nano-optoelectronic devices}
\label{tuning}

%
Current metasurfaces and nanophotonic devices provide efficient light manipulation, detection/emission, and wavefront shaping, but their function so far has largely remained static. At the same time, novel and upcoming technologies demand active control over light fields for LiDAR, AR/VR and other wearables, optical communications, and LiFi. Resonant light-matter interactions lie at the heart of nanophotonics devices and metasurfaces. However, the tunability of plasmon and Mie resonances in metallic and dielectric particles is limited or very challenging. 
In contrast, 2D semiconductors have emerged as a promising material platform that exhibits remarkable tunability in their optical properties. 
The strong exciton resonance in monolayer semiconductors, in particular, offers a uniquely strong light-matter interaction that is tunable in both amplitude and/or energy, which opens new pathways to explore active/tunable optical elements and nanophotonic devices.
While initial demonstrations of such tunable metasurfaces are promising~\cite{VandeGroep2020,Li2023,Li2023beamsteering,Hoekstra2025}, several key challenges hinder progress toward the widespread application of exciton resonance tuning.
Here, we briefly review the physical mechanisms underlying the strong excitonic tunability, and highlight the key challenges in the field.

%
Tuning mechanisms in 2D nano-optoelectronic devices can be broadly categorized into static tuning, achieved by structural design (including thickness, twist angle in heterostructures, and dielectric environment), and dynamic tuning, which relies on external parameters. 
While static tuning determines the fundamental optical properties of a material, dynamic tuning provides real-time control and adaptability, making it the most impactful for technological applications.

One of the most effective dynamic tuning mechanisms is electrostatic gating, which modulates the carrier density (Fermi level).
High free-carrier concentrations lead to electrostatic screening of the exciton field lines, enhanced exciton-electron scattering, and the formation of charged trions. 
Together, these effects enable efficient modulation of exciton transitions, affecting both incoherent photoluminescence and coherent light scattering~\cite{Yu2017}. 
Beyond simple gating, external electric and magnetic fields offer additional knobs to manipulate exciton states. 
In particular, out-of-plane electric fields can induce a quantum-confined Stark effect, shifting exciton resonances by up to several hundred meV, often with minimal linewidth broadening, especially in interlayer excitons, whose intrinsic dipole moments align with the external electric field~\cite{Jauregui2019}.
Similarly, magnetic fields cause Zeeman splitting of excitonic states, with pronounced effects on excited states ({\it e.g.}, 2s, 3s, 4s), reaching several tens of meV~\cite{Stier2018}.

Mechanical strain, in contrast, provides a route toward spatially localized and reversible control. 
By deforming the crystal lattice, strain engineering modifies the band structure and exciton binding energy, enabling excitonic shifts of up to several hundred meV~\cite{Aslan2018}. 
Complementing these electrical and mechanical approaches, thermal and optical modulation offer alternative pathways for exciton control.
Temperature affects exciton linewidth and amplitudes through exciton-phonon interactions~\cite{Henriquez-Guerra2023}. 
Meanwhile, optical excitation provides ultrafast control through Pauli blocking and carrier-induced broadening.
Pump-probe experiments have demonstrated blue shifts and linewidth changes on femtosecond timescales, highlighting the potential for all-optical switching.

\begin{figure}
\includegraphics[width=1\textwidth]{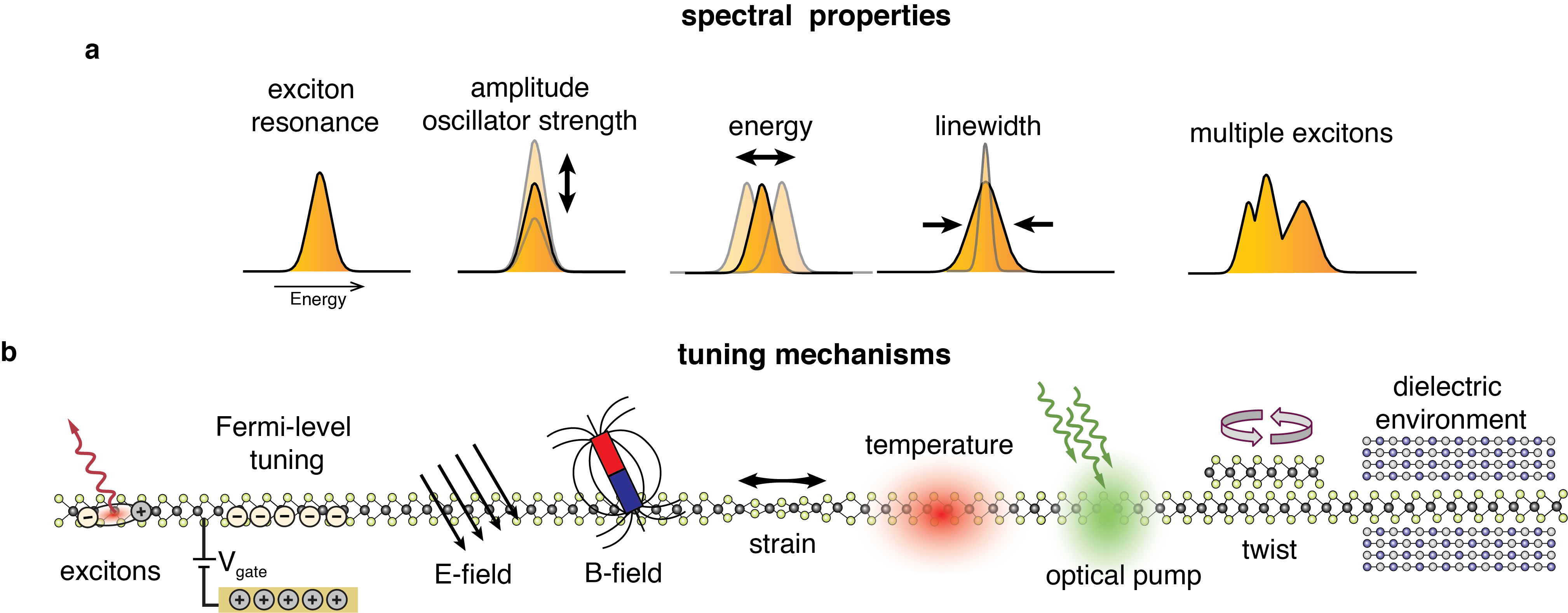}
    \caption[]{Schematic of (a) tunable spectral properties of exciton resonances and (b) experimentally available tuning mechanisms.}
\label{poster}
\end{figure}

These tuning mechanisms, ranging from electrostatic gating and strain to optical and magnetic control, establish a rich toolbox for modulating light–matter interactions in atomically thin semiconductors. 
Yet, realizing their full potential in scalable nanophotonic systems requires overcoming critical integration and performance challenges, which we discuss next.


A major challenge in the field of 2D excitonic nanophotonics lies in efficient integration. 
While the tunability of exciton-light interactions in monolayer 2D semiconductors is well established, their absolute optical efficiency remains limited due to the inherently weak light-matter interaction at the atomic scale~\cite{VandeGroep2020}.
Overcoming this requires integrating 2D semiconductors with resonant nanophotonic architectures, metasurfaces, waveguides, or cavities, to enhance modulation efficiency, beam steering, and wavefront control. 
However, such integration presents additional challenges, including precise material transfer, alignment on nanostructured substrates, and degradation during nanofabrication~\cite{Li2023}.

A related concern is the uniformity of excitonic response across large areas.
Most demonstrations use small exfoliated flakes encapsulated in hexagonal boron nitride (hBN), where structural and dielectric homogeneity ensures consistent performance.
Scaling to larger areas, however, introduces inhomogeneous broadening, spatial variations in defect densities, and physical imperfections such as cracks and wrinkles.
These factors compromise device reproducibility and performance; for instance, cracks hinder carrier transport and wrinkles perturb strain-tuning profiles. 
Dielectric disorder in the substrate further amplifies optical non-uniformity~\cite{Raja2019}.

An additional hurdle is achieving simultaneous amplitude and phase modulation. 
While electrostatic gating can modulate exciton oscillator strength and thus emission amplitude, isolated monolayers offer limited phase tuning~\cite{vandegroep2023}.
Embedding TMDs in optical cavities can enhance phase response, as recent studies show ~\cite{Datta2020, Li2024, Lynch2025}, but independent and broadband control over both amplitude and phase remains elusive, yet essential for free-form metasurfaces design.
Designing reliable excitonic devices also hinges on comprehensive documentation of materials properties.
While theoretical databases like C2DB~\cite{Haastrup2018,Gjerding2021} and CRYSP provide useful predictions, experimental datasets remain incomplete.
Unresolved questions, such as exciton decay channels and lifetimes in specific TMDs, limit the predictive accuracy and engineering reliability of 2D-based photonic components. 
%

Beyond incomplete datasets, another major bottleneck lies in the intrinsic spectral limitations of excitonic resonances.
%
%
Due to their large binding energies and discrete energy levels, exciton linewidths are inherently narrow, restricting their applicability in tunable lasers, modulators, and photodetectors that require broader spectral tunability. 
External tuning mechanisms, such as electric or magnetic fields and strain, provide only limited shifts in exciton energy, keeping their operation within a narrow spectral range~\cite{Stier2016, Lu2017}. 
While exciton-polaritons extend the spectral tunability, they remain constrained by the Rabi splitting. 
A promising avenue for overcoming this limitation involves 2D heterostructures, particularly interlayer excitons, where tuning is achieved through interlayer distance control and twist-angle engineering~\cite{Li2020, Lin2021}. 
Additionally, employing multiple TMDs can expand the operational spectral range by leveraging different excitonic states. 
A well-characterized library of material and heterostructure properties is essential for developing broadband 2D optoelectronic devices.

Finally, response speed and compact integration remain key bottlenecks. 
Tuning bandwidths range from kHz to GHz depending on the mechanism, constrained by carrier mobility, contact resistance, and device architecture. 
While electric field modulation is CMOS-compatible, integration of 2D materials into CMOS foundries is still an open challenge. 
Back-end-of-line approaches~\cite{Goossens2017} and MEMS-based strain tuning offer promising alternatives.
In contrast, methods relying on magnetic fields or temperature are harder to miniaturize. 
All-optical approaches, particularly when integrated into Si or InP photonics, may eventually offer a scalable route toward compact, high-speed, and CMOS-compatible excitonic devices.

\section*{Heterostructure engineering and integration challenges}
\label{heterostructures}
%

To address these spectral and functional limitations, heterostructure engineering offers a promising route forward. 
By vertically stacking distinct 2D materials into vdW heterostructures, one can access new excitonic states, such as interlayer excitons, and finely tune their energies via twist angle, interlayer separation, and dielectric environment. 
These atomically precise assemblies not only broaden the optical tunability beyond what monolayers allow, but also introduce new degrees of freedom for device functionality, including long-lived dipolar excitons and momentum-dependent optical selection rules.

\begin{figure}[h!]
\centering 
\includegraphics[width=0.9\textwidth]{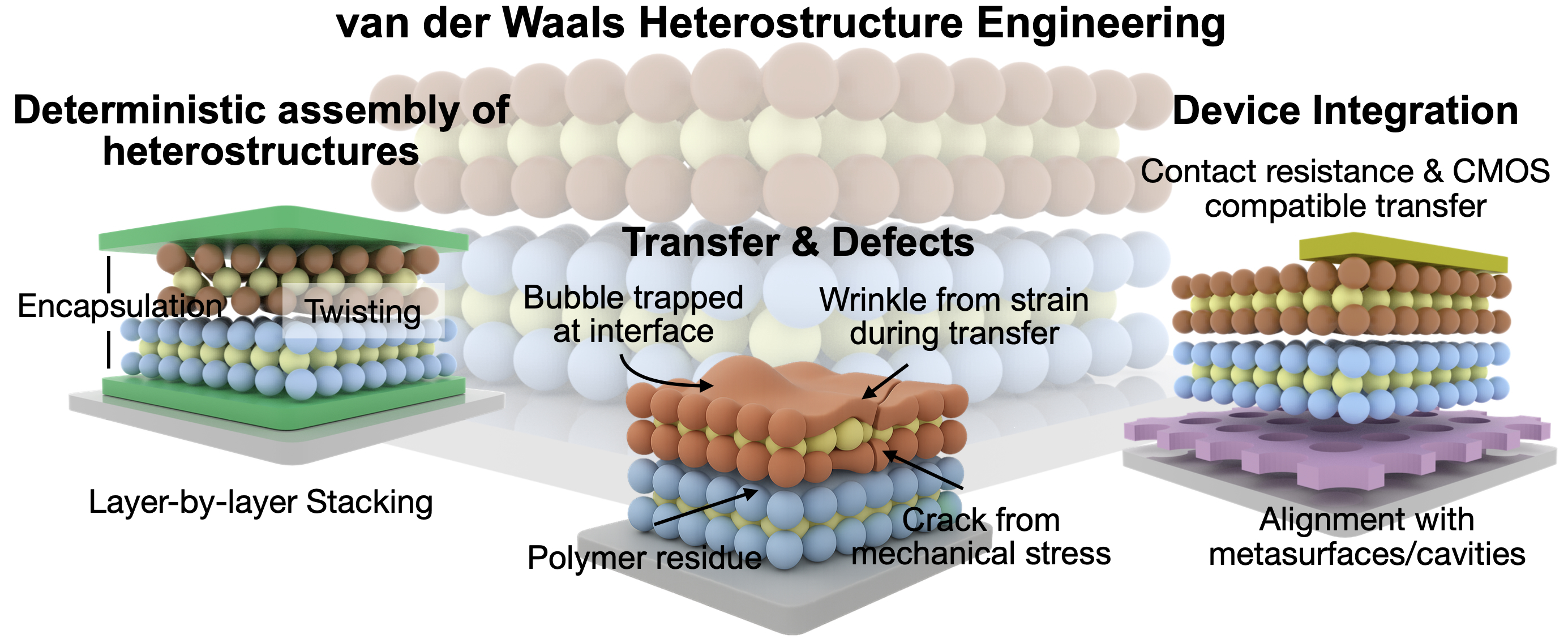}
\caption{Schematic illustration of key challenges in the engineering of vdW heterostructures. These include interfacial contamination and polymer residues affecting adhesion and contact resistance, twist-angle control, strain introduced during transfer, alignment with photonic structures such as metasurfaces or cavities, and compatibility with CMOS back-end processing.}
\label{heterostructure}
\end{figure}

Unlike traditional bulk materials, which are constrained by lattice matching and prone to interfacial defects that degrade performance, vdW heterostructures are assembled without chemical bonding, allowing sharp, defect-free interfaces between materials like TMDs, graphene, and hBN~\cite{novoselov20162d, Fan2022, Yu2021, Li2016, https://doi.org/10.1002/smll.202107059, Wurstbauer2017}. 
This distinctive composition allows precise control over interlayer interactions, making vdW heterostructures highly suitable for applications that require engineered light-matter interactions~\cite{Britnell2013, Wurstbauer2017}. 
The ability to control composition, twist angle, and stacking order at the atomic level~\cite{Du_S_23} opens new possibilities for dynamic photonic applications, such as optical modulators~\cite{Kaushik20, Guo2020}, light sources~\cite{Withers2015, Hwang2005,Lin_NC_24} (including single-photon emitters~\cite{Chuang2024, Yu2024, Parto2021, Rosenberger2019}), tunable polaritons~\cite{Hu_S_23}, and spectrometers~\cite{Yoon_S_22,Uddin_NC_24}. 
Initial demonstrations using vdW heterostructures for tunable nanophotonics have shown promise, yet significant challenges remain in the efficient integration and scalability of these materials for practical device applications. 
In this Section, we outline the structural and functional benefits of vdW heterostructures for nanophotonics and critically assess the key limitations and opportunities toward their incorporation into integrated photonic systems. 
Figure~\ref{heterostructure} provides a visual summary of the physical and technological constraints encountered during deterministic assembly of vdW heterostructures.
Moreover, compatibility with CMOS back-end processes imposes stringent constraints on thermal budgets and interfacial quality, adding further complexity to deterministic stacking strategies.

Developing reliable fabrication techniques to ensure high-quality heterostructures remains challenging. 
Producing high-quality layers, especially at scale, involves advanced techniques such as Chemical vapor deposition (CVD), which enables controlled growth of large-area, monolayer materials~\cite{Shen2018, Xia2014, Mandyam2020}. 
However, achieving uniform, defect-free layers over large areas remains difficult, particularly as the process must be scalable for commercial applications. 
Recent work emphasizes the importance of precise growth control to avoid grain boundaries and thickness fluctuations that can affect device performance~\cite{Kim2023}. 
Following controlled growth through techniques like CVD, transferring these monolayers onto target substrates without introducing defects or contaminants presents additional challenges. 
Both wet and dry transfer methods require optimization to minimize issues like polymer residue from adhesives or mechanical damage during handling~\cite{jainMinimizingResiduesStrain2018a, castellanos-gomezDeterministicTransferTwodimensional2014b, linGentleTransferMethod2016}. 
The formation of bubbles and wrinkles during the transfer process also creates non-uniform interfaces, which complicates reproducibility~\cite{haleyHeatedAssemblyTransfer2021, jainMinimizingResiduesStrain2018a}. 
Recent advances, such as vdW assembly using silicon nitride membranes, have shown significant promise, reportedly improving moir\'e superlattice homogeneity by an order of magnitude~\cite{wangCleanAssemblyVan2023}.

Furthermore, as device dimensions are miniaturized to scales comparable to or smaller than the wavelength of light being sensed, additional challenges arise in both the fabrication process and in controlling optical responses. 
Scaling down device dimensions while maintaining the necessary stacking precision requires advanced nanofabrication techniques and precise control over interlayer alignment. 
Upscaling processes to produce large arrays of such devices while preserving alignment remains an open area for further research and development.

The integration of complex heterostructures, such as moir\'e configurations, introduces new functional capabilities that heavily rely on precise control over stacking order, twist angle, and material compatibility~\cite{Behura2021, reganEmergingExcitonPhysics2022a}. 
Moir\'e patterns can localize interlayer excitons, as demonstrated in materials like TMDs and bilayer graphene, where control over stacking angles enables tunable miniband structures~\cite{Montblanch2021, Blundo2024}. 
These minibands open possibilities for applications in infrared and terahertz sensing. 
Studies have shown that angle-controlled bilayer graphene aligned with hBN supports miniband formation suitable for infrared applications~\cite{Deng2020}, while magic-angle twisted bilayer graphene has demonstrated distinctive superlattice minibands that can be diagnosed through infrared spectroscopy~\cite{Ulstrup2020, li2024infraredspectroscopydiagnosingsuperlattice}. 
Additionally, twisted graphene heterostructures have exhibited giant, ultra-broadband photoconductivity, expanding their potential in broadband photodetection~\cite{Agarwal2023}. 
Further examples include moir\'e engineering in WS\textsubscript{2}/WSe\textsubscript{2} heterostructures, where the twist angle precisely tunes interlayer excitonic properties, enabling applications in tunable photonic and optoelectronic devices~\cite{gogoiLayerRotationAngleDependentExcitonic2019a, barmanTwistDependentTuningExcitonic2022}.

The choice of materials and their combinations in heterostructures is crucial for achieving desired electronic and optical properties. 
Materials must be selected not only for their individual characteristics but also for their chemical and structural compatibility when stacked. 
For instance, combining graphene with TMDs leverages graphene's high mobility and TMDs' strong light-matter interactions, leading to hybrid structures with enhanced functionality. 
Similarly, monolayer TMD superlattices with dielectric spacers can be used for optimal light absorption ~\cite{Kumar2021, Elrafei2024superlattices, Lynch2025}
The alignment and compatibility of lattice constants and interfacial bonding are key factors that determine the performance and stability of heterostructures~\cite{Hussain2022}.

Ultimately, demonstrating the practical viability of vdW heterostructure-based devices in real-world applications requires a careful balance of scalability, reliability, and performance. 
Key performance parameters include achieving an optimal trade-off between gain and bandwidth, on-off ratio, reducing contact resistance, and ensuring long-term operational stability in field-effect devices~\cite{Duan2014, Gong2014, Smith2013}. 
While vdW heterostructures offer enhanced functionalities, translating these capabilities into robust, scalable devices that maintain performance over extended use periods requires further research. 
Addressing issues such as thermal management, environmental stability, and interfacial degradation will be crucial for the commercial adoption of vdW-based technologies in sectors such as flexible electronics, high-speed photonics, and quantum information processing~\cite{Wang2015, Hou2020}.

Advances in vdW heterostructures are paving the way for a range of optoelectronic applications that extend beyond existing technologies. 
For instance, tailored bandgaps and controlled carrier dynamics could lead to advanced photodetectors with enhanced sensitivity and selectivity~\cite{Agarwal2023}. 
Likewise, next-generation photovoltaic technologies~\cite{Gong2014, Hong2014}, including flexible and transparent solar cells, stand to benefit from these materials. 
Their potential extends to energy storage~\cite{Teshome2017, Blackstone2021}, biochemical sensing~\cite{Oh2021}, wearable electronics~\cite{Lin2019}, and imaging systems~\cite{9976044, Zhang2022}. 
The unique quantum properties of 2D materials embedded in vdW heterostructures may also drive innovations in quantum computing and secure quantum communication~\cite{Kavokin2022, Pal2022}. 
Extreme control over light emission, enabled by interlayer coupling and tunable band alignment, could revolutionize LED and laser technologies for future data communication and lighting applications. 
Furthermore, their high specific surface area and chemical adaptability make these heterostructures attractive for photoelectrochemical processes, where surface modifications achievable through defect engineering or catalyst deposition enable fine-tuned reactivity for catalysis and sensing~\cite{Khaidar2024, Voiry2016}.

Beyond traditional vertical or lateral stacking, recent advances allow 2D crystals to be folded, rolled, or twisted, creating intricate three-dimensional geometries with unique interface properties~\cite{zhaoHighorderSuperlatticesRolling2021,Ouyang2022}. 
These 3D vdW heterostructures require novel fabrication techniques and specialized probing methods, such as near-field optical techniques~\cite{Fox2023}, cathodoluminescence (CL)~\cite{Bonnet2021, Nayak2019}, and electron energy loss spectroscopy (EELS)~\cite{Susarla2021,vanHeijst2024}, to investigate their complex internal structures. 
Such architectures hold significant potential for integrated optoelectronic and photonic systems, where interface control is paramount. 
Moreover, vdW heterostructures can be engineered through phase conversion, leveraging materials with multiple stable or metastable phases such as certain TMDs and tin chalcogenides (e.g., SnS, SnS$_2$). 
By adjusting the chemical potential of chalcogen elements, phase transitions can be induced in these layered materials, leading to heterostructures with unique interface-driven functionalities that arise from structural conflicts during phase transformations~\cite{Kim2023-a}. 

Overcoming challenges in scalable, reliable material synthesis and heterostructure fabrication remains essential for achieving reproducible results and enabling industrial-scale production. 
New approaches, like CVD for high-quality films, strain engineering, and solvent-free transfer techniques, are under active investigation to enhance material quality and consistency. 
Future research will continue to focus on developing interdisciplinary methods to integrate these materials into functional devices and to ensure compatibility with CMOS technology, requiring innovations not only in fabrication but also in theoretical models that can accurately predict interlayer interactions and optimize device performance.

\section*{Valleytronic nanophotonics}
\label{valleytronic}
\begin{figure}[b!]
\centering 
\includegraphics[width=0.6\textwidth]{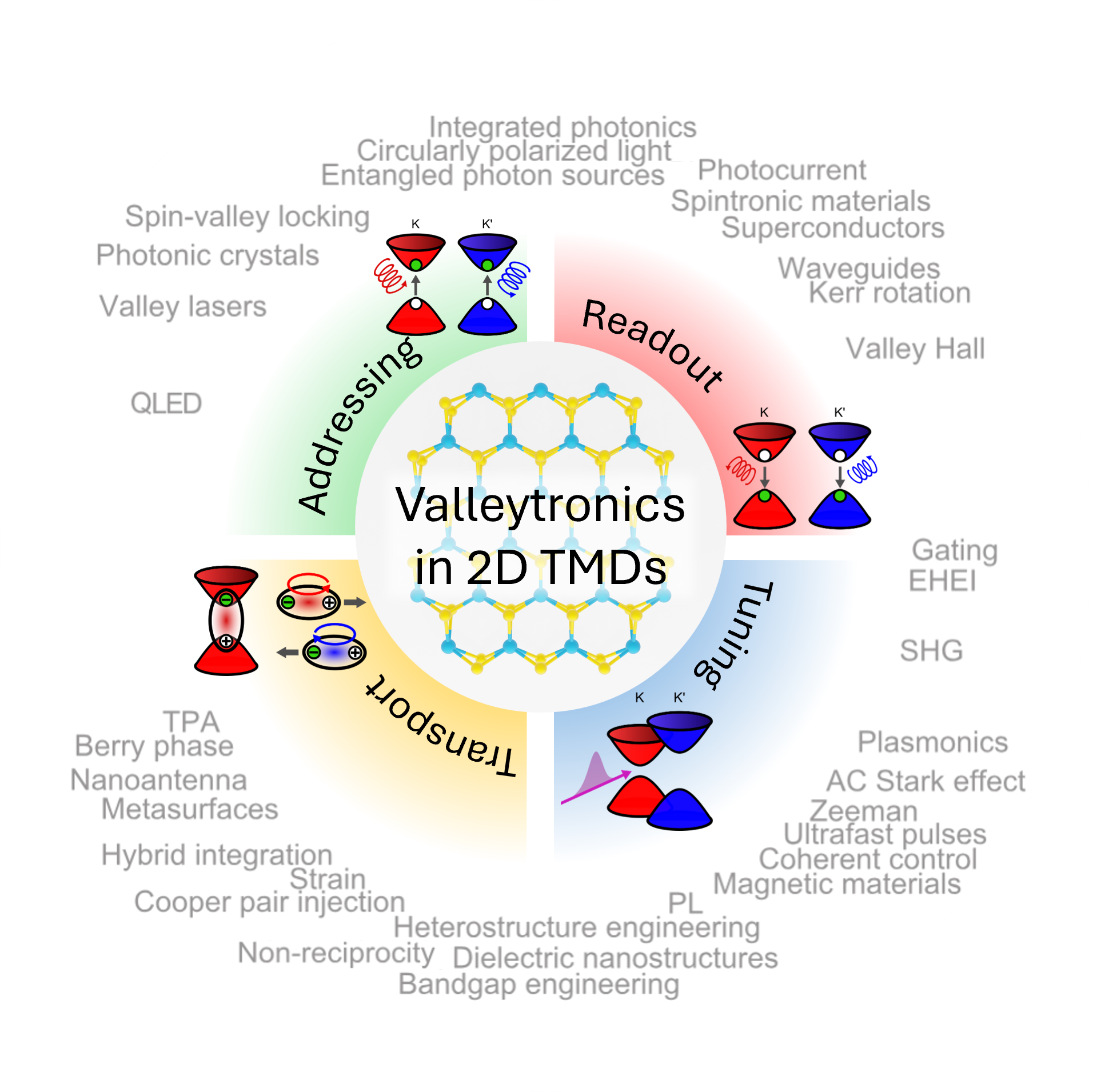}
\caption{Conceptual landscape of valleytronics in 2D TMDs. The valley degree of freedom in 2D TMDs offers opportunities for novel quantum photonic and optoelectronic functionalities. Key pillars of valleytronic control -- addressing, readout, tunability, and transport -- are illustrated at the core, each enabled by a diverse range of mechanisms and hybrid platforms as represented by outer keywords.}
\label{valleytronics}
\end{figure}

%
%
Valleytronic functionalities leverage the control of electronic valleys as an additional quantum degree of freedom, offering new possibilities for information processing and quantum photonics.
While the concept is longstanding, practical implementation was initially limited by the lack of systems with measurable valley contrast.
The advent of 2D materials, especially monolayer TMDs, has transformed this field by offering intrinsic valley contrast and a direct band gap at the inequivalent $K$ and $K'$ points~\cite{xiao2012coupled}. 
Valley-contrasting optical selection rules enable circularly polarized light to address specific valleys, with the resulting circularly polarized photoluminescence encoding the valley of origin -- an effect demonstrated in numerous photoluminescence experiments~\cite{cao2012valley,zeng2012valley,Mak2012PL}.
Although many valleytronic experiments still rely on complex and costly experimental setups, significant progress has been made over the past decade in addressing, manipulating, and detecting valley states. Furthermore, the optical addressability of the valley degree of freedom in 2D TMDs offers exciting prospects for integration with nanophotonic platforms, such as resonant nanoantennas, metasurfaces, integrated photonic circuits, and structured light beams. 
These hybrid systems open new avenues for valleytronic functionalities, enhanced light–matter interactions, and potential spin-valley-photon interfaces, with far-reaching implications for both fundamental research and emerging technologies. 
In this Section, we highlight the key challenges and emerging opportunities in the field of valleytronics. The main pillars of valleytronics in 2D TMDs are summarized in Fig.~\ref{valleytronics}.

Valley depolarization remains one of the central challenges in valleytronics. 
While early theories predicted long valley coherence times~\cite{morpurgo2006intervalley}, experiments revealed fast exciton recombination and decoherence as well as short valley lifetimes, typically in the range of a few to tens of picoseconds~\cite{PhysRevB.90.161302, robert2016exciton, moody2016exciton}. 
This discrepancy has driven extensive efforts to uncover the mechanisms behind valley depolarization.
The identified key contributors include phonon- and defect-mediated intervalley scattering, with zone-corner acoustic phonons and Eliott-Yafet spin-flip playing a major role~\cite{bae2022k,lin2022phonon,jeong2020valley}. 
Material-specific features, such as Rashba-type mixing in MoSe\textsubscript{2} and MoTe\textsubscript{2}, can further accelerate depolarization~\cite{yang2020exciton}.
At cryogenic temperatures, the electron-hole exchange interaction (EHEI), governed by the Maialle-Silva-Sham mechanism, sets a fundamental limit for valley coherence times~\cite{ye2017optical,pattanayak2022steady}. 
It largely depends on the band structure: in MoSe\textsubscript{2}, spin-protected against EHEI negative trions exhibit longer valley lifetimes~\cite{schaibley2016valleytronics}, while in WSe\textsubscript{2}, the trion fine structure facilitates intervalley scattering via singlet-triplet conversion~\cite{zipfel2020light,Zhumagulov2022}.
Mitigating EHEI through electrical gating~\cite{dey2017gate,zhang2022prolonging,siao2025electrostatic} or band structure engineering~\cite{rivera2016valley,liu2020room,an2023strain} remains a key research direction. 
Overall, the multifaceted nature of valley depolarization highlights the need for a deeper understanding of spin-valley photophysics to enable robust valleytronic functionality.


The ability to efficiently read out valley states with high sensitivity and minimal disturbance is key to both fundamental studies and device applications. 
The current gold standard -- polarization-resolved photoluminescence~\cite{Mak2012PL} or time-resolved Kerr rotations~\cite{Yang2015Kerr} -- suffers from drawbacks: PL is inherently destructive and slow, while Kerr rotation measurements can perturb the system due to intense and resonant probe pulses. 
Nonlinear optical techniques have recently emerged as powerful alternatives.
In particular, valley polarization can be read out via polarization rotation of the second harmonic signal~\cite{Herrmann2023SHG, Ho2020SHG, Mouchliadis2021SHG}. 
This approach resembles nonlinear Kerr rotation that, in comparison to its linear counterpart, offers enhanced sensitivity and background-free signals in transparent spectral regions~\cite{Koerkamp1996NLOKerr, Matsubara2012NLOKerr,Shuang2023NLOKerr}.
Valley imbalance can also be inferred from deviations in SHG power scaling between circularly and linearly polarized excitation~\cite{Herrmann2024SHG}.
Furthermore, circularly polarized, nonresonant femtosecond pulses, commonly used for SHG, can lift the valley degeneracy via the AC optical Stark effect~\cite{kim2014ultrafast,Herrmann2023SHG}.
This ultrafast tuning method provides an attractive alternative to conventional valley Zeeman splitting, which requires strong magnetic fields.
In the future, extending SHG-based approaches to more complex ultrafast and time-resolved schemes could unlock deeper insights into valley dynamics, while nonlinear excitation methods, such as resonant two-photon absorption, may enable selective valley control~\cite{Wang2015SHG}.

In addition to Kerr rotation and harmonic generation, photocurrent-based techniques are gaining attention as non-invasive measurements of valley-specific dynamics~\cite{ma2023photocurrent,yin2022tunable}.
These methods, relying on various optically induced current generation processes, allow us to probe the quantum geometric tensor~\cite{morimoto2016topological}, whose real and imaginary parts are known as  quantum metric and Berry curvature, respectively.
Such measurements may offer a route to resolve the underlying topology of valley states, complementing traditional optical readout approaches.


Beyond improved readout schemes, integrating 2D TMDs with nanophotonic structures offers powerful opportunities to control their valley degree of freedom. 
Architectures such as single nanoantennas~\cite{wen2020steering,zheng2023electron}, metallic or dielectric metasurfaces~\cite{li2018tailoring,liu2023controlling}, photonic crystals~\cite{wang2020routing}, and waveguides~\cite{sun2019separation} can be designed to boost valley polarization, facilitate coupling to a specific valley, or induce valley-dependent optical responses. 
For instance, valley-polarized emission has been directionally routed in waveguides via optical spin-momentum locking~\cite{Chervy2018, Gong2018}. 
However, achieving strong and robust valley contrast in such hybrid systems remains challenging. 
The reason is that TMDs typically trade off quantum efficiency and valley polarization: materials with high quantum yield often suffer from low polarization, and vice versa. 
Nanophotonic strategies seek to overcome this limitation via Purcell-enhanced emission, directional outcoupling, and optimized excitation schemes.
Simultaneously, care must
be taken to preserve the purity of circular polarization used for excitation and/or readout~\cite{Liu2023, Raziman2019,bucher2024influence}.

The interaction between excitonic dynamics and nanophotonic effects further complicates the picture. 
Strain, doping, or exciton diffusion and annihilation, all influenced by the local photonic environment, can obscure the origin of observed valley contrasts. 
Existing models, such as rotating dipole approximations, fall short in capturing these many-body and transport effects, highlighting the need for unified frameworks that combine photonic design with realistic excitonic physics~\cite{Raziman2022}.
Another key open question is how to disentangle intrinsic changes in valley polarization from those induced by the nanostructures themselves. 
Progress will require closer integration of theoretical, computational, and experimental approaches to predict and control hybrid valleytronic-nanophotonic behavior~\cite{bucher2024influence}.

Looking ahead, valleytronic nanophotonics holds promise for enabling advanced functionalities such as room-temperature operation, robust non-reciprocal components, and scalable on-chip architectures. 
The recent demonstration of valley-polarized lasing marks a key milestone~\cite{duan2023valley}, highlighting the potential of nanophotonic enhancement in active valleytronic devices. 
Valley polarization offers immense potential for realizing non-reciprocal nanophotonic components such as optical isolators~\cite{guddala2021all}. 
By combining enhanced light–matter interaction in metasurfaces or waveguides with optical pumping or spin injection from conventional spintronic materials, such devices could achieve directional control without relying on bulky magnets. 
Integrating 2D semiconductors into these platforms may thus enable compact, magnet-free isolation schemes with high degrees of tunability. 
More broadly, strategies such as spin injection or valley tuning could be coupled with nanophotonic resonators based on spintronic or magnetic materials to further amplify valley-dependent optical effects.

Finally, another exciting direction involves the integration of 2D TMDs with superconducting materials.
Such hybrid systems offer a compelling platform for advanced entangled-photon sources.
The underlying principle is that electrically injected Cooper pairs can radiatively recombine in a monolayer TMD, transferring their spin entanglement to valley-entangled photon pairs.
The atomically thin geometry ensures full superconducting proximity, suppressing parasitic emission, and enabling high-fidelity photon generation.
This paves the way for compact, electrically driven valley-selective quantum LEDs (QLEDs), bridging superconducting and photonic platforms for scalable quantum information technologies.

\section*{Quantum photonics}
\label{Quantum photonics}
%

%
Quantum photonics drives quantum technology progress by generating, manipulating, and detecting quantum light for computing, sensing, and secure communication~\cite{turunen2022quantum}. 
It requires precise photon-state control for efficient quantum information transfer and faces challenges in preserving coherence, entanglement, and superposition amid environmental disruptions. 
The field also explores how photons interact with many-body quantum systems, revealing emergent phenomena like polariton engineering. These quasiparticles create new possibilities for quantum-level light manipulation, enabling next-generation technologies in information processing. 
As quantum photonics rapidly evolves, it promises transformative advances in various applications.

2D materials, especially graphene and TMDs, offer transformative opportunities in quantum photonics through their remarkable electronic and optical properties~\cite{turunen2022quantum,reserbat2021quantum,gonzalez2024light}. 
Their atomically thin geometry enables strong light-matter interactions, high exciton binding energies, and direct bandgaps across diverse electronic phases (e.g.,superconductors, insulators, semiconductors, and metals) within a single platform. 
These 2D materials can be seamlessly integrated into fiber- or silicon waveguide-based photonic systems, or employed directly as dielectric media for guiding and confining light~\cite{Cui_NS_22,Hu_NC_17}. 
Further, polaritons in these materials exhibit extreme confinement ---demonstrated by orders-of-magnitude in-plane compression in graphene plasmons and hBN phonon polaritons--- leading to unprecedented control of quantum emitters and ultrafast compact on-chip devices~\cite{Yang_AM_2025,Guo_AM_23}. 
These unique features open new pathways for computing~\cite{Zhang_SA_22,Zhang_APL_23}, communications, and sensing~\cite{Cui_SA_2025,Das_LSA_2025,Wu_AM_22,Hu_NC_19}, heralding significant potential in next-generation quantum photonic technologies.


\begin{figure}
    \centering
	\includegraphics[width=1\textwidth]{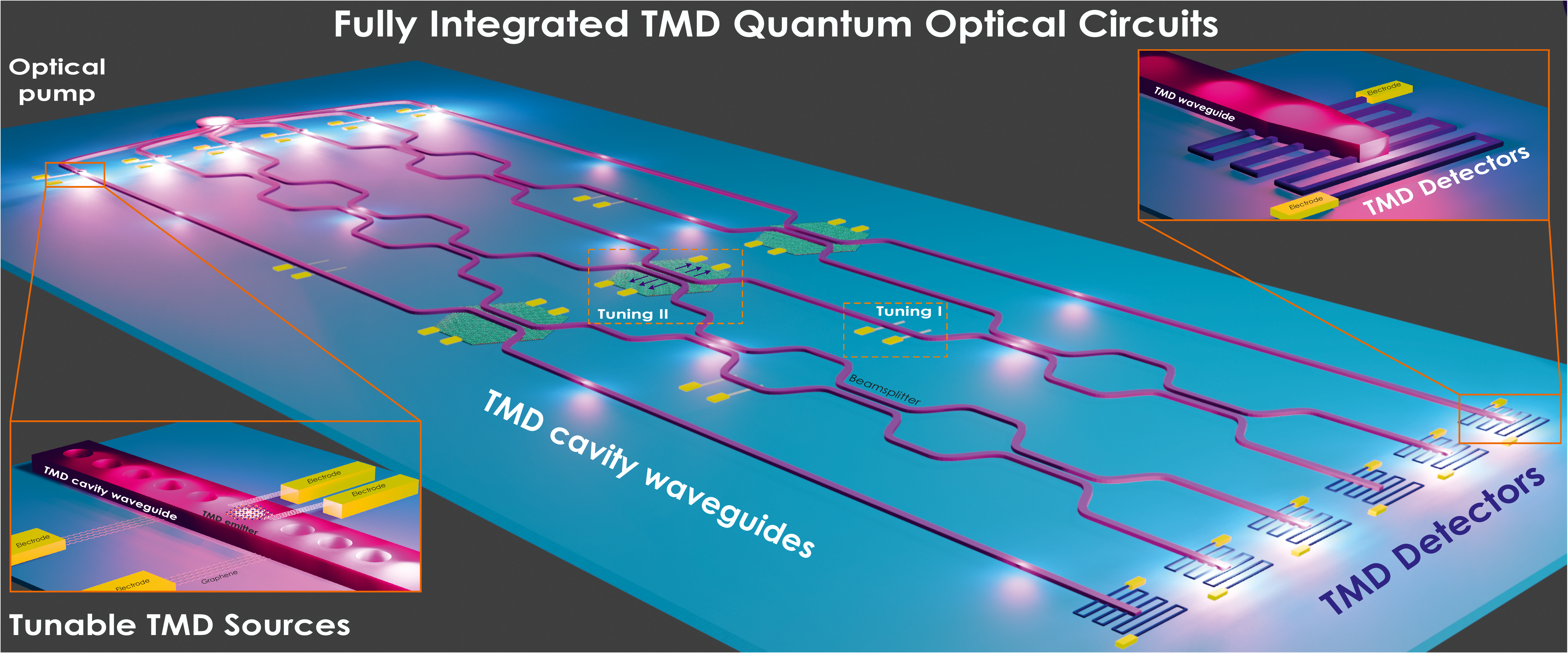}
    \caption{Vision of quantum circuits with 2D materials for applications, indicating key quantum devices, such as sources (inset), gates, and detectors(inset).}
    \label{fig:enter-label}
\end{figure}

%
In photonic quantum information, qubits are carried by photons emitted by single-photon sources (SPSs)~\cite{shields2007semiconductor}, which must offer high purity, efficiency, and indistinguishability for reliable operation in quantum computing, cryptography, and communication~\cite{zhong2020quantum,o2009photonic, wang2020integrated, maring2024versatile}. 
Recently, TMD-based quantum emitters have emerged as a promising platform for deterministic single-photon generation~\cite{srivastava2015optically, tonndorf2015single, he2015single, montblanch2023layered, micevic2022demand, kumar2015strain,Du_S_23}, thanks to their vdW nature (enabling facile exfoliation and stacking) and sub-nanometer thickness (enhancing light extraction and electrical integration)~\cite{novoselov2005two,Withers2015,radisavljevic2011single,munkhbat2020electrical,wang2021stacking}. 
Single-photon emission in TMDs spans from the visible to telecom wavelengths~\cite{he2015single, klein2019site, tran2016quantum, palacios2017large, yu2021site, zhao2021site} and has been integrated with photonic micro- and nanostructures for improved emission efficiency~\cite{iff2021purcell, errando-herranz_resonance_2021, tonndorf2017chip, sortino2020dielectric}.

%
%
Despite these promising developments, the indistinguishability of single photons emitted from TMD-based sources remains a significant bottleneck for their implementation in scalable quantum photonic circuits.\cite{drawer2023monolayer} Reported indistinguishability values remain low, typically around $\sim$2\%, far below the levels required for quantum interference-based protocols. A main issue lies in the considerable deviation of the emission linewidth from the transform-limited regime, where the photon coherence is solely determined by the radiative lifetime of the emitter.\cite{dastidar2022quantum, So+2025, senellart2017high} TMD quantum emitters exhibit radiative lifetimes ranging from sub-nanosecond to several tens of nanoseconds,\cite{yu2021site, klein2019site, srivastava2015optically, kumar2015strain, paralikis2024tailoring, Piccinini2025, tonndorf2015single} corresponding to transform-limit linewidths ($W_\text{rad}$) between 0.03 and 4.4 $\mu$eV. In contrast, experimentally observed linewidths ($W_\text{exp}$) are often broadened to hundreds or even thousands of $\mu$eV,\cite{srivastava2015optically, Parto2021DefectK, Piccinini2025, branny2017deterministic} primarily due to phonon-assisted dephasing and spectral diffusion. Among these, spectral fluctuations from dynamic charge noise in the emitter’s environment have been identified as the dominant contributor to decoherence.\cite{vannucci2024single} Several mitigation strategies have been explored to address this challenge, including hBN encapsulation, which passivates surface defects and reduces environmental disorder, and electrostatic biasing, which stabilizes the local charge landscape.\cite{chakraborty2019electrical, lenferink2022tunable, howarth2024electroluminescent, kim2019position, klein2019site}
Although near-lifetime-limited emission and high indistinguishability have yet to be realized, TMD-based SPSs already meet the purity and brightness requirements for quantum key distribution (e.g., BB84), where indistinguishability is not strictly necessary. 
Consequently, refining charge stabilization remains the crucial next step toward unlocking the full potential of vdW materials and their heterostructures~\cite{Du_S_23,Du_NM_24} for scalable quantum photonic technologies.

%
In addition to hosting single-photon quantum emitters, 2D materials also provide unique opportunities for generating photon pairs, taking advantage of their non-centrosymmetric structures for second-order (\(\chi^{(2)}\)) processes~\cite{Du_NRP_21} and exploiting symmetry-independent third-order (\(\chi^{(3)}\)) nonlinearities~\cite{Autere_AM_18,Saynatjoki_NC_17,Kim_AM_17}.
For example, the point symmetry of a monolayer TMD such as MoS$_2$ is $\mathrm{D}_{\mathrm{3h}}$. This point symmetry group includes a combination of a three-fold rotational symmetry around the $z$-axis ($\mathrm{C}_\mathrm{3}$) and horizontal mirror plane, along with vertical mirror planes. 
As such, there are multiple nonzero elements of the $\chi^{(2)}$ nonlinear susceptibility tensor. 
In addition, the high refractive index of TMD materials can reach $n>4$, which is beneficial for photonic nanostructures and is linked to a high nonlinear susceptibility. 
Furthermore, the strong out-of-plane anisotropy could offer new opportunities for phase matching in waveguides or by twisting of stacked layers~\cite{xu2022towards, Hong_PRL_23}. 
Finally, the nonlinear susceptibility is strongly enhanced by excitonic resonances, resulting in an effective \(\chi^{(2)}\) enhancement of several thousand times~\cite{vermeulen2023post}.

These unique properties of 2D materials have led to a surge of interest in using 2D TMDs for photon-pair generation through the nonlinear process of spontaneous parametric down-conversion (SPDC). 
However, multiple constraints have been found when operating close to the excitonic resonances~\cite{marini2018constraints}. 
The key drawback has been the strong fluorescence of the materials that resulted in large background emissions from the TMDs and the inability to detect the correlations of the SPDC photons. 
This challenge has recently been circumvented using wide-bandgap 2D layered salts, such as NbOCl$_2$~\cite{guo2023ultrathin}. 
Although the wide band gap of NbOCl$_2$ reduced the coincidence background, the small volume of the materials resulted in negligible photon-pair rates. 
Therefore, multilayer 2D materials were required to detect the photon pairs with $g^{(2)}>2$, as required for a quantum light source~\cite{guo2023ultrathin}. 
In this respect, the stacking of the layers in a vdW material is of paramount importance. 
While 2H-type stacking induces a center of symmetry—causing the $\chi^{(2)}$ tensor components to vanish—3R stacking preserves the non-centrosymmetric nature of the material, ensuring that $\chi^{(2)}$ remains nonzero.
Moreover, the twisted stacking configuration introduces an additional degree of freedom, enabling both enhanced nonlinear optical responses —such as SPDC, thanks to an effectively increased crystal length— and refined control over the second-order nonlinear response~\cite{Du_M_20,Hong_PRL_23}. 
Recent works have explored this approach to demonstrate SPDC and polarization entanglement in 3R-stacked MoS$_2$ and WS$_2$ crystals~\cite{weissflog2024tunable, feng2024polarization}.

These first demonstrations have opened a plethora of new opportunities for photon-pair generation. 
These include possibilities for realizing quantum hyperentanglement or path-polarization entanglement in 2D-material photonic circuits. 
Finally, we believe that these advances will lead to the development of ultrathin devices for quantum sensing and imaging. 
Important future opportunities include integrating such quantum sources with cavities to enhance the generation efficiency. 
For example, we can envision the integration of SPDC sources with various integrated platforms (e.g., Si photonics chips~\cite{Vincent_LAM_23} or fiber-based platforms~\cite{Zuo_NN_20,Chen_NP_19}), or metasurfaces operating in free space, which could push the rates to practical application values. 
In the longer term, merging quantum sources and nonlinear down-conversion on a single chip would enable quantum state translation into telecommunication wavelengths, crucial for quantum communications and other quantum applications. 
The easily tunable nonlinear optical response of 2D materials (e.g., via optical control~\cite{Wang_AP_21,Zhang_LSA_22,Akkanen_am_2022}, strain~\cite{Liang_NL_17}, electric fields~\cite{Dai_AN_20}, or other physical methods ~\cite{Sun_NP_16}) may unlock new functionalities not achievable with conventional bulk nonlinear optical crystals. 
Ultimately, integrating these sources with trapped-exciton-based logic gates may lead to full-scale quantum computing architectures.

%
Detection of quantum light at power levels of about 10$^{-19}$ J is highly challenging~\cite{silberhorn2007detecting}, requiring extremely high detection efficiency, low dark-count rates, and high net gain. 
Commercial detectors, such as avalanche photodiodes and photomultiplier tubes based on 3D bulk materials, are commonly used in quantum light technologies. 
2D materials—due to their strong, broadband absorption and high electron mobility (e.g., 15000 cm$^2$/Vs for graphene~\cite{novoselov20162d})— have driven extensive development of 2D photodetectors~\cite{abdullah2024recent,Koppens2014}.
For example, a graphene/MoS$_2$ vdW hybrid photodetector has been employed for photon-counting via the photo-gating effect~\cite{Roy_AM_2018}. 
Leveraging a large surface-to-volume ratio and high sensitivity to localized trap states, this device exhibits high optical gain and low noise, enabling single-photon detection at \(80\) K. 
Meanwhile, superconducting states can also be exploited for single-photon detection, which was recently demonstrated at \(1550\) nm using a graphene-based Josephson junction~\cite{Walsh_Science_2021}, where graphene is sandwiched between two superconducting layers to effectively couple photons via dissipative surface plasmons. 
This finding lays the foundation for developing single-photon detectors and imaging devices based on 2D superconducting materials (including unconventional superconductors of stacked graphene and TMDs). More recent work~\cite{metuh2025singlephotondetectionsuperconductingniobium}  demonstrated the first single-photon-sensitive superconducting nanowire detector using nanostructured few-layer NbSe$_{2}$ nanowires, enabled by precise hBN encapsulation and etching. This positions nanoengineered 2D superconductors as promising candidates for ultrathin, efficient superconducting nanowire single-photon detectors in quantum photonics.

%
Quantum sensing utilizes a quantum system, quantum property, or quantum phenomenon to perform precise measurements of various weak signals. 
A wide range of solid-state quantum sensors based on spin defects embedded in diamond and silicon carbide have been successfully demonstrated and developed~\cite{Quantum_sensing}. 
However, quantum sensors embedded in 3D host materials are still limited in their ability to closely interact with external objects and are more challenging to integrate with other materials. 
In contrast, the inherently high surface-to-volume ratio in 2D materials allows their defects to interact with the external environment more effectively, thereby providing natural advantages for quantum sensing.
Indeed, quantum sensing in 2D materials—especially in hBN—has recently emerged as a promising platform for various applications. 
Spin defects in hBN enable precise measurements of physical quantities such as temperature (sensitivity: \(3.82 \textrm{ K/Hz}^{0.5}\)), pressure (sensitivity: \(17.5106 \textrm{ Pa/Hz}^{0.5}\)), magnetic fields (sensitivity: \(85.1~\mu \textrm{T/Hz}^{0.5}\)), and liquid ions (sensitivity: \(10^{-18} \textrm{ mol/Hz}^{0.5}\)), with sensor volumes on the order of a cubic sub-micrometer~\cite{Gottscholl_bn_2021,Lyu_NL_2022,Rizzato_NC_2023,Robertson_ACS_Nano_2023,Gao_NL_2021,Sasaki_APL_2023,Gao_ACS_Photonics_2023}. 
Similarly to the development of vacancies in diamond, single-spin centers in hBN have recently been demonstrated for vertical nanoscale magnetometry, achieving sensitivities in the sub-\(\mathrm{\mu} \textrm{T/Hz}^{0.5}\) range at room temperature~\cite{gilardoni2024singlespinhexagonalboron}.
hBN has also been integrated with fiber optics for quantum sensing, highlighting the high flexibility of these 2D quantum sensors compared to traditional diamond-based systems~\cite{Moon_AOM_2024}. 
Operating at room temperature, these spin-defect 2D quantum sensors constitute a versatile platform for nanoscale sensing under a wide range of conditions and hold great promise for \textit{in-situ} measurements. 
For example, because hBN is normally an insulator, a 2D quantum sensor in hBN could synchronously monitor electronic processes or internal temperatures in 2D electronic devices.

While experimental advances in 2D-based quantum photonics have been remarkable, realizing the full potential of these systems requires a deeper theoretical understanding. 
Accurately modelling excitonic interactions, light–matter coupling, and many-body effects is essential to predict device behaviour, guide materials engineering, and identify fundamental limitations. 
Next, we discuss the theoretical foundations that are critical to support and accelerate the development of scalable and tunable photonic and quantum devices based on 2D semiconductors.

\section*{Theoretical foundations}
\label{theory}
%

The design of nanophotonic systems crucially depends on both the nanoscale geometry and response function of the material -- the dielectric susceptibility. 

Given the atomic thickness of the materials used in many devices, it is generally a good approximation to describe their optical response using a frequency- and wave-vector-dependent surface conductivity $\sigma(\kpar,\omega)$. In most nanophotonic scenarios, the $\kpar$ dependence can be ignored (local approximation) and a frequency-dependent conductivity $\sigma(\omega)$ suffices to describe the linear optical response of these materials \cite{G14,G14_2}. The dependence on the in-plane wave vector $\kpar$, however, can become important when spatial features of the order of the inverse Fermi wave vector are involved. This applies to structures with nanoscale lateral extension, or when the optical fields under consideration are either scattered by small particles or generated by compact emitters (e.g., quantum dots or Raman-active molecules) placed near the 2D material. A full description of such features remains an important open challenge.

In extended layers, the optical response of 2D materials can be represented in terms of their Fresnel coefficients,
which describe the reflection and transmission of s- and p-polarized light between media on either side of the 2D material \cite{GP16}.
Poles in the Fresnel coefficients signal the presence of surface polaritons, which typically exhibit small in-plane wavelengths compared with the free-space light wavelength at the same frequency \cite{G25}. In such cases, one can adopt a quasistatic approximation ($c\to\infty$), yielding $k_{\perp j}=i\kpar$ and p-polarized polaritons that directly follow the dispersion relation $\kpar=i\omega(\epsilon_2+\epsilon_1)/4\pi\sigma$. Here, $\epsilon_1$ and $\epsilon_2$ are the permittivities of the media surrounding the polariton. This expression can be applied to explain a wide range of polaritonic behavior in 2D materials, including mode hybridization in thin (relative to the polariton wavelength) heterostructures formed by stacking different layers, where the overall surface conductivity can be approximated by the sum of the individual conductivities, $\sigma=\sum_j\sigma_j$, in the so-called zero-thickness approximation (ZTA) \cite{G25}.

In a nanostructured environment using extended layers, the local approximation can, in many circumstances, be invoked to continue using $\sigma(\omega)$ to describe the 2D material. The optical response can then be obtained either analytically for relatively simple geometries \cite{G14,G14_2} or, more generally, via numerical electromagnetic solvers, where the conductivity enters through the boundary conditions under the ZTA. Within such a classical framework, one may use either the measured local surface conductivity $\sigma(\omega)$ or a Lorentzian fit to excitonic features extracted from experiments \cite{LCZ14}. This approach yields accurate predictions for many properties of 2D semiconductors embedded in nanostructures, as validated by comparisons with measured electron energy-loss spectra \cite{WSA24}.

A central opportunity lies in extending the classical picture to a fully quantum mechanical treatment. Some of the key advantages of 2D semiconductors in nanophotonic applications arise directly from quantum mechanical effects that fundamentally shape their optical response. For example, the strong optical absorption and bright photoluminescence of monolayer TMDs result from the steady-state formation and decay of 2D excitons that are strongly quantum-confined within the monolayer. While such resonant light-matter interaction can be leveraged in nanophotonic devices and metasurfaces as a tunable optical resonance (see Section active 2D nano-optoelectronic devices), the underlying physical mechanism is distinctly different from conventional plasmon or Mie resonances, which are not affected by decoherence. In contrast, even excitons that are generated through resonant excitation (i.e. no thermalization involved) are subject to exciton-phonon scattering, which results in dephasing and re-emitted light that is only partially coherent. 

In current descriptions of nanophotonic systems taking into account their quantum nature, the surface conductivity of extended 2D semiconductors can be obtained using {\it ab initio} methods based on density functional theory (DFT), supplemented by the so-called GW and plasmon-pole approximations \cite{H65,ORR02} to amend the Kohn-Sham eigenenergies and yield reasonably accurate gap energies. In addition, the Bethe-Salpeter equation, which describes electron-hole pairs in the system, allows for the inclusion of some exciton effects \cite{RL00,PPD04}. As an example, a detailed calculation of the nonlocal optical conductivity of common TMDs is presented in Ref.~\citenum{WPI23}, applied to explain electron tunneling in the presence of semiconductor monolayers. Similarly, well-developed models exist that capture the observed 2D conductivity of graphene and TMDs~\cite{Li2018}.

These approaches, however, still ignore much of the complex, many-body aspects of interacting quasi-particles, which ultimately underlie the optical response. For example, excitonic resonances require demanding calculations to be described\cite{SKR17,AJM18}. These can rely for example on simplified few-band models and effective Bloch equations \cite{AJM18}, or on first-principles (DFT) based methods in a perturbative fashion \cite{WIR22,WIB22}. Even in these approximate treatments, that can account, for example, for quantum mechanical dephasing within the Lorentzian line shapes describing excitonic light-matter interaction~\cite{Scuri2018,Epstein2020}, other aspects, like the nonzero center-of-mass momentum of excitons, can typically not be accounted for. 

Established quantum theories that describe the electronic and excitonic properties of 2D semiconductors thus employ assumptions that ignore aspects of the photonic environment, which may in fact shape the functionality of nanophotonic devices. Devising methods for fully incorporating the existing quantum theory of semiconductors\cite{Lindberg1988,Stroucken2013,Selig2016,Selig2018} into the description of 2D excitonic devices, predicting the role of quantum states in macro-scale electronic and optical parameters, is thus a key challenge going forward. 

\begin{figure}
    \centering
    \includegraphics[width=0.8\linewidth]{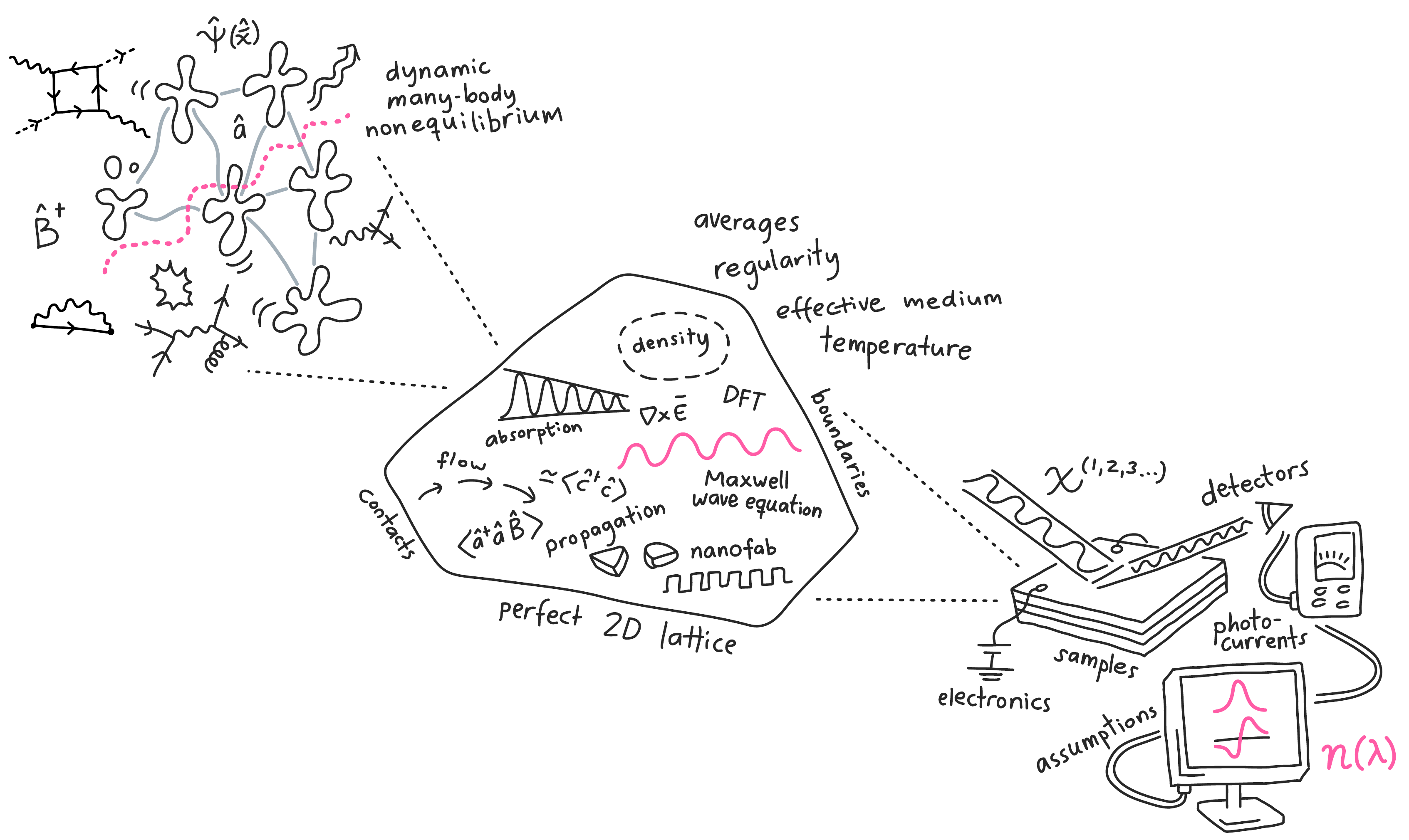}
    \caption{Schematic view of microscopic quantum properties through the lens of macroscopic measurement. The left side indicates the quantum mechanical particles and interactions that underly the macroscopic photonic responses whose experimental probing is indicated on the right side. The middle indicates the array of models and theories connecting them, all of which necessarily rely on approximations, whose elimination provides both a major challenge and a prime opportunity for the field.} 
    \label{fig:theoryScalesIllustration}
\end{figure}

A unified quantum theory that combines first-principle models with a comprehensive description of the nanoscale light-matter interaction (Fig.~\ref{fig:theoryScalesIllustration}) would enable experimental validation of quantum models of 2D semiconductors. It would pave the way to more powerful prediction and analytical tools, including detailed descriptions of the optical response beyond the linear response theory for nonlinear 2D nanophotonics.

Interestingly, the nonlinear response of 2D semiconductors is remarkably strong when normalized to the material volume \cite{Maduro2022,Bolhuis2020}. In particular, nonlinear effects are substantially enhanced by excitonic resonances \cite{SKR17,AJM18}. In the context of nonlinear nanophotonics, the response of a 2D material can be modeled using effective nonlinear surface susceptibilities, which are incorporated into the response of a nanostructure using perturbation theory. This approach is commonly employed to interpret experiments, in which the susceptibilities are either treated as adjustable parameters or extracted from optical measurements.

Because the observation of nonlinear effects typically requires intense optical fields, the electronic band populations can be substantially modified, leading to an interesting interplay between ultrafast carrier dynamics and the nonlinear response. Such phenomena have been experimentally explored as a means to control harmonic generation in a pump-probe fashion. Due to the complexity of this behavior, heuristic approximations have been employed to reduce the computational demand of the simulations \cite{WIR22,WIB22}. In general, the interplay between elastic and inelastic processes in ultrafast carrier dynamics remains an unsolved problem, both from first-principles and from rigorous phenomenological perspectives. This sets a key challenge for the field, as it is an essential ingredient in a comprehensive understanding of the nonlinear optical response, particularly in 2D semiconductors.

More generally, typical theoretical descriptions of 2D quantum materials assume idealized conditions that do not reflect the full experimental reality. To make quantitative predictions for practical configurations within devices, for example, finite-size effects and interactions with a substrate are essential. This becomes especially important when accounting for the interference of light emitted from the material of interest with, for example, substrate reflections~\cite{vandegroep2023,Scuri2018}. Similarly, the effect of inhomogeneities such as charge puddles, local defects, and material roughness on the observed light field will require dedicated theoretical tools and techniques to model.

At the same time, typical experiments may be unable to differentiate between effects that appear entirely distinct in theoretical models. Separate predictions for intensities of coherent and incoherent radiation, for example, require dedicated tools to disentangle in a single observation of emitted intensity in any realistic experiment, even if the difficulty of separating the pump field from emitted radiation can be overcome. Similarly, a momentum-resolved computation or one including multiple radiative and non-radiative channels may offer physical insight into the microscopic processes underlying light-matter interaction in 2D devices~\cite{Fitzgerald:24}, but can only be connected to experimental observation if benchmarked against quantities beyond system averages such as photoluminescence and radiative rates.


One also needs to include edge effects in the description of mesoscopic patterns ranging between the atomic and wavelength scales. This is particularly important for the modeling of devices where electronic contacts are made at the edges (see section heterostructure engineering), as well as atomic-scale defects. Bulk interfaces in 2D materials, in particular those formed by stacking~\cite{Kennes2021Moire} or functionalization~\cite{Bhimanapati2015Recent,Greten2024Dipolar}, are already well studied, motivated by their practical applications. However, while there exist specialized theories that address edge or defect-related~\cite{Gjerding2021, Sajid2020Single, Yazyev2010Emergence, Ghorashi2024Highly, Huang2022Carbon, Vinichenko2017Accurate,Sajid2020V} phenomena, there remains potential for integration into more holistic models of realistic devices.

Specifically on the mesoscopic scale, a consequence of miniaturization is that photonic devices~\cite{Kurman2020Tunable,Yan20202d} (see section heterostructure engineering) have entered a regime in which the spatial extent of electronic wave functions can approach or exceed the scale of the nanopatterned electric fields. Furthermore, momentum-resolved descriptions reveal that electronic excitations become delocalized or may propagate through the material~\cite{Alpeggiani2014Semiclassical,Lundeberg2017Tuning,Rosner2016Two-dimensional}. In these cases, a local approximation, where the dipole density at a position is determined only by the electric field at that same location, becomes inadequate. 

Going beyond the modelling of individual nanophotonic experiments, an even greater challenge lies in the inclusion of locally tunable parameters and out-of-equilibrium conditions accessible in practical realizations. For example, control over the local density of electrons can be obtained by local gating or nanopatterning, and similarly, local gradients or variations in strain, as well as applied electric and magnetic fields, may be purposely designed and controlled in nanodevices. Alternatively, dynamically driven systems (pump-probe, or Floquet) add additional challenges to the exciton dynamics that require time-resolved descriptions of the material properties. Such external tuning parameters can significantly impact both the outgoing radiation and internal exciton dynamics of nanophotonic devices, for example, causing (local) enhancements of the Purcell factor as well as exciton resonance (as described in section active 2D nano-optoelectronic devices).

Connecting microscopic and macroscopic theories within numerical tools, such that they can be applied to calculate measurable quantities that are directly relevant for existing and envisioned experimental configurations and devices, is a crucial overarching challenge for the field.

\section*{Scaling towards real-world photonic applications}
\label{scaling}
%

To take full advantage of the opportunities that 2D semiconductors offer for active devices, heterostructure engineering, valleytronics, or quantum photonics, their implementation needs to become scalable as well. However, the transition of 2D semiconductors from laboratory research to industrial-scale photonic applications faces practical challenges. The requirements and primary bottlenecks of 2D materials for photonics vary depending on the specific applications, which can be classified into four categories: nanophotonics, optoelectronic devices, integrated photonics, and energy conversion applications. Specifically, the 2D semiconductors employed for nanophotonic components such as metasurfaces and photonic crystals typically require lateral sizes below 100 micrometers due to the use of resonant elements. These applications require a high crystal quality for a high quantum efficiency or a narrow exciton linewidth. Optoelectronic devices based on 2D semiconductors, including modulators, photodetectors, LEDs, sensors, and spectrometers, can have lateral dimensions of up to tens or hundreds of micrometers. These devices benefit particularly from vdW heterostructures and appropriate encapsulation. Photonic integrated circuits can integrate 2D-semiconductor-based active devices with classical Si or SiN or other integrated photonic platforms. Alternatively, photonic circuits with all passive and active components made of 2D semiconductors are also being explored. These applications require wafer-scale fabrication of device arrays with high yield and reproducibility, as well as low losses in 2D semiconductor-based circuits. Energy conversion applications, including photocatalysis, photovoltaics, and fiber-based saturable absorbers, have lower crystal quality requirements and can even benefit from the presence of defects. In this Section, we describe the three key challenges for the translation of 2D semiconductors to the photonics industry: 1) Wafer-scale crystal growth and transfer methods; 2) Reproducible interface quality and passivation; 3) Device scalability. Many of these challenges overlap with those in electronics. Therefore, industrial advances in 2D semiconductor electronics can directly benefit photonics. 

\begin{figure}
\centering
\includegraphics[width=0.75\textwidth]{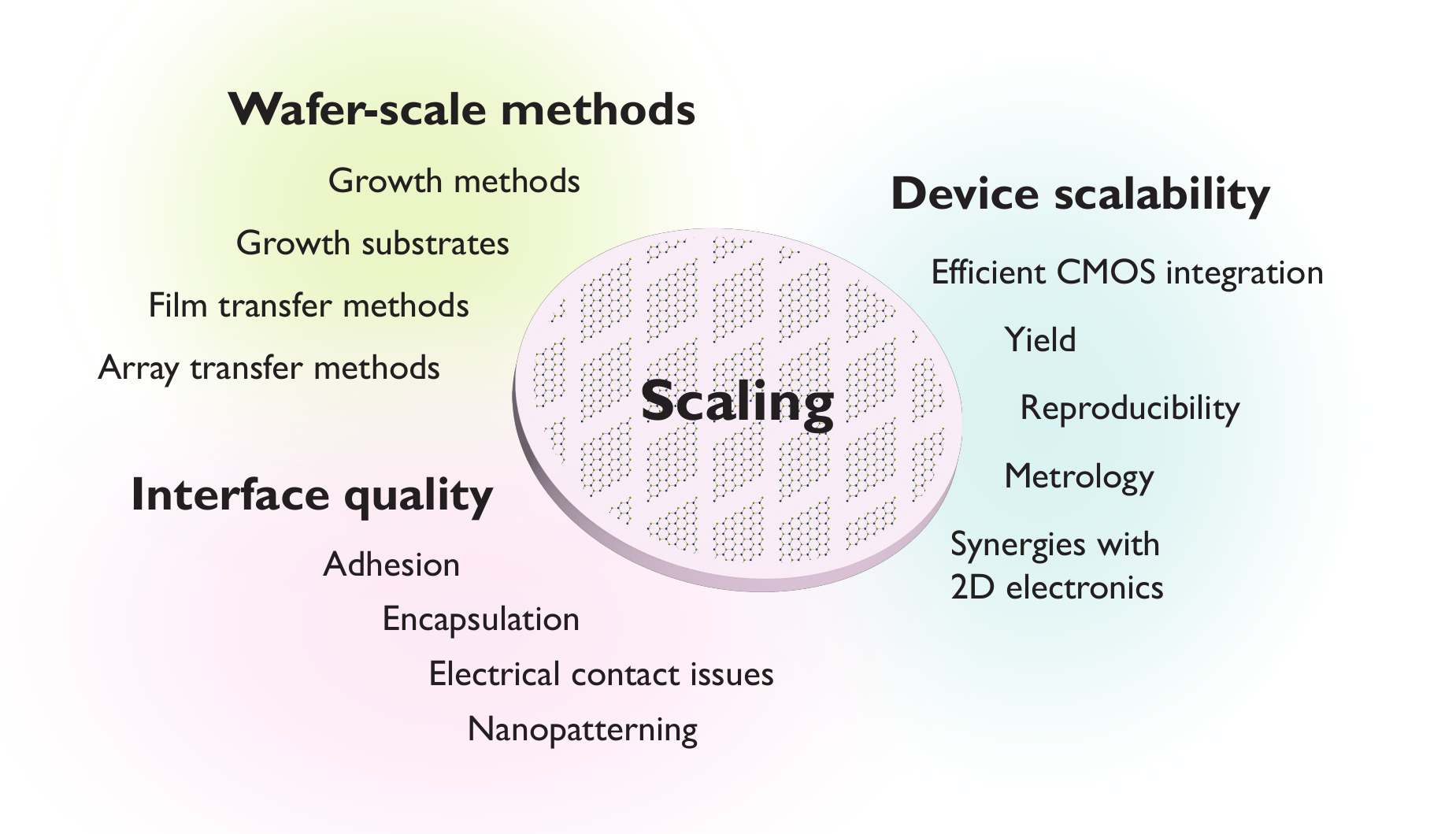}
    \caption[]{Challenges and opportunities for the scalability of 2D semiconductors for photonics applications.}
\label{scaling}
\end{figure}

%
%

The crystal quality of 2D materials, which is largely given by the chosen growth method, is a critical factor in determining device performance. Nevertheless, developing methods to grow large-area monolayers with high quantum efficiency remains a challenge. Techniques such as CVD, molecular beam epitaxy (MBE), and atomic layer deposition (ALD) still require further optimization to achieve uniform quality films across entire wafers. CVD is a scalable method for synthesizing 2D materials with near-intrinsic quality. The CVD growth of high-quality, wafer-scale monolayer MoS\textsubscript{2} on an 8-inch sapphire substrate was recently demonstrated~\cite{Yu2024_2}, although the material contained domains with antiparallel alignments. However, the high temperatures required for CVD make it incompatible with direct growth on preprocessed silicon substrates. This limitation has been mitigated using metal-organic precursors that vaporize at lower temperatures\cite{Lee2020}. Combining top-down and bottom-up methods, such as lithography and CVD, enables controllable wafer-scale growth\cite{Han2015}. MBE offers superior crystallinity and controllability of the film thickness and composition, and has proven effective for the growth of monoelemental materials like Te. However, this technique, in addition to its high cost, exhibits limitations when applied to S- and Se-based 2D materials. Large-area ALD growth is better suited for applications requiring large-area deposition and low defect density, such as photocatalysis. Finally, Au-assisted exfoliation of millimeter-scale monolayers is a viable approach for producing high-quality, medium-sized S- and Se-based 2D semiconductors, though it can suffer from strain and cracking during film transfer. For energy conversion and photocatalysis applications, the large-area deposition of active TMD layers, without restriction to monolayer thickness, is essential. Direct deposition of homogeneous TMD films with controlled thickness in the few-layer regime has been achieved by physical deposition (ion beam sputtering) of the precursor film from stoichiometric TMD targets or transition metal targets, followed by thermal annealing in an atmosphere rich in S~\cite{Ferrando2023, Martella2017}.
Following this approach, vertical stacks of dissimilar TMD layers have also been demonstrated.
These large-area vdW heterostructures feature type-II band alignment, which enables photoconversion (photovoltaic effect) and enhances photocatalytic efficiency by increasing the lifetime of photogenerated carriers, which are spatially separated at the junction~\cite{Gardella2024}.

%
The growth substrate, in addition to the growth method, can also have a major impact on the quality of the crystal and, by extension, the device. Silicon is widely used in industry because of its well-established processing techniques. However, its lattice mismatch with the hexagonal lattice of TMDs often leads to the formation of defects that act as charge carrier traps, which in turn adversely affect the photoluminescence behavior~\cite{Yan2018}. 
Alternatively, substrates with lattice structures more compatible with MoS\textsubscript{2}, such as c-plane sapphire, have been utilized to facilitate epitaxial growth and minimize defect density. Beyond sapphire, Xia \textit{et al.}~\cite{Xia2023} reported the growth of monolayer MoS\textsubscript{2} on a 12-inch fused silica substrate. Using a thin Al\textsubscript{2}O\textsubscript{3} seeding layer and optimizing the delivery of metal oxide precursors, they achieved precise control over domain alignment and film quality. The successful fabrication of 37 field-effect transistors (FETs) from the resulting wafer underscores the potential for scalability and industrial viability of the process.
%
One last aspect of growth that demands attention is increasing spectral coverage. Currently available 2D semiconductors, particularly TMDs, are limited in effectively covering the visible-to-near-infrared spectrum, exhibiting high quantum efficiency only at discrete wavelengths. Although TMD alloys can extend the accessible range, achieving more continuous spectral coverage will require the integration of additional materials.

Scalable transfer techniques are another major need because monolayer 2D materials are typically grown on a substrate that differs from the one used in final device applications. In particular, the high temperature required for CVD often necessitates a transfer process from growth to device substrates. Wet transfer methods, which use polymeric carriers like poly(methyl methacrylate) (PMMA) or polycarbonate, mechanically support the 2D material during transfer. This approach allows placement on target substrates, including CMOS-compatible wafers, but introduces issues such as defects, wrinkles, unintentional doping, strain, and polymer residue, which can degrade the electronic and optical properties of the material~\cite{Song2021}.
Efforts to mitigate these problems include bubble delamination, advanced cleaning methods, and the use of alternative carriers such as paraffin. However, these strategies have not fully overcome the associated challenges. Dry transfer techniques, which use rollers, laminators, or hot presses, avoid the submersion of substrates in liquids and also allow for the reuse of metallic growth substrates. 
Although beneficial for integrating 2D materials into certain devices, these methods often lead to microcracks, wrinkles, and reduced charge carrier mobility due to residual contamination. Mechanical cleaning techniques help address localized contaminants and restore material properties on the micrometer scale. For example, delaminating single-crystal CVD graphene from copper foil using hBN stamps can preserve nearly intrinsic graphene properties, though this method is currently limited to small areas.

Instead of starting with a transferred 2D film, an alternative approach to produce multiple devices in parallel is fabricating arrays of components on a convenient substrate and then transferring them to a target device wafer. Such array transfer techniques offer a promising route for scalable processing. To date, a variety of vdW array integration strategies have been reported~\cite{Mannix2022, Liu2022, Yang2023, Huang2022, Mootheri2021}. These methods primarily manipulate interface adhesion between the transfer stamp, the 2D film, and the target substrate, often by tuning the viscosity and mechanical stiffness of the stamps. In addition to using adhesion layer materials such as self-assembled monolayers~\cite{Nguyen2023}, future transfer approaches may leverage monolayer or device encapsulation to better preserve material quality during transfer.
Large-area transfer can introduce non-uniform strain across the film, which could require working with smaller dies to maintain strain under control.

Improving the adhesion of 2D materials to bulk 3D substrates is crucial for device stability and performance, which requires advances in new adhesive approaches and surface treatments. Annealing is usually applied to improve interface quality, but thermal budget limitations must be taken into account; for example, temperatures should not exceed 450 °C for back-end-of-line CMOS-compatible processing.

For the encapsulation of 2D semiconductors, such as sandwiching in hBN or capping with Al$_2$O$_3$ deposited by ALD, both heterogeneous compatibility and interface quality are crucial. Another critical issue is the Schottky barrier generated at the TMD-electrode interface, where contact resistance can be reduced by tunnel barriers, contact-area doping, or optimizing contact geometry~\cite{jung2019transferred}.

Effective patterning strategies to create atomically precise nanostructures in 2D semiconductors, using bottom-up or top-down approaches, are an active area of research. A key challenge is minimizing defect formation during fabrication. Thermal scanning probe lithography (t-SPL), applied to large-area TMDs grown via physical deposition, has emerged as a promising method for the deterministic fabrication of nanostructure arrays and nanocircuits~\cite{Giordano2023}. Alternatively, laser interference lithography enables the scalable production of periodic TMD nanoarrays over wafer-scale areas, both on flat and nanostructured substrates~\cite{Bhatnagar2021}. There are also opportunities to develop techniques to directly grow 2D nanostructures on photonic components in integrated circuits. This approach could streamline device fabrication by avoiding transfer steps with the potential to simultaneously enhance light-matter interaction using electromagnetic waveguiding and confinement.

The slow transfer processes currently used are not suitable for the fast-paced CMOS manufacturing environment. Therefore, streamlining these processes and ensuring compatibility with existing CMOS technologies is crucial for efficiency in CMOS integration. To take advantage of 2D materials in commercial devices, they must be integrated into established semiconductor fabrication processes, which offer the benefits of low-cost and high-volume manufacturing on large silicon substrates. 
However, several challenges remain to achieve this integration. CVD is a scalable method for synthesizing 2D materials with near-intrinsic quality. However, the high temperatures required for CVD make it incompatible with direct growth on preprocessed silicon substrates, necessitating a transfer process from growth to device substrates. Wafer-scale transfer with high reproducibility is needed, and enhancing the speed and accuracy of these methods would help to bridge the gap between the lab and industry. Despite significant progress, no existing transfer methods are fully compatible with industrial-scale manufacturing while preserving the high quality of 2D materials as grown on their substrates ---an essential requirement for many applications~\cite{quellmalz2021large}. Furthermore, effectively stacking fabricated 2D layers on a large scale remains a significant challenge~\cite{Liao_NC_20}.

Device yield refers to the proportion of devices that function correctly, meeting the specifications and tolerance limits, out of the total devices tested. It serves as a critical measure of the quality of the fabrication process and the maturity of integrated devices. Device-to-device variability, on the other hand, represents the variation in key device parameters, such as carrier mobility or gate oxide leakage in FETs, responsivity and optoelectronic bandwidths in photodetectors, and switching voltages in memristors. These variations are typically assessed using the coefficient of variance, which is the ratio of the standard deviation to the mean. Both yield and variability are influenced by defects introduced during the fabrication process, such as those occurring during material synthesis, storage, transfer, patterning, and material deposition. In 2D-material-based devices, intrinsic defects include vacancies, impurities, atomic misalignments, strained bonds, cracks, wrinkles, and thickness inconsistencies. In contrast, extrinsic defects arise from environmental interactions that affect adhesion and compatibility with surrounding materials. Minor defects may alter device performance and increase variability within acceptable limits, but more severe defects can lead to device failure, reducing the yield~\cite{lanza2020yield}. Achieving high reproducibility in device performance is essential to ensure functional reliability in practical applications. Addressing variability in fabrication, particularly in heterostructures, is critical for applications requiring tunability.

Quick and easy metrology techniques for 2D materials in industry are required for process development using specific metrics, particularly to compare interface quality. The characterization of contaminants on the few-nanometer scale is thus needed. As they are non-destructive and fast, optical techniques could prove scalable when tailored to the metrology needs of 2D materials in an industrial environment, including Raman scattering, dark-field scattering, nonlinear microscopy, or photoluminescence spectroscopy~\cite{Raja2019}, lifetime~\cite{eizagirre2019}, and fluctuation imaging~\cite{Godiksen2020}. The difficulties in large-area growth and transfer imply the requirement for characterization at the wafer scale of monolayer character, continuity, and defects. Similarly to electronics, metrology is also needed for characterizing nanostructures and patterned devices. Electron microscopy will also play an important role in characterizing defects and nanostructures.

Exploiting the synergies with 2D-semiconductor electronics will be fundamental for their success in photonics. 2D semiconductors are considered promising candidates for ultra-scaled MOSFETs, which alleviate the critical issue of off-state current leakage in silicon transistors with gate lengths below 10 nm. At this scale, MoS$_2$ transistors are predicted to exhibit a leakage current of more than two orders of magnitude lower than their silicon counterparts and show less degradation in carrier mobility as channel thickness decreases. Moreover, the vertical stacking of 2D materials creates vdW heterostructures, introducing novel material properties that arise from interactions between the stacked layers. Driven by this strong motivation for next-generation devices, developments in the TMD electronics industry are poised to spill over to the application prospects of 2D semiconductors in photonics. In summary, scalability for applications in photonics relies on developments in uniformity across large areas because of its impact on device reliability and performance. Transfer processes are central, and advances in high-throughput methods will be welcome developments. Standardization of fabrication and metrology are necessary steps to reduce costs and improve yield to realize the promise of 2D semiconductors in photonics.

\section*{Conclusions \& outlook}
\label{conclusion}
2D semiconductors have established themselves as a promising platform for active and tunable nanophotonic and quantum photonic devices. 
Their strong excitonic resonances, dynamic tunability, and compatibility with heterogeneous integration offer compelling opportunities for future technologies.
However, realizing the full potential of 2D materials requires addressing critical challenges across multiple fronts.

Fundamental limitations in light–matter interaction strength and optical efficiency must be mitigated through improved device architectures, such as resonant nanostructures and hybrid systems. 
Scalable synthesis techniques must overcome inhomogeneities to ensure reproducibility and uniformity over large areas. 
Integration of 2D materials into complex photonic and electronic platforms demands advances in fabrication, alignment, and material handling to preserve optical quality. 
Furthermore, a deeper theoretical understanding of excitonic dynamics, many-body interactions, and valley degrees of freedom will be essential to guide device design and predict performance limits.

Looking ahead, interdisciplinary approaches combining materials science, nanofabrication, device engineering, and theoretical modelling will be crucial. 
Emerging areas such as electrically tunable metasurfaces, valleytronic circuits, quantum light sources, and hybrid photonic–electronic integration represent particularly promising directions. 
As fabrication techniques mature and theoretical models become more predictive, 2D semiconductor-based photonics is poised to transition from proof-of-concept demonstrations to scalable technologies with impact in quantum communications, imaging, sensing, and information processing.

Continued progress will require coordinated efforts across academia, industry, and national laboratories to bridge fundamental discoveries with system-level applications.
By harnessing the unique properties of 2D materials and overcoming current bottlenecks, the field is well-positioned to shape the future landscape of photonics and quantum technologies.

\section*{Acknowledgements}
This perspective is the result of a Lorentz workshop that acknowledges funding from the Lorentz Center, the Dutch Research Council (NWO) through the Lorentz Center and Vidi project (VI.Vidi.203.027), Leiden University, the University of Amsterdam, and Ghent University through the financial support of the European Research Council (ERC) under the European Union’s Horizon 2020 Research and Innovation Program (Grant Agreement 948804, CHANSON). Z.F., G.S., A.B. and I.S. acknowledge funding by the Deutsche Forschungsgemeinschaft (DFG, German Research Foundation), Project-ID: 437527638 – IRTG 2675 (Meta-Active). S.C.-B. acknowledge financial support from the Dutch Research Council (NWO) via a Vidi Grant (VI.Vidi.213.159).

\section*{Author contributions statement}
All authors contributed to the writing of the manuscript, coordinated by the corresponding authors.



\clearpage

\bibliography{main}

\begin{thebibliography}{100}
\urlstyle{rm}
\expandafter\ifx\csname url\endcsname\relax
  \def\url#1{\texttt{#1}}\fi
\expandafter\ifx\csname urlprefix\endcsname\relax\def\urlprefix{URL }\fi
\expandafter\ifx\csname doiprefix\endcsname\relax\def\doiprefix{DOI: }\fi
\providecommand{\bibinfo}[2]{#2}
\providecommand{\eprint}[2][]{\url{#2}}

\bibitem{radisavljevic2011single}
\bibinfo{author}{Radisavljevic, B.}, \bibinfo{author}{Radenovic, A.}, \bibinfo{author}{Brivio, J.}, \bibinfo{author}{Giacometti, V.} \& \bibinfo{author}{Kis, A.}
\newblock \bibinfo{journal}{\bibinfo{title}{Single-layer {MoS}$_2$ transistors}}.
\newblock {\emph{\JournalTitle{Nature Nanotechnology}}} \textbf{\bibinfo{volume}{6}}, \bibinfo{pages}{147--150}, \doiprefix\url{10.1038/nnano.2010.279} (\bibinfo{year}{2011}).

\bibitem{srivastava2015optically}
\bibinfo{author}{Srivastava, A.} \emph{et~al.}
\newblock \bibinfo{journal}{\bibinfo{title}{Optically active quantum dots in monolayer {WSe}$_2$}}.
\newblock {\emph{\JournalTitle{Nature Nanotechnology}}} \textbf{\bibinfo{volume}{10}}, \bibinfo{pages}{491--496}, \doiprefix\url{10.1038/nnano.2015.60} (\bibinfo{year}{2015}).

\bibitem{novoselov2005two}
\bibinfo{author}{Novoselov, K.~S.} \emph{et~al.}
\newblock \bibinfo{journal}{\bibinfo{title}{Two-dimensional atomic crystals}}.
\newblock {\emph{\JournalTitle{Proceedings of the National Academy of Sciences}}} \textbf{\bibinfo{volume}{102}}, \bibinfo{pages}{10451--10453}, \doiprefix\url{10.1073/pnas.0502848102} (\bibinfo{year}{2005}).

\bibitem{Sun_NP_16}
\bibinfo{author}{Sun, Z.}, \bibinfo{author}{Martinez, A.} \& \bibinfo{author}{Wang, F.}
\newblock \bibinfo{journal}{\bibinfo{title}{Optical modulators with {2D} layered materials}}.
\newblock {\emph{\JournalTitle{Nature Photonics}}} \textbf{\bibinfo{volume}{10}}, \bibinfo{pages}{227--238}, \doiprefix\url{10.1038/nphoton.2016.15} (\bibinfo{year}{2016}).

\bibitem{Dai_AN_20}
\bibinfo{author}{Dai, Y.} \emph{et~al.}
\newblock \bibinfo{journal}{\bibinfo{title}{Electrical control of interband resonant nonlinear optics in monolayer {MoS}$_2$}}.
\newblock {\emph{\JournalTitle{ACS Nano}}} \textbf{\bibinfo{volume}{14}}, \bibinfo{pages}{8442--8448}, \doiprefix\url{10.1021/acsnano.0c02642} (\bibinfo{year}{2020}).

\bibitem{Liang_NL_17}
\bibinfo{author}{Liang, J.} \emph{et~al.}
\newblock \bibinfo{journal}{\bibinfo{title}{Monitoring local strain vector in atomic-layered {MoSe}$_2$ by second-harmonic generation}}.
\newblock {\emph{\JournalTitle{Nano Letters}}} \textbf{\bibinfo{volume}{17}}, \bibinfo{pages}{7539--7543}, \doiprefix\url{10.1021/acs.nanolett.7b03476} (\bibinfo{year}{2017}).

\bibitem{Du_NRP_21}
\bibinfo{author}{Du, L.} \emph{et~al.}
\newblock \bibinfo{journal}{\bibinfo{title}{Engineering symmetry breaking in {2D} layered materials}}.
\newblock {\emph{\JournalTitle{Nature Reviews Physics}}} \textbf{\bibinfo{volume}{3}}, \bibinfo{pages}{193--206}, \doiprefix\url{10.1038/s42254-020-00276-0} (\bibinfo{year}{2021}).

\bibitem{wang2021stacking}
\bibinfo{author}{Wang, S.} \emph{et~al.}
\newblock \bibinfo{journal}{\bibinfo{title}{Stacking-engineered heterostructures in transition metal dichalcogenides}}.
\newblock {\emph{\JournalTitle{Advanced Materials}}} \textbf{\bibinfo{volume}{33}}, \bibinfo{pages}{2005735}, \doiprefix\url{10.1002/adma.202005735} (\bibinfo{year}{2021}).

\bibitem{Guo_AM_23}
\bibinfo{author}{Guo, X.} \emph{et~al.}
\newblock \bibinfo{journal}{\bibinfo{title}{Polaritons in van der {Waals} heterostructures}}.
\newblock {\emph{\JournalTitle{Advanced Materials}}} \textbf{\bibinfo{volume}{35}}, \bibinfo{pages}{2201856}, \doiprefix\url{10.1002/adma.202201856} (\bibinfo{year}{2023}).

\bibitem{xiao2012coupled}
\bibinfo{author}{Xiao, D.}, \bibinfo{author}{Liu, G.-B.}, \bibinfo{author}{Feng, W.}, \bibinfo{author}{Xu, X.} \& \bibinfo{author}{Yao, W.}
\newblock \bibinfo{journal}{\bibinfo{title}{Coupled spin and valley physics in monolayers of {MoS}$_2$ and other group-{VI} dichalcogenides}}.
\newblock {\emph{\JournalTitle{Physical Review Letters}}} \textbf{\bibinfo{volume}{108}}, \bibinfo{pages}{196802}, \doiprefix\url{10.1103/PhysRevLett.108.196802} (\bibinfo{year}{2012}).

\bibitem{Mak2012PL}
\bibinfo{author}{Mak, K.~F.}, \bibinfo{author}{He, K.}, \bibinfo{author}{Shan, J.} \& \bibinfo{author}{Heinz, T.~F.}
\newblock \bibinfo{journal}{\bibinfo{title}{Control of valley polarization in monolayer {MoS}$_2$ by optical helicity}}.
\newblock {\emph{\JournalTitle{Nature Nanotechnology}}} \textbf{\bibinfo{volume}{7}}, \bibinfo{pages}{494--498} (\bibinfo{year}{2012}).

\bibitem{cao2012valley}
\bibinfo{author}{Cao, T.} \emph{et~al.}
\newblock \bibinfo{journal}{\bibinfo{title}{Valley-selective circular dichroism of monolayer molybdenum disulphide}}.
\newblock {\emph{\JournalTitle{Nature Communications}}} \textbf{\bibinfo{volume}{3}}, \bibinfo{pages}{887}, \doiprefix\url{10.1038/ncomms1882} (\bibinfo{year}{2012}).

\bibitem{Wang_AP_21}
\bibinfo{author}{Wang, Y.} \emph{et~al.}
\newblock \bibinfo{journal}{\bibinfo{title}{Giant all-optical modulation of second-harmonic generation mediated by dark excitons}}.
\newblock {\emph{\JournalTitle{ACS Photonics}}} \textbf{\bibinfo{volume}{8}}, \bibinfo{pages}{2320--2328}, \doiprefix\url{10.1021/acsphotonics.1c00466} (\bibinfo{year}{2021}).

\bibitem{he2015single}
\bibinfo{author}{He, Y.-M.} \emph{et~al.}
\newblock \bibinfo{journal}{\bibinfo{title}{Single quantum emitters in monolayer semiconductors}}.
\newblock {\emph{\JournalTitle{Nature Nanotechnology}}} \textbf{\bibinfo{volume}{10}}, \bibinfo{pages}{497--502}, \doiprefix\url{10.1038/nnano.2015.75} (\bibinfo{year}{2015}).

\bibitem{iff2021purcell}
\bibinfo{author}{Iff, O.} \emph{et~al.}
\newblock \bibinfo{journal}{\bibinfo{title}{Purcell-enhanced single photon source based on a deterministically placed {WSe}$_2$ monolayer quantum dot in a circular {B}ragg grating cavity}}.
\newblock {\emph{\JournalTitle{Nano Letters}}} \textbf{\bibinfo{volume}{21}}, \bibinfo{pages}{4715--4720}, \doiprefix\url{10.1021/acs.nanolett.1c00978} (\bibinfo{year}{2021}).

\bibitem{zhao2021site}
\bibinfo{author}{Zhao, H.}, \bibinfo{author}{Pettes, M.~T.}, \bibinfo{author}{Zheng, Y.} \& \bibinfo{author}{Htoon, H.}
\newblock \bibinfo{journal}{\bibinfo{title}{Site-controlled telecom-wavelength single-photon emitters in atomically-thin {MoTe}$_2$}}.
\newblock {\emph{\JournalTitle{Nature Communications}}} \textbf{\bibinfo{volume}{12}}, \bibinfo{pages}{6753}, \doiprefix\url{10.1038/s41467-021-27033-w} (\bibinfo{year}{2021}).

\bibitem{shields2007semiconductor}
\bibinfo{author}{Shields, A.~J.}
\newblock \bibinfo{journal}{\bibinfo{title}{Semiconductor quantum light sources}}.
\newblock {\emph{\JournalTitle{Nature Photonics}}} \textbf{\bibinfo{volume}{1}}, \bibinfo{pages}{215--223}, \doiprefix\url{10.1038/nphoton.2007.46} (\bibinfo{year}{2007}).

\bibitem{wang2020integrated}
\bibinfo{author}{Wang, J.}, \bibinfo{author}{Sciarrino, F.}, \bibinfo{author}{Laing, A.} \& \bibinfo{author}{Thompson, M.~G.}
\newblock \bibinfo{journal}{\bibinfo{title}{Integrated photonic quantum technologies}}.
\newblock {\emph{\JournalTitle{Nature Photonics}}} \textbf{\bibinfo{volume}{14}}, \bibinfo{pages}{273--284}, \doiprefix\url{10.1038/s41566-019-0532-1} (\bibinfo{year}{2020}).

\bibitem{maring2024versatile}
\bibinfo{author}{Maring, N.} \emph{et~al.}
\newblock \bibinfo{journal}{\bibinfo{title}{A versatile single-photon-based quantum computing platform}}.
\newblock {\emph{\JournalTitle{Nature Photonics}}} \textbf{\bibinfo{volume}{18}}, \bibinfo{pages}{603--609}, \doiprefix\url{10.1038/s41566-024-01403-4} (\bibinfo{year}{2024}).

\bibitem{Liao_NC_20}
\bibinfo{author}{Liao, M.} \emph{et~al.}
\newblock \bibinfo{journal}{\bibinfo{title}{Precise control of the interlayer twist angle in large scale {MoS}$_2$ homostructures}}.
\newblock {\emph{\JournalTitle{Nature Communications}}} \textbf{\bibinfo{volume}{11}}, \bibinfo{pages}{2153}, \doiprefix\url{10.1038/s41467-020-16056-4} (\bibinfo{year}{2020}).

\bibitem{Hu_NC_17}
\bibinfo{author}{Hu, D.} \emph{et~al.}
\newblock \bibinfo{journal}{\bibinfo{title}{Probing optical anisotropy of nanometer-thin van der {Waals} microcrystals by near-field imaging}}.
\newblock {\emph{\JournalTitle{Nature Communications}}} \textbf{\bibinfo{volume}{8}}, \bibinfo{pages}{1471}, \doiprefix\url{10.1038/s41467-017-01580-7} (\bibinfo{year}{2017}).

\bibitem{Vincent_LAM_23}
\bibinfo{author}{Pelgrin, V.}, \bibinfo{author}{Yoon, H.~H.}, \bibinfo{author}{Cassan, E.} \& \bibinfo{author}{Sun, Z.}
\newblock \bibinfo{journal}{\bibinfo{title}{Hybrid integration of {2D} materials for on-chip nonlinear photonics}}.
\newblock {\emph{\JournalTitle{Light: Advanced Manufacturing}}} \textbf{\bibinfo{volume}{4}}, \bibinfo{pages}{311}, \doiprefix\url{10.37188/lam.2023.014} (\bibinfo{year}{2023}).

\bibitem{Du_NM_24}
\bibinfo{author}{Du, L.} \emph{et~al.}
\newblock \bibinfo{journal}{\bibinfo{title}{Nonlinear physics of moir\'{e} superlattices}}.
\newblock {\emph{\JournalTitle{Nature Materials}}} \textbf{\bibinfo{volume}{23}}, \bibinfo{pages}{1179--1192}, \doiprefix\url{10.1038/s41563-024-01951-8} (\bibinfo{year}{2024}).

\bibitem{Lin_NC_24}
\bibinfo{author}{Lin, Q.} \emph{et~al.}
\newblock \bibinfo{journal}{\bibinfo{title}{Moiré-engineered light-matter interactions in {MoS}$_2$/{WSe}$_2$ heterobilayers at room temperature}}.
\newblock {\emph{\JournalTitle{Nature Communications}}} \textbf{\bibinfo{volume}{15}}, \bibinfo{pages}{8762}, \doiprefix\url{10.1038/s41467-024-53083-x} (\bibinfo{year}{2024}).

\bibitem{Akkanen_am_2022}
\bibinfo{author}{Akkanen, S.-T.~M.}, \bibinfo{author}{Fernandez, H.~A.} \& \bibinfo{author}{Sun, Z.}
\newblock \bibinfo{journal}{\bibinfo{title}{Optical modification of {2D} materials: Methods and applications}}.
\newblock {\emph{\JournalTitle{Advanced Materials}}} \textbf{\bibinfo{volume}{34}}, \bibinfo{pages}{2110152}, \doiprefix\url{10.1002/adma.202110152} (\bibinfo{year}{2022}).

\bibitem{Du_S_23}
\bibinfo{author}{Du, L.} \emph{et~al.}
\newblock \bibinfo{journal}{\bibinfo{title}{Moiré photonics and optoelectronics}}.
\newblock {\emph{\JournalTitle{Science}}} \textbf{\bibinfo{volume}{379}}, \bibinfo{pages}{eadg0014}, \doiprefix\url{doi:10.1126/science.adg0014} (\bibinfo{year}{2023}).

\bibitem{Yang2015Kerr}
\bibinfo{author}{Yang, L.} \emph{et~al.}
\newblock \bibinfo{journal}{\bibinfo{title}{Long-lived nanosecond spin relaxation and spin coherence of electrons in monolayer {MoS}$_2$ and {WS}$_2$}}.
\newblock {\emph{\JournalTitle{Nature Physics}}} \textbf{\bibinfo{volume}{11}}, \bibinfo{pages}{830--834}, \doiprefix\url{10.1038/nphys3419} (\bibinfo{year}{2015}).

\bibitem{Herrmann2023SHG}
\bibinfo{author}{Herrmann, P.} \emph{et~al.}
\newblock \bibinfo{journal}{\bibinfo{title}{Nonlinear all-optical coherent generation and read-out of valleys in atomically thin semiconductors}}.
\newblock {\emph{\JournalTitle{Small}}} \textbf{\bibinfo{volume}{19}}, \bibinfo{pages}{2301126}, \doiprefix\url{10.1002/smll.202301126} (\bibinfo{year}{2023}).

\bibitem{Hu_S_23}
\bibinfo{author}{Hu, H.} \emph{et~al.}
\newblock \bibinfo{journal}{\bibinfo{title}{Gate-tunable negative refraction of mid-infrared polaritons}}.
\newblock {\emph{\JournalTitle{Science}}} \textbf{\bibinfo{volume}{379}}, \bibinfo{pages}{558--561}, \doiprefix\url{doi:10.1126/science.adf1251} (\bibinfo{year}{2023}).

\bibitem{Uddin_NC_24}
\bibinfo{author}{Uddin, M.~G.} \emph{et~al.}
\newblock \bibinfo{journal}{\bibinfo{title}{Broadband miniaturized spectrometers with a van der {Waals} tunnel diode}}.
\newblock {\emph{\JournalTitle{Nature Communications}}} \textbf{\bibinfo{volume}{15}}, \bibinfo{pages}{571}, \doiprefix\url{10.1038/s41467-024-44702-8} (\bibinfo{year}{2024}).

\bibitem{VandeGroep2020}
\bibinfo{author}{van~de Groep, J.} \emph{et~al.}
\newblock \bibinfo{journal}{\bibinfo{title}{{Exciton resonance tuning of an atomically thin lens}}}.
\newblock {\emph{\JournalTitle{Nature Photonics}}} \textbf{\bibinfo{volume}{14}}, \bibinfo{pages}{426--430}, \doiprefix\url{10.1038/s41566-020-0624-y} (\bibinfo{year}{2020}).

\bibitem{Li2023}
\bibinfo{author}{Li, Q.} \emph{et~al.}
\newblock \bibinfo{journal}{\bibinfo{title}{{A Purcell-enabled monolayer semiconductor free-space optical modulator}}}.
\newblock {\emph{\JournalTitle{Nature Photonics}}} \bibinfo{pages}{897–903}, \doiprefix\url{10.1038/s41566-023-01250-9} (\bibinfo{year}{2023}).

\bibitem{Li2023beamsteering}
\bibinfo{author}{Li, M.}, \bibinfo{author}{Hail, C.~U.}, \bibinfo{author}{Biswas, S.} \& \bibinfo{author}{Atwater, H.~A.}
\newblock \bibinfo{journal}{\bibinfo{title}{Excitonic beam steering in an active van der {Waals} metasurface}}.
\newblock {\emph{\JournalTitle{Nano Letters}}} \textbf{\bibinfo{volume}{23}}, \bibinfo{pages}{2771--2777}, \doiprefix\url{10.1021/acs.nanolett.3c00032} (\bibinfo{year}{2023}).

\bibitem{Hoekstra2025}
\bibinfo{author}{Hoekstra, T.} \& \bibinfo{author}{Groep, J. V.~D.}
\newblock \bibinfo{journal}{\bibinfo{title}{Electrically tunable strong coupling in a hybrid-{2D} excitonic metasurface for optical modulation}}.
\newblock {\emph{\JournalTitle{arXiv:2502.12132}}}  (\bibinfo{year}{2025}).

\bibitem{Yu2017}
\bibinfo{author}{Yu, Y.} \emph{et~al.}
\newblock \bibinfo{journal}{\bibinfo{title}{Giant gating tunability of optical refractive index in transition metal dichalcogenide monolayers}}.
\newblock {\emph{\JournalTitle{Nano Letters}}} \textbf{\bibinfo{volume}{17}}, \bibinfo{pages}{3613--3618}, \doiprefix\url{10.1021/acs.nanolett.7b00768} (\bibinfo{year}{2017}).

\bibitem{Jauregui2019}
\bibinfo{author}{Jauregui, L.~A.} \emph{et~al.}
\newblock \bibinfo{journal}{\bibinfo{title}{{Electrical control of interlayer exciton dynamics in atomically thin heterostructures}}}.
\newblock {\emph{\JournalTitle{Science}}} \textbf{\bibinfo{volume}{366}}, \bibinfo{pages}{870--875}, \doiprefix\url{10.1126/science.aaw4194} (\bibinfo{year}{2019}).

\bibitem{Stier2018}
\bibinfo{author}{Stier, A.~V.} \emph{et~al.}
\newblock \bibinfo{journal}{\bibinfo{title}{{Magnetooptics of exciton {R}ydberg states in a monolayer semiconductor}}}.
\newblock {\emph{\JournalTitle{Physical Review Letters}}} \textbf{\bibinfo{volume}{120}}, \bibinfo{pages}{057405}, \doiprefix\url{10.1103/physrevlett.120.057405} (\bibinfo{year}{2018}).

\bibitem{Aslan2018}
\bibinfo{author}{Aslan, B.}, \bibinfo{author}{Deng, M.} \& \bibinfo{author}{Heinz, T.~F.}
\newblock \bibinfo{journal}{\bibinfo{title}{Strain tuning of excitons in monolayer {WSe}$_2$}}.
\newblock {\emph{\JournalTitle{Physical Review B}}} \textbf{\bibinfo{volume}{98}}, \bibinfo{pages}{115308}, \doiprefix\url{10.1103/PhysRevB.98.115308} (\bibinfo{year}{2018}).

\bibitem{Henriquez-Guerra2023}
\bibinfo{author}{Henr{\'{i}}quez-Guerra, E.} \emph{et~al.}
\newblock \bibinfo{journal}{\bibinfo{title}{{Large Biaxial Compressive Strain Tuning of Neutral and Charged Excitons in Single-Layer Transition Metal Dichalcogenides}}}.
\newblock {\emph{\JournalTitle{ACS Applied Materials and Interfaces}}} \textbf{\bibinfo{volume}{15}}, \bibinfo{pages}{57369--57378}, \doiprefix\url{10.1021/acsami.3c13281} (\bibinfo{year}{2023}).

\bibitem{Raja2019}
\bibinfo{author}{Raja, A.} \emph{et~al.}
\newblock \bibinfo{journal}{\bibinfo{title}{Dielectric disorder in two-dimensional materials}}.
\newblock {\emph{\JournalTitle{Nature Nanotechnology}}} \textbf{\bibinfo{volume}{14}}, \bibinfo{pages}{832--837}, \doiprefix\url{10.1038/s41565-019-0520-0} (\bibinfo{year}{2019}).

\bibitem{vandegroep2023}
\bibinfo{author}{van~de Groep, J.}, \bibinfo{author}{Li, Q.}, \bibinfo{author}{Song, J.-H.}, \bibinfo{author}{Kik, P.~G.} \& \bibinfo{author}{Brongersma, M.~L.}
\newblock \bibinfo{journal}{\bibinfo{title}{{Impact of substrates and quantum effects on exciton line shapes of {2D} semiconductors at room temperature}}}.
\newblock {\emph{\JournalTitle{Nanophotonics}}} \textbf{\bibinfo{volume}{12}}, \bibinfo{pages}{3291}, \doiprefix\url{10.1515/nanoph-2023-0193} (\bibinfo{year}{2023}).

\bibitem{Datta2020}
\bibinfo{author}{Datta, I.} \emph{et~al.}
\newblock \bibinfo{journal}{\bibinfo{title}{Low-loss composite photonic platform based on 2{D} semiconductor monolayers}}.
\newblock {\emph{\JournalTitle{Nature Photonics}}} \textbf{\bibinfo{volume}{14}}, \bibinfo{pages}{256–262}, \doiprefix\url{10.1038/s41566-020-0590-4} (\bibinfo{year}{2020}).

\bibitem{Li2024}
\bibinfo{author}{Li, M.}, \bibinfo{author}{Michaeli, L.} \& \bibinfo{author}{Atwater, H.~A.}
\newblock \bibinfo{journal}{\bibinfo{title}{{Electrically Tunable Topological Singularities in Excitonic Two-Dimensional Heterostructures for Wavefront Manipulation}}}.
\newblock {\emph{\JournalTitle{ACS Photonics}}} \textbf{\bibinfo{volume}{11}}, \bibinfo{pages}{3554–3562}, \doiprefix\url{10.1021/acsphotonics.4c00397} (\bibinfo{year}{2024}).

\bibitem{Lynch2025}
\bibinfo{author}{Lynch, J.} \emph{et~al.}
\newblock \bibinfo{journal}{\bibinfo{title}{Full 2$\pi$ phase modulation using exciton-polaritons in a two-dimensional superlattice}}.
\newblock {\emph{\JournalTitle{Device}}} \textbf{\bibinfo{volume}{3}}, \bibinfo{pages}{100639}, \doiprefix\url{10.1016/j.device.2024.100639} (\bibinfo{year}{2025}).

\bibitem{Haastrup2018}
\bibinfo{author}{Haastrup, S.} \emph{et~al.}
\newblock \bibinfo{journal}{\bibinfo{title}{The computational {2D} materials database: High-throughput modeling and discovery of atomically thin crystals}}.
\newblock {\emph{\JournalTitle{2D Materials}}} \textbf{\bibinfo{volume}{5}}, \bibinfo{pages}{042002}, \doiprefix\url{10.1088/2053-1583/aacfc1} (\bibinfo{year}{2018}).

\bibitem{Gjerding2021}
\bibinfo{author}{Gjerding, M.~N.} \emph{et~al.}
\newblock \bibinfo{journal}{\bibinfo{title}{Recent progress of the computational {2D} materials database ({C2DB})}}.
\newblock {\emph{\JournalTitle{2D Materials}}} \textbf{\bibinfo{volume}{8}}, \bibinfo{pages}{044002}, \doiprefix\url{10.1088/2053-1583/ac1059} (\bibinfo{year}{2021}).

\bibitem{Stier2016}
\bibinfo{author}{Stier, A.~V.}, \bibinfo{author}{Wilson, N.~P.}, \bibinfo{author}{Clark, G.}, \bibinfo{author}{Xu, X.} \& \bibinfo{author}{Crooker, S.~A.}
\newblock \bibinfo{journal}{\bibinfo{title}{Probing the influence of dielectric environment on excitons in monolayer {WSe}$_2$: Insight from high magnetic fields}}.
\newblock {\emph{\JournalTitle{Nano Letters}}} \textbf{\bibinfo{volume}{16}}, \bibinfo{pages}{7054--7060}, \doiprefix\url{10.1021/acs.nanolett.6b03276} (\bibinfo{year}{2016}).

\bibitem{Lu2017}
\bibinfo{author}{Lu, X.} \& \bibinfo{author}{Yang, L.}
\newblock \bibinfo{journal}{\bibinfo{title}{{Stark effect of doped two-dimensional transition metal dichalcogenides}}}.
\newblock {\emph{\JournalTitle{Applied Physics Letters}}} \textbf{\bibinfo{volume}{111}}, \bibinfo{pages}{193104}, \doiprefix\url{10.1063/1.5004413} (\bibinfo{year}{2017}).

\bibitem{Li2020}
\bibinfo{author}{Li, L.} \emph{et~al.}
\newblock \bibinfo{journal}{\bibinfo{title}{Wavelength-tunable interlayer exciton emission at the near-infrared region in van der {Waals} semiconductor heterostructures}}.
\newblock {\emph{\JournalTitle{Nano Letters}}} \textbf{\bibinfo{volume}{20}}, \bibinfo{pages}{3361--3368}, \doiprefix\url{10.1021/acs.nanolett.0c00258} (\bibinfo{year}{2020}).

\bibitem{Lin2021}
\bibinfo{author}{Lin, K.-Q.} \emph{et~al.}
\newblock \bibinfo{journal}{\bibinfo{title}{Twist-angle engineering of excitonic quantum interference and optical nonlinearities in stacked {2D} semiconductors}}.
\newblock {\emph{\JournalTitle{Nature Communications}}} \textbf{\bibinfo{volume}{12}}, \bibinfo{pages}{1553}, \doiprefix\url{10.1038/s41467-021-21547-z} (\bibinfo{year}{2021}).

\bibitem{Goossens2017}
\bibinfo{author}{Goossens, S.} \emph{et~al.}
\newblock \bibinfo{journal}{\bibinfo{title}{Broadband image sensor array based on graphene-{CMOS} integration}}.
\newblock {\emph{\JournalTitle{Nature Photonics}}} \textbf{\bibinfo{volume}{11}}, \bibinfo{pages}{366--371}, \doiprefix\url{10.1038/nphoton.2017.75} (\bibinfo{year}{2017}).

\bibitem{novoselov20162d}
\bibinfo{author}{Novoselov, K.~S.}, \bibinfo{author}{Mishchenko, A.}, \bibinfo{author}{Carvalho, A.} \& \bibinfo{author}{Castro~Neto, A.}
\newblock \bibinfo{journal}{\bibinfo{title}{{2D} materials and van der {Waals} heterostructures}}.
\newblock {\emph{\JournalTitle{Science}}} \textbf{\bibinfo{volume}{353}}, \bibinfo{pages}{aac9439}, \doiprefix\url{10.1126/science.aac9439} (\bibinfo{year}{2016}).

\bibitem{Fan2022}
\bibinfo{author}{Fan, K.}, \bibinfo{author}{Averitt, R.~D.} \& \bibinfo{author}{Padilla, W.~J.}
\newblock \bibinfo{journal}{\bibinfo{title}{Active and tunable nanophotonic metamaterials}}.
\newblock {\emph{\JournalTitle{Nanophotonics}}} \textbf{\bibinfo{volume}{11}}, \bibinfo{pages}{3769–3803}, \doiprefix\url{10.1515/nanoph-2022-0188} (\bibinfo{year}{2022}).

\bibitem{Yu2021}
\bibinfo{author}{Yu, X.}, \bibinfo{author}{Wang, X.}, \bibinfo{author}{Zhou, F.}, \bibinfo{author}{Qu, J.} \& \bibinfo{author}{Song, J.}
\newblock \bibinfo{journal}{\bibinfo{title}{{2D} van der {Waals} heterojunction nanophotonic devices: From fabrication to performance}}.
\newblock {\emph{\JournalTitle{Advanced Functional Materials}}} \textbf{\bibinfo{volume}{31}}, \bibinfo{pages}{2104260}, \doiprefix\url{10.1002/adfm.202104260} (\bibinfo{year}{2021}).

\bibitem{Li2016}
\bibinfo{author}{Li, M.-Y.}, \bibinfo{author}{Chen, C.-H.}, \bibinfo{author}{Shi, Y.} \& \bibinfo{author}{Li, L.-J.}
\newblock \bibinfo{journal}{\bibinfo{title}{Heterostructures based on two-dimensional layered materials and their potential applications}}.
\newblock {\emph{\JournalTitle{Materials Today}}} \textbf{\bibinfo{volume}{19}}, \bibinfo{pages}{322–335}, \doiprefix\url{10.1016/j.mattod.2015.11.003} (\bibinfo{year}{2016}).

\bibitem{https://doi.org/10.1002/smll.202107059}
\bibinfo{author}{Li, J.} \emph{et~al.}
\newblock \bibinfo{journal}{\bibinfo{title}{Controllable preparation of {2D} vertical van der {Waals} heterostructures and superlattices for functional applications}}.
\newblock {\emph{\JournalTitle{Small}}} \textbf{\bibinfo{volume}{18}}, \bibinfo{pages}{2107059}, \doiprefix\url{10.1002/smll.202107059} (\bibinfo{year}{2022}).

\bibitem{Wurstbauer2017}
\bibinfo{author}{Wurstbauer, U.}, \bibinfo{author}{Miller, B.}, \bibinfo{author}{Parzinger, E.} \& \bibinfo{author}{Holleitner, A.~W.}
\newblock \bibinfo{journal}{\bibinfo{title}{Light–matter interaction in transition metal dichalcogenides and their heterostructures}}.
\newblock {\emph{\JournalTitle{Journal of Physics D: Applied Physics}}} \textbf{\bibinfo{volume}{50}}, \bibinfo{pages}{173001}, \doiprefix\url{10.1088/1361-6463/aa5f81} (\bibinfo{year}{2017}).

\bibitem{Britnell2013}
\bibinfo{author}{Britnell, L.} \emph{et~al.}
\newblock \bibinfo{journal}{\bibinfo{title}{Strong light-matter interactions in heterostructures of atomically thin films}}.
\newblock {\emph{\JournalTitle{Science}}} \textbf{\bibinfo{volume}{340}}, \bibinfo{pages}{1311–1314}, \doiprefix\url{10.1126/science.1235547} (\bibinfo{year}{2013}).

\bibitem{Kaushik20}
\bibinfo{author}{Kaushik, V.}, \bibinfo{author}{Rajput, S.} \& \bibinfo{author}{Kumar, M.}
\newblock \bibinfo{journal}{\bibinfo{title}{Broadband optical modulation in a zinc-oxide-based heterojunction via optical lifting}}.
\newblock {\emph{\JournalTitle{Optics Letters}}} \textbf{\bibinfo{volume}{45}}, \bibinfo{pages}{363--366}, \doiprefix\url{10.1364/OL.379257} (\bibinfo{year}{2020}).

\bibitem{Guo2020}
\bibinfo{author}{Guo, X.} \emph{et~al.}
\newblock \bibinfo{journal}{\bibinfo{title}{Efficient all‐optical plasmonic modulators with atomically thin van der {Waals} heterostructures}}.
\newblock {\emph{\JournalTitle{Advanced Materials}}} \textbf{\bibinfo{volume}{32}}, \bibinfo{pages}{1907105}, \doiprefix\url{10.1002/adma.201907105} (\bibinfo{year}{2020}).

\bibitem{Withers2015}
\bibinfo{author}{Withers, F.} \emph{et~al.}
\newblock \bibinfo{journal}{\bibinfo{title}{Light-emitting diodes by band-structure engineering in van der {Waals} heterostructures}}.
\newblock {\emph{\JournalTitle{Nature Materials}}} \textbf{\bibinfo{volume}{14}}, \bibinfo{pages}{301–306}, \doiprefix\url{10.1038/nmat4205} (\bibinfo{year}{2015}).

\bibitem{Hwang2005}
\bibinfo{author}{Hwang, D.-K.} \emph{et~al.}
\newblock \bibinfo{journal}{\bibinfo{title}{p-{ZnO}/n-{GaN} heterostructure {ZnO} light-emitting diodes}}.
\newblock {\emph{\JournalTitle{Applied Physics Letters}}} \textbf{\bibinfo{volume}{86}}, \bibinfo{pages}{222101}, \doiprefix\url{10.1063/1.1940736} (\bibinfo{year}{2005}).

\bibitem{Chuang2024}
\bibinfo{author}{Chuang, H.-J.} \emph{et~al.}
\newblock \bibinfo{journal}{\bibinfo{title}{Enhancing single photon emission purity via design of van der {Waals} heterostructures}}.
\newblock {\emph{\JournalTitle{Nano Letters}}} \textbf{\bibinfo{volume}{24}}, \bibinfo{pages}{5529–5535}, \doiprefix\url{10.1021/acs.nanolett.4c00683} (\bibinfo{year}{2024}).

\bibitem{Yu2024}
\bibinfo{author}{Yu, Y.} \emph{et~al.}
\newblock \bibinfo{journal}{\bibinfo{title}{Tunable single-photon emitters in {2D} materials}}.
\newblock {\emph{\JournalTitle{Nanophotonics}}} \textbf{\bibinfo{volume}{13}}, \bibinfo{pages}{3615–3629}, \doiprefix\url{10.1515/nanoph-2024-0050} (\bibinfo{year}{2024}).

\bibitem{Parto2021}
\bibinfo{author}{Parto, K.}, \bibinfo{author}{Azzam, S.~I.}, \bibinfo{author}{Banerjee, K.} \& \bibinfo{author}{Moody, G.}
\newblock \bibinfo{journal}{\bibinfo{title}{Defect and strain engineering of monolayer {WSe}$_2$ enables site-controlled single-photon emission up to 150 {K}}}.
\newblock {\emph{\JournalTitle{Nature Communications}}} \textbf{\bibinfo{volume}{12}}, \bibinfo{pages}{3585}, \doiprefix\url{10.1038/s41467-021-23709-5} (\bibinfo{year}{2021}).

\bibitem{Rosenberger2019}
\bibinfo{author}{Rosenberger, M.~R.} \emph{et~al.}
\newblock \bibinfo{journal}{\bibinfo{title}{Quantum calligraphy: Writing single-photon emitters in a two-dimensional materials platform}}.
\newblock {\emph{\JournalTitle{ACS Nano}}} \textbf{\bibinfo{volume}{13}}, \bibinfo{pages}{904–912}, \doiprefix\url{10.1021/acsnano.8b08730} (\bibinfo{year}{2019}).

\bibitem{Yoon_S_22}
\bibinfo{author}{Yoon, H.~H.} \emph{et~al.}
\newblock \bibinfo{journal}{\bibinfo{title}{Miniaturized spectrometers with a tunable van der {Waals} junction}}.
\newblock {\emph{\JournalTitle{Science}}} \textbf{\bibinfo{volume}{378}}, \bibinfo{pages}{296--299}, \doiprefix\url{doi:10.1126/science.add8544} (\bibinfo{year}{2022}).

\bibitem{Shen2018}
\bibinfo{author}{Shen, P.-C.} \emph{et~al.}
\newblock \bibinfo{journal}{\bibinfo{title}{{CVD} technology for {2-D} materials}}.
\newblock {\emph{\JournalTitle{IEEE Transactions on Electron Devices}}} \textbf{\bibinfo{volume}{65}}, \bibinfo{pages}{4040--4052}, \doiprefix\url{10.1109/TED.2018.2866390} (\bibinfo{year}{2018}).

\bibitem{Xia2014}
\bibinfo{author}{Xia, J.} \emph{et~al.}
\newblock \bibinfo{journal}{\bibinfo{title}{{CVD} synthesis of large-area, highly crystalline {MoSe}$_2$ atomic layers on diverse substrates and application to photodetectors}}.
\newblock {\emph{\JournalTitle{Nanoscale}}} \textbf{\bibinfo{volume}{6}}, \bibinfo{pages}{8949}, \doiprefix\url{10.1039/c4nr02311k} (\bibinfo{year}{2014}).

\bibitem{Mandyam2020}
\bibinfo{author}{Mandyam, S.~V.}, \bibinfo{author}{Kim, H.~M.} \& \bibinfo{author}{Drndić, M.}
\newblock \bibinfo{journal}{\bibinfo{title}{Large area few-layer {TMD} film growths and their applications}}.
\newblock {\emph{\JournalTitle{Journal of Physics: Materials}}} \textbf{\bibinfo{volume}{3}}, \bibinfo{pages}{024008}, \doiprefix\url{10.1088/2515-7639/ab82b3} (\bibinfo{year}{2020}).

\bibitem{Kim2023}
\bibinfo{author}{Kim, H.~W.}
\newblock \bibinfo{journal}{\bibinfo{title}{Recent progress in the role of grain boundaries in two-dimensional transition metal dichalcogenides studied using scanning tunneling microscopy/spectroscopy}}.
\newblock {\emph{\JournalTitle{Applied Microscopy}}} \textbf{\bibinfo{volume}{53}}, \doiprefix\url{10.1186/s42649-023-00088-3} (\bibinfo{year}{2023}).

\bibitem{jainMinimizingResiduesStrain2018a}
\bibinfo{author}{Jain, A.} \emph{et~al.}
\newblock \bibinfo{journal}{\bibinfo{title}{Minimizing residues and strain in {2D} materials transferred from {PDMS}}}.
\newblock {\emph{\JournalTitle{Nanotechnology}}} \textbf{\bibinfo{volume}{29}}, \bibinfo{pages}{265203}, \doiprefix\url{10.1088/1361-6528/aabd90} (\bibinfo{year}{2018}).

\bibitem{castellanos-gomezDeterministicTransferTwodimensional2014b}
\bibinfo{author}{Castellanos-Gomez, A.} \emph{et~al.}
\newblock \bibinfo{journal}{\bibinfo{title}{Deterministic transfer of two-dimensional materials by all-dry viscoelastic stamping}}.
\newblock {\emph{\JournalTitle{2D Materials}}} \textbf{\bibinfo{volume}{1}}, \bibinfo{pages}{011002}, \doiprefix\url{10.1088/2053-1583/1/1/011002} (\bibinfo{year}{2014}).

\bibitem{linGentleTransferMethod2016}
\bibinfo{author}{Lin, J.}, \bibinfo{author}{Lin, Y.-C.}, \bibinfo{author}{Wang, X.}, \bibinfo{author}{Xie, L.} \& \bibinfo{author}{Suenaga, K.}
\newblock \bibinfo{journal}{\bibinfo{title}{Gentle transfer method for water- and acid/alkali-sensitive {2D} materials for ({S}){TEM} study}}.
\newblock {\emph{\JournalTitle{APL Materials}}} \textbf{\bibinfo{volume}{4}}, \bibinfo{pages}{116108}, \doiprefix\url{10.1063/1.4967938} (\bibinfo{year}{2016}).

\bibitem{haleyHeatedAssemblyTransfer2021}
\bibinfo{author}{Haley, K.~L.} \emph{et~al.}
\newblock \bibinfo{journal}{\bibinfo{title}{Heated assembly and transfer of van der {Waals} heterostructures with common nail polish}}.
\newblock {\emph{\JournalTitle{Nanomanufacturing}}} \textbf{\bibinfo{volume}{1}}, \bibinfo{pages}{49--56}, \doiprefix\url{10.3390/nanomanufacturing1010005} (\bibinfo{year}{2021}).

\bibitem{wangCleanAssemblyVan2023}
\bibinfo{author}{Wang, W.} \emph{et~al.}
\newblock \bibinfo{journal}{\bibinfo{title}{Clean assembly of van der {Waals} heterostructures using silicon nitride membranes}}.
\newblock {\emph{\JournalTitle{Nature Electronics}}} \textbf{\bibinfo{volume}{6}}, \bibinfo{pages}{981--990}, \doiprefix\url{10.1038/s41928-023-01075-y} (\bibinfo{year}{2023}).

\bibitem{Behura2021}
\bibinfo{author}{Behura, S.~K.} \emph{et~al.}
\newblock \bibinfo{journal}{\bibinfo{title}{Moir\'{e} physics in twisted van der {Waals} heterostructures of {2D} materials}}.
\newblock {\emph{\JournalTitle{Emergent Materials}}} \textbf{\bibinfo{volume}{4}}, \bibinfo{pages}{813–826}, \doiprefix\url{10.1007/s42247-021-00270-x} (\bibinfo{year}{2021}).

\bibitem{reganEmergingExcitonPhysics2022a}
\bibinfo{author}{Regan, E.~C.} \emph{et~al.}
\newblock \bibinfo{journal}{\bibinfo{title}{Emerging exciton physics in transition metal dichalcogenide heterobilayers}}.
\newblock {\emph{\JournalTitle{Nature Reviews Materials}}} \textbf{\bibinfo{volume}{7}}, \bibinfo{pages}{778--795}, \doiprefix\url{10.1038/s41578-022-00440-1} (\bibinfo{year}{2022}).

\bibitem{Montblanch2021}
\bibinfo{author}{Montblanch, A. R.-P.} \emph{et~al.}
\newblock \bibinfo{journal}{\bibinfo{title}{Confinement of long-lived interlayer excitons in {WS}$_2$/{WSe}$_2$ heterostructures}}.
\newblock {\emph{\JournalTitle{Communications Physics}}} \textbf{\bibinfo{volume}{4}}, \doiprefix\url{10.1038/s42005-021-00625-0} (\bibinfo{year}{2021}).

\bibitem{Blundo2024}
\bibinfo{author}{Blundo, E.} \emph{et~al.}
\newblock \bibinfo{journal}{\bibinfo{title}{Localisation-to-delocalisation transition of moir\'{e} excitons in {WSe}$_2$/{MoSe}$_2$ heterostructures}}.
\newblock {\emph{\JournalTitle{Nature Communications}}} \textbf{\bibinfo{volume}{15}}, \bibinfo{pages}{1057}, \doiprefix\url{10.1038/s41467-024-44739-9} (\bibinfo{year}{2024}).

\bibitem{Deng2020}
\bibinfo{author}{Deng, B.} \emph{et~al.}
\newblock \bibinfo{journal}{\bibinfo{title}{Strong mid-infrared photoresponse in small-twist-angle bilayer graphene}}.
\newblock {\emph{\JournalTitle{Nature Photonics}}} \textbf{\bibinfo{volume}{14}}, \bibinfo{pages}{549–553}, \doiprefix\url{10.1038/s41566-020-0644-7} (\bibinfo{year}{2020}).

\bibitem{Ulstrup2020}
\bibinfo{author}{Ulstrup, S.} \emph{et~al.}
\newblock \bibinfo{journal}{\bibinfo{title}{Direct observation of minibands in a twisted graphene/{WS}$_2$ bilayer}}.
\newblock {\emph{\JournalTitle{Science Advances}}} \textbf{\bibinfo{volume}{6}}, \bibinfo{pages}{eaay6104}, \doiprefix\url{10.1126/sciadv.aay6104} (\bibinfo{year}{2020}).

\bibitem{li2024infraredspectroscopydiagnosingsuperlattice}
\bibinfo{author}{Li, G.} \emph{et~al.}
\newblock \bibinfo{journal}{\bibinfo{title}{Infrared spectroscopy for diagnosing superlattice minibands in twisted bilayer graphene near the magic angle}}.
\newblock {\emph{\JournalTitle{Nano Letters}}} \textbf{\bibinfo{volume}{24}}, \bibinfo{pages}{15956--15963}, \doiprefix\url{10.1021/acs.nanolett.4c02853} (\bibinfo{year}{2024}).

\bibitem{Agarwal2023}
\bibinfo{author}{Agarwal, H.} \emph{et~al.}
\newblock \bibinfo{journal}{\bibinfo{title}{Ultra-broadband photoconductivity in twisted graphene heterostructures with large responsivity}}.
\newblock {\emph{\JournalTitle{Nature Photonics}}} \textbf{\bibinfo{volume}{17}}, \bibinfo{pages}{1047–1053}, \doiprefix\url{10.1038/s41566-023-01291-0} (\bibinfo{year}{2023}).

\bibitem{gogoiLayerRotationAngleDependentExcitonic2019a}
\bibinfo{author}{Gogoi, P.~K.} \emph{et~al.}
\newblock \bibinfo{journal}{\bibinfo{title}{Layer {Rotation}-{Angle}-{Dependent} {Excitonic} {Absorption} in van der {Waals} {Heterostructures} {Revealed} by {Electron} {Energy} {Loss} {Spectroscopy}}}.
\newblock {\emph{\JournalTitle{ACS Nano}}} \textbf{\bibinfo{volume}{13}}, \bibinfo{pages}{9541--9550}, \doiprefix\url{10.1021/acsnano.9b04530} (\bibinfo{year}{2019}).

\bibitem{barmanTwistDependentTuningExcitonic2022}
\bibinfo{author}{Barman, P.~K.} \emph{et~al.}
\newblock \bibinfo{journal}{\bibinfo{title}{Twist-{Dependent} {Tuning} of {Excitonic} {Emissions} in {Bilayer} {WSe}$_{\textrm{2}}$}}.
\newblock {\emph{\JournalTitle{ACS Omega}}} \textbf{\bibinfo{volume}{7}}, \bibinfo{pages}{6412--6418}, \doiprefix\url{10.1021/acsomega.1c07219} (\bibinfo{year}{2022}).

\bibitem{Kumar2021}
\bibinfo{author}{Kumar, P.} \emph{et~al.}
\newblock \bibinfo{journal}{\bibinfo{title}{Light–matter coupling in large-area van der {W}aals superlattices}}.
\newblock {\emph{\JournalTitle{Nature Nanotechnology}}} \textbf{\bibinfo{volume}{17}}, \bibinfo{pages}{182–189}, \doiprefix\url{10.1038/s41565-021-01023-x} (\bibinfo{year}{2021}).

\bibitem{Elrafei2024superlattices}
\bibinfo{author}{Elrafei, S.~A.}, \bibinfo{author}{Heijnen, L.~M.}, \bibinfo{author}{Godiksen, R.~H.} \& \bibinfo{author}{Curto, A.~G.}
\newblock \bibinfo{journal}{\bibinfo{title}{Monolayer semiconductor superlattices with high optical absorption}}.
\newblock {\emph{\JournalTitle{ACS Photonics}}} \textbf{\bibinfo{volume}{11}}, \bibinfo{pages}{2587--2594}, \doiprefix\url{10.1021/acsphotonics.4c00277} (\bibinfo{year}{2024}).

\bibitem{Hussain2022}
\bibinfo{author}{Hussain, G.}, \bibinfo{author}{Asghar, M.}, \bibinfo{author}{Waqas~Iqbal, M.}, \bibinfo{author}{Ullah, H.} \& \bibinfo{author}{Autieri, C.}
\newblock \bibinfo{journal}{\bibinfo{title}{Exploring the structural stability, electronic and thermal attributes of synthetic {2D} materials and their heterostructures}}.
\newblock {\emph{\JournalTitle{Applied Surface Science}}} \textbf{\bibinfo{volume}{590}}, \bibinfo{pages}{153131}, \doiprefix\url{10.1016/j.apsusc.2022.153131} (\bibinfo{year}{2022}).

\bibitem{Duan2014}
\bibinfo{author}{Duan, X.} \emph{et~al.}
\newblock \bibinfo{journal}{\bibinfo{title}{Lateral epitaxial growth of two-dimensional layered semiconductor heterojunctions}}.
\newblock {\emph{\JournalTitle{Nature Nanotechnology}}} \textbf{\bibinfo{volume}{9}}, \bibinfo{pages}{1024–1030}, \doiprefix\url{10.1038/nnano.2014.222} (\bibinfo{year}{2014}).

\bibitem{Gong2014}
\bibinfo{author}{Gong, Y.} \emph{et~al.}
\newblock \bibinfo{journal}{\bibinfo{title}{Vertical and in-plane heterostructures from {WS}$_2$/{MoS}$_2$ monolayers}}.
\newblock {\emph{\JournalTitle{Nature Materials}}} \textbf{\bibinfo{volume}{13}}, \bibinfo{pages}{1135–1142}, \doiprefix\url{10.1038/nmat4091} (\bibinfo{year}{2014}).

\bibitem{Smith2013}
\bibinfo{author}{Smith, J.~T.}, \bibinfo{author}{Franklin, A.~D.}, \bibinfo{author}{Farmer, D.~B.} \& \bibinfo{author}{Dimitrakopoulos, C.~D.}
\newblock \bibinfo{journal}{\bibinfo{title}{Reducing contact resistance in graphene devices through contact area patterning}}.
\newblock {\emph{\JournalTitle{ACS Nano}}} \textbf{\bibinfo{volume}{7}}, \bibinfo{pages}{3661–3667}, \doiprefix\url{10.1021/nn400671z} (\bibinfo{year}{2013}).

\bibitem{Wang2015}
\bibinfo{author}{Wang, B.}, \bibinfo{author}{Eichfield, S.~M.}, \bibinfo{author}{Wang, D.}, \bibinfo{author}{Robinson, J.~A.} \& \bibinfo{author}{Haque, M.~A.}
\newblock \bibinfo{journal}{\bibinfo{title}{In situ degradation studies of two-dimensional {WSe}$_2$–graphene heterostructures}}.
\newblock {\emph{\JournalTitle{Nanoscale}}} \textbf{\bibinfo{volume}{7}}, \bibinfo{pages}{14489–14495}, \doiprefix\url{10.1039/c5nr03357h} (\bibinfo{year}{2015}).

\bibitem{Hou2020}
\bibinfo{author}{Hou, L.}, \bibinfo{author}{Zhang, Q.}, \bibinfo{author}{Shautsova, V.} \& \bibinfo{author}{Warner, J.~H.}
\newblock \bibinfo{journal}{\bibinfo{title}{Operational limits and failure mechanisms in all-{2D} van der {Waals} vertical heterostructure devices with long-lived persistent electroluminescence}}.
\newblock {\emph{\JournalTitle{ACS Nano}}} \textbf{\bibinfo{volume}{14}}, \bibinfo{pages}{15533–15543}, \doiprefix\url{10.1021/acsnano.0c06153} (\bibinfo{year}{2020}).

\bibitem{Hong2014}
\bibinfo{author}{Hong, X.} \emph{et~al.}
\newblock \bibinfo{journal}{\bibinfo{title}{Ultrafast charge transfer in atomically thin {MoS}$_2$/{WS}$_2$ heterostructures}}.
\newblock {\emph{\JournalTitle{Nature Nanotechnology}}} \textbf{\bibinfo{volume}{9}}, \bibinfo{pages}{682–686}, \doiprefix\url{10.1038/nnano.2014.167} (\bibinfo{year}{2014}).

\bibitem{Teshome2017}
\bibinfo{author}{Teshome, T.} \& \bibinfo{author}{Datta, A.}
\newblock \bibinfo{journal}{\bibinfo{title}{Two-dimensional graphene–gold interfaces serve as robust templates for dielectric capacitors}}.
\newblock {\emph{\JournalTitle{ACS Applied Materials and Interfaces}}} \textbf{\bibinfo{volume}{9}}, \bibinfo{pages}{34213–34220}, \doiprefix\url{10.1021/acsami.7b09360} (\bibinfo{year}{2017}).

\bibitem{Blackstone2021}
\bibinfo{author}{Blackstone, C.} \& \bibinfo{author}{Ignaszak, A.}
\newblock \bibinfo{journal}{\bibinfo{title}{Van der {Waals} heterostructures—recent progress in electrode materials for clean energy applications}}.
\newblock {\emph{\JournalTitle{Materials}}} \textbf{\bibinfo{volume}{14}}, \bibinfo{pages}{3754}, \doiprefix\url{10.3390/ma14133754} (\bibinfo{year}{2021}).

\bibitem{Oh2021}
\bibinfo{author}{Oh, S.-H.} \emph{et~al.}
\newblock \bibinfo{journal}{\bibinfo{title}{Nanophotonic biosensors harnessing van der {Waals} materials}}.
\newblock {\emph{\JournalTitle{Nature Communications}}} \textbf{\bibinfo{volume}{12}}, \bibinfo{pages}{3824}, \doiprefix\url{10.1038/s41467-021-23564-4} (\bibinfo{year}{2021}).

\bibitem{Lin2019}
\bibinfo{author}{Lin, Z.}, \bibinfo{author}{Huang, Y.} \& \bibinfo{author}{Duan, X.}
\newblock \bibinfo{journal}{\bibinfo{title}{Van der {Waals} thin-film electronics}}.
\newblock {\emph{\JournalTitle{Nature Electronics}}} \textbf{\bibinfo{volume}{2}}, \bibinfo{pages}{378–388}, \doiprefix\url{10.1038/s41928-019-0301-7} (\bibinfo{year}{2019}).

\bibitem{9976044}
\bibinfo{author}{Liu, J.} \emph{et~al.}
\newblock \bibinfo{journal}{\bibinfo{title}{Low dark-current {V}$_{2}${CT}$_{x}$/n-{Si} van der {Waals} {S}chottky photodiode for {H}adamard single-pixel imaging}}.
\newblock {\emph{\JournalTitle{IEEE Electron Device Letters}}} \textbf{\bibinfo{volume}{44}}, \bibinfo{pages}{285--288}, \doiprefix\url{10.1109/LED.2022.3227764} (\bibinfo{year}{2023}).

\bibitem{Zhang2022}
\bibinfo{author}{Zhang, L.} \emph{et~al.}
\newblock \bibinfo{journal}{\bibinfo{title}{Gate-tunable photovoltaic behavior and polarized image sensor based on all‐{2D} {TaIrTe}$_4$/{MoS}$_2$ {van der Waals Schottky} diode}}.
\newblock {\emph{\JournalTitle{Advanced Electronic Materials}}} \textbf{\bibinfo{volume}{8}}, \bibinfo{pages}{2200551}, \doiprefix\url{10.1002/aelm.202200551} (\bibinfo{year}{2022}).

\bibitem{Kavokin2022}
\bibinfo{author}{Kavokin, A.} \emph{et~al.}
\newblock \bibinfo{journal}{\bibinfo{title}{Polariton condensates for classical and quantum computing}}.
\newblock {\emph{\JournalTitle{Nature Reviews Physics}}} \textbf{\bibinfo{volume}{4}}, \bibinfo{pages}{435–451}, \doiprefix\url{10.1038/s42254-022-00447-1} (\bibinfo{year}{2022}).

\bibitem{Pal2022}
\bibinfo{author}{Pal, A.} \emph{et~al.}
\newblock \bibinfo{journal}{\bibinfo{title}{Quantum-engineered devices based on {2D} materials for next-generation information processing and storage}}.
\newblock {\emph{\JournalTitle{Advanced Materials}}} \textbf{\bibinfo{volume}{35}}, \bibinfo{pages}{2109894}, \doiprefix\url{10.1002/adma.202109894} (\bibinfo{year}{2022}).

\bibitem{Khaidar2024}
\bibinfo{author}{Khaidar, D.~M.}, \bibinfo{author}{Isahak, W. N. R.~W.}, \bibinfo{author}{Ramli, Z. A.~C.} \& \bibinfo{author}{Ahmad, K.~N.}
\newblock \bibinfo{journal}{\bibinfo{title}{Transition metal dichalcogenides-based catalysts for {CO}$_{2}$ conversion: An updated review}}.
\newblock {\emph{\JournalTitle{International Journal of Hydrogen Energy}}} \textbf{\bibinfo{volume}{68}}, \bibinfo{pages}{35–50}, \doiprefix\url{10.1016/j.ijhydene.2024.04.220} (\bibinfo{year}{2024}).

\bibitem{Voiry2016}
\bibinfo{author}{Voiry, D.}, \bibinfo{author}{Yang, J.} \& \bibinfo{author}{Chhowalla, M.}
\newblock \bibinfo{journal}{\bibinfo{title}{Recent strategies for improving the catalytic activity of {2D} {TMD} nanosheets toward the hydrogen evolution reaction}}.
\newblock {\emph{\JournalTitle{Advanced Materials}}} \textbf{\bibinfo{volume}{28}}, \bibinfo{pages}{6197--6206}, \doiprefix\url{10.1002/adma.201505597} (\bibinfo{year}{2016}).

\bibitem{zhaoHighorderSuperlatticesRolling2021}
\bibinfo{author}{Zhao, B.} \emph{et~al.}
\newblock \bibinfo{journal}{\bibinfo{title}{High-order superlattices by rolling up van der {Waals} heterostructures}}.
\newblock {\emph{\JournalTitle{Nature}}} \textbf{\bibinfo{volume}{591}}, \bibinfo{pages}{385--390}, \doiprefix\url{10.1038/s41586-021-03338-0} (\bibinfo{year}{2021}).

\bibitem{Ouyang2022}
\bibinfo{author}{Ouyang, D.}, \bibinfo{author}{Zhang, N.}, \bibinfo{author}{Li, Y.} \& \bibinfo{author}{Zhai, T.}
\newblock \bibinfo{journal}{\bibinfo{title}{Emerging nonplanar van der {Waals} nanoarchitectures from {2D} allotropes for optoelectronics}}.
\newblock {\emph{\JournalTitle{Advanced Functional Materials}}} \textbf{\bibinfo{volume}{33}}, \bibinfo{pages}{2208321}, \doiprefix\url{10.1002/adfm.202208321} (\bibinfo{year}{2022}).

\bibitem{Fox2023}
\bibinfo{author}{Fox, C.}, \bibinfo{author}{Mao, Y.}, \bibinfo{author}{Zhang, X.}, \bibinfo{author}{Wang, Y.} \& \bibinfo{author}{Xiao, J.}
\newblock \bibinfo{journal}{\bibinfo{title}{Stacking order engineering of two-dimensional materials and device applications}}.
\newblock {\emph{\JournalTitle{Chemical Reviews}}} \textbf{\bibinfo{volume}{124}}, \bibinfo{pages}{1862–1898}, \doiprefix\url{10.1021/acs.chemrev.3c00618} (\bibinfo{year}{2023}).

\bibitem{Bonnet2021}
\bibinfo{author}{Bonnet, N.} \emph{et~al.}
\newblock \bibinfo{journal}{\bibinfo{title}{Nanoscale modification of {WS}$_2$ trion emission by its local electromagnetic environment}}.
\newblock {\emph{\JournalTitle{Nano Letters}}} \textbf{\bibinfo{volume}{21}}, \bibinfo{pages}{10178–10185}, \doiprefix\url{10.1021/acs.nanolett.1c02600} (\bibinfo{year}{2021}).

\bibitem{Nayak2019}
\bibinfo{author}{Nayak, G.} \emph{et~al.}
\newblock \bibinfo{journal}{\bibinfo{title}{Cathodoluminescence enhancement and quenching in type-{I} van der {Waals} heterostructures: Cleanliness of the interfaces and defect creation}}.
\newblock {\emph{\JournalTitle{Physical Review Materials}}} \textbf{\bibinfo{volume}{3}}, \bibinfo{pages}{114001}, \doiprefix\url{10.1103/physrevmaterials.3.114001} (\bibinfo{year}{2019}).

\bibitem{Susarla2021}
\bibinfo{author}{Susarla, S.} \emph{et~al.}
\newblock \bibinfo{journal}{\bibinfo{title}{Mapping modified electronic levels in the moir\'{e} patterns in {MoS}$_2$/{WSe}$_2$ using low-loss {EELS}}}.
\newblock {\emph{\JournalTitle{Nano Letters}}} \textbf{\bibinfo{volume}{21}}, \bibinfo{pages}{4071–4077}, \doiprefix\url{10.1021/acs.nanolett.1c00984} (\bibinfo{year}{2021}).

\bibitem{vanHeijst2024}
\bibinfo{author}{van Heijst, S.~E.}, \bibinfo{author}{Bolhuis, M.}, \bibinfo{author}{Brokkelkamp, A.}, \bibinfo{author}{Sangers, J. J.~M.} \& \bibinfo{author}{Conesa-Boj, S.}
\newblock \bibinfo{journal}{\bibinfo{title}{Heterostrain-driven bandgap increase in twisted {WS}$_2$: A nanoscale study}}.
\newblock {\emph{\JournalTitle{Advanced Functional Materials}}} \textbf{\bibinfo{volume}{34}}, \bibinfo{pages}{2307893}, \doiprefix\url{10.1002/adfm.202307893} (\bibinfo{year}{2024}).

\bibitem{Kim2023-a}
\bibinfo{author}{Kim, J.~H.}, \bibinfo{author}{Sung, H.} \& \bibinfo{author}{Lee, G.-H.}
\newblock \bibinfo{journal}{\bibinfo{title}{Phase engineering of two‐dimensional transition metal dichalcogenides}}.
\newblock {\emph{\JournalTitle{Small Science}}} \textbf{\bibinfo{volume}{4}}, \bibinfo{pages}{2300093}, \doiprefix\url{10.1002/smsc.202300093} (\bibinfo{year}{2023}).

\bibitem{zeng2012valley}
\bibinfo{author}{Zeng, H.}, \bibinfo{author}{Dai, J.}, \bibinfo{author}{Yao, W.}, \bibinfo{author}{Xiao, D.} \& \bibinfo{author}{Cui, X.}
\newblock \bibinfo{journal}{\bibinfo{title}{Valley polarization in {MoS}$_2$ monolayers by optical pumping}}.
\newblock {\emph{\JournalTitle{Nature Nanotechnology}}} \textbf{\bibinfo{volume}{7}}, \bibinfo{pages}{490--493}, \doiprefix\url{10.1038/nnano.2012.95} (\bibinfo{year}{2012}).

\bibitem{morpurgo2006intervalley}
\bibinfo{author}{Morpurgo, A.} \& \bibinfo{author}{Guinea, F.}
\newblock \bibinfo{journal}{\bibinfo{title}{Intervalley scattering, long-range disorder, and effective time-reversal symmetry breaking in graphene}}.
\newblock {\emph{\JournalTitle{Physical Review Letters}}} \textbf{\bibinfo{volume}{97}}, \bibinfo{pages}{196804}, \doiprefix\url{10.1103/PhysRevLett.97.196804} (\bibinfo{year}{2006}).

\bibitem{PhysRevB.90.161302}
\bibinfo{author}{Zhu, C.~R.} \emph{et~al.}
\newblock \bibinfo{journal}{\bibinfo{title}{Exciton valley dynamics probed by {K}err rotation in {WSe}$_2$ monolayers}}.
\newblock {\emph{\JournalTitle{Physical Review B}}} \textbf{\bibinfo{volume}{90}}, \bibinfo{pages}{161302}, \doiprefix\url{10.1103/PhysRevB.90.161302} (\bibinfo{year}{2014}).

\bibitem{robert2016exciton}
\bibinfo{author}{Robert, C.} \emph{et~al.}
\newblock \bibinfo{journal}{\bibinfo{title}{Exciton radiative lifetime in transition metal dichalcogenide monolayers}}.
\newblock {\emph{\JournalTitle{Physical Review B}}} \textbf{\bibinfo{volume}{93}}, \bibinfo{pages}{205423}, \doiprefix\url{10.1103/PhysRevB.93.205423} (\bibinfo{year}{2016}).

\bibitem{moody2016exciton}
\bibinfo{author}{Moody, G.}, \bibinfo{author}{Schaibley, J.} \& \bibinfo{author}{Xu, X.}
\newblock \bibinfo{journal}{\bibinfo{title}{Exciton dynamics in monolayer transition metal dichalcogenides}}.
\newblock {\emph{\JournalTitle{Journal of the Optical Society of America B}}} \textbf{\bibinfo{volume}{33}}, \bibinfo{pages}{C39--C49}, \doiprefix\url{10.1364/JOSAB.33.000C39} (\bibinfo{year}{2016}).

\bibitem{bae2022k}
\bibinfo{author}{Bae, S.} \emph{et~al.}
\newblock \bibinfo{journal}{\bibinfo{title}{{K}-point longitudinal acoustic phonons are responsible for ultrafast intervalley scattering in monolayer {MoSe}$_2$}}.
\newblock {\emph{\JournalTitle{Nature Communications}}} \textbf{\bibinfo{volume}{13}}, \bibinfo{pages}{4279}, \doiprefix\url{10.1038/s41467-022-32008-6} (\bibinfo{year}{2022}).

\bibitem{lin2022phonon}
\bibinfo{author}{Lin, Z.} \emph{et~al.}
\newblock \bibinfo{journal}{\bibinfo{title}{Phonon-limited valley polarization in transition-metal dichalcogenides}}.
\newblock {\emph{\JournalTitle{Physical Review Letters}}} \textbf{\bibinfo{volume}{129}}, \bibinfo{pages}{027401}, \doiprefix\url{10.1103/PhysRevLett.129.027401} (\bibinfo{year}{2022}).

\bibitem{jeong2020valley}
\bibinfo{author}{Jeong, T.-Y.} \emph{et~al.}
\newblock \bibinfo{journal}{\bibinfo{title}{Valley depolarization in monolayer transition-metal dichalcogenides with zone-corner acoustic phonons}}.
\newblock {\emph{\JournalTitle{Nanoscale}}} \textbf{\bibinfo{volume}{12}}, \bibinfo{pages}{22487--22494}, \doiprefix\url{10.1039/D0NR04761A} (\bibinfo{year}{2020}).

\bibitem{yang2020exciton}
\bibinfo{author}{Yang, M.} \emph{et~al.}
\newblock \bibinfo{journal}{\bibinfo{title}{Exciton valley depolarization in monolayer transition-metal dichalcogenides}}.
\newblock {\emph{\JournalTitle{Physical Review B}}} \textbf{\bibinfo{volume}{101}}, \bibinfo{pages}{115307}, \doiprefix\url{10.1103/PhysRevB.101.115307} (\bibinfo{year}{2020}).

\bibitem{ye2017optical}
\bibinfo{author}{Ye, Z.}, \bibinfo{author}{Sun, D.} \& \bibinfo{author}{Heinz, T.~F.}
\newblock \bibinfo{journal}{\bibinfo{title}{Optical manipulation of valley pseudospin}}.
\newblock {\emph{\JournalTitle{Nature Physics}}} \textbf{\bibinfo{volume}{13}}, \bibinfo{pages}{26--29}, \doiprefix\url{10.1038/nphys3891} (\bibinfo{year}{2017}).

\bibitem{pattanayak2022steady}
\bibinfo{author}{Pattanayak, A.~K.} \emph{et~al.}
\newblock \bibinfo{journal}{\bibinfo{title}{A steady-state approach for studying valley relaxation using an optical vortex beam}}.
\newblock {\emph{\JournalTitle{Nano Letters}}} \textbf{\bibinfo{volume}{22}}, \bibinfo{pages}{4712–4717}, \doiprefix\url{10.1021/acs.nanolett.2c00824} (\bibinfo{year}{2022}).

\bibitem{schaibley2016valleytronics}
\bibinfo{author}{Schaibley, J.~R.} \emph{et~al.}
\newblock \bibinfo{journal}{\bibinfo{title}{Valleytronics in {2D} materials}}.
\newblock {\emph{\JournalTitle{Nature Reviews Materials}}} \textbf{\bibinfo{volume}{1}}, \bibinfo{pages}{16055}, \doiprefix\url{10.1038/natrevmats.2016.55} (\bibinfo{year}{2016}).

\bibitem{zipfel2020light}
\bibinfo{author}{Zipfel, J.} \emph{et~al.}
\newblock \bibinfo{journal}{\bibinfo{title}{Light--matter coupling and non-equilibrium dynamics of exchange-split trions in monolayer {WS}$_2$}}.
\newblock {\emph{\JournalTitle{Journal of Chemical Physics}}} \textbf{\bibinfo{volume}{153}}, \bibinfo{pages}{034706}, \doiprefix\url{10.1063/5.0012721} (\bibinfo{year}{2020}).

\bibitem{Zhumagulov2022}
\bibinfo{author}{Zhumagulov, Y.~V.}, \bibinfo{author}{Vagov, A.}, \bibinfo{author}{Gulevich, D.~R.} \& \bibinfo{author}{Perebeinos, V.}
\newblock \bibinfo{journal}{\bibinfo{title}{Electrostatic and environmental control of the trion fine structure in transition metal dichalcogenide monolayers}}.
\newblock {\emph{\JournalTitle{Nanomaterials}}} \textbf{\bibinfo{volume}{12}}, \bibinfo{pages}{3728}, \doiprefix\url{10.3390/nano12213728} (\bibinfo{year}{2022}).

\bibitem{dey2017gate}
\bibinfo{author}{Dey, P.} \emph{et~al.}
\newblock \bibinfo{journal}{\bibinfo{title}{Gate-controlled spin-valley locking of resident carriers in {WSe}$_2$ monolayers}}.
\newblock {\emph{\JournalTitle{Physical Review Letters}}} \textbf{\bibinfo{volume}{119}}, \bibinfo{pages}{137401}, \doiprefix\url{10.1103/PhysRevLett.119.137401} (\bibinfo{year}{2017}).

\bibitem{zhang2022prolonging}
\bibinfo{author}{Zhang, Q.} \emph{et~al.}
\newblock \bibinfo{journal}{\bibinfo{title}{Prolonging valley polarization lifetime through gate-controlled exciton-to-trion conversion in monolayer molybdenum ditelluride}}.
\newblock {\emph{\JournalTitle{Nature Communications}}} \textbf{\bibinfo{volume}{13}}, \bibinfo{pages}{4101}, \doiprefix\url{10.1038/s41467-022-31672-y} (\bibinfo{year}{2022}).

\bibitem{siao2025electrostatic}
\bibinfo{author}{Siao, J.-Y.}, \bibinfo{author}{Lin, H.-L.}, \bibinfo{author}{Lin, T.-C.}, \bibinfo{author}{Chu, Y.-H.} \& \bibinfo{author}{Lin, M.-T.}
\newblock \bibinfo{journal}{\bibinfo{title}{Electrostatic modulation of valley polarization via a single-contact method in monolayer {WSe}$_2$ for valleytronic devices}}.
\newblock {\emph{\JournalTitle{ACS Applied Nano Materials}}} \textbf{\bibinfo{volume}{8}}, \bibinfo{pages}{7520--7529}, \doiprefix\url{10.1021/acsanm.4c07356} (\bibinfo{year}{2025}).

\bibitem{rivera2016valley}
\bibinfo{author}{Rivera, P.} \emph{et~al.}
\newblock \bibinfo{journal}{\bibinfo{title}{Valley-polarized exciton dynamics in a {2D} semiconductor heterostructure}}.
\newblock {\emph{\JournalTitle{Science}}} \textbf{\bibinfo{volume}{351}}, \bibinfo{pages}{688--691}, \doiprefix\url{10.1126/science.aac7820} (\bibinfo{year}{2016}).

\bibitem{liu2020room}
\bibinfo{author}{Liu, S.} \emph{et~al.}
\newblock \bibinfo{journal}{\bibinfo{title}{Room-temperature valley polarization in atomically thin semiconductors via chalcogenide alloying}}.
\newblock {\emph{\JournalTitle{ACS Nano}}} \textbf{\bibinfo{volume}{14}}, \bibinfo{pages}{9873--9883}, \doiprefix\url{10.1021/acsnano.0c02703} (\bibinfo{year}{2020}).

\bibitem{an2023strain}
\bibinfo{author}{An, Z.} \emph{et~al.}
\newblock \bibinfo{journal}{\bibinfo{title}{Strain control of exciton and trion spin-valley dynamics in monolayer transition metal dichalcogenides}}.
\newblock {\emph{\JournalTitle{Physical Review B}}} \textbf{\bibinfo{volume}{108}}, \bibinfo{pages}{L041404}, \doiprefix\url{10.1103/PhysRevB.108.L041404} (\bibinfo{year}{2023}).

\bibitem{Ho2020SHG}
\bibinfo{author}{Ho, Y.~W.} \emph{et~al.}
\newblock \bibinfo{journal}{\bibinfo{title}{Measuring valley polarization in two-dimensional materials with second-harmonic spectroscopy}}.
\newblock {\emph{\JournalTitle{ACS Photonics}}} \textbf{\bibinfo{volume}{7}}, \bibinfo{pages}{925--931}, \doiprefix\url{10.1021/acsphotonics.0c00174} (\bibinfo{year}{2020}).

\bibitem{Mouchliadis2021SHG}
\bibinfo{author}{Mouchliadis, L.} \emph{et~al.}
\newblock \bibinfo{journal}{\bibinfo{title}{Probing valley population imbalance in transition metal dichalcogenides via temperature-dependent second harmonic generation imaging}}.
\newblock {\emph{\JournalTitle{npj 2D Materials and Applications}}} \textbf{\bibinfo{volume}{5}}, \bibinfo{pages}{6}, \doiprefix\url{10.1038/s41699-020-00183-z} (\bibinfo{year}{2021}).

\bibitem{Koerkamp1996NLOKerr}
\bibinfo{author}{Koerkamp, M.~G.} \& \bibinfo{author}{Rasing, T.}
\newblock \bibinfo{journal}{\bibinfo{title}{Giant nonlinear {K}err effects}}.
\newblock {\emph{\JournalTitle{Journal of Magnetism and Magnetic Materials}}} \textbf{\bibinfo{volume}{156}}, \bibinfo{pages}{213--214}, \doiprefix\url{10.1016/0304-8853(95)00844-6} (\bibinfo{year}{1996}).

\bibitem{Matsubara2012NLOKerr}
\bibinfo{author}{Matsubara, M.}, \bibinfo{author}{Schmehl, A.}, \bibinfo{author}{Mannhart, J.}, \bibinfo{author}{Schlom, D.~G.} \& \bibinfo{author}{Fiebig, M.}
\newblock \bibinfo{journal}{\bibinfo{title}{Giant third-order magneto-optical rotation in ferromagnetic {EuO}}}.
\newblock {\emph{\JournalTitle{Physical Review B}}} \textbf{\bibinfo{volume}{86}}, \bibinfo{pages}{195127}, \doiprefix\url{10.1103/PhysRevB.86.195127} (\bibinfo{year}{2012}).

\bibitem{Shuang2023NLOKerr}
\bibinfo{author}{Wu, S.} \emph{et~al.}
\newblock \bibinfo{journal}{\bibinfo{title}{Extrinsic nonlinear {K}err rotation in topological materials under a magnetic field}}.
\newblock {\emph{\JournalTitle{ACS Nano}}} \textbf{\bibinfo{volume}{17}}, \bibinfo{pages}{18905--18913}, \doiprefix\url{10.1021/acsnano.3c04153} (\bibinfo{year}{2023}).

\bibitem{Herrmann2024SHG}
\bibinfo{author}{Herrmann, P.} \emph{et~al.}
\newblock \bibinfo{journal}{\bibinfo{title}{Nonlinear valley selection rules and all-optical probe of broken time-reversal symmetry in monolayer {WSe}$_2$}}.
\newblock {\emph{\JournalTitle{Nature Photonics}}} \textbf{\bibinfo{volume}{19}}, \bibinfo{pages}{300--306}, \doiprefix\url{10.1038/s41566-024-01591-z} (\bibinfo{year}{2025}).

\bibitem{kim2014ultrafast}
\bibinfo{author}{Kim, J.} \emph{et~al.}
\newblock \bibinfo{journal}{\bibinfo{title}{Ultrafast generation of pseudo-magnetic field for valley excitons in {WSe}$_2$ monolayers}}.
\newblock {\emph{\JournalTitle{Science}}} \textbf{\bibinfo{volume}{346}}, \bibinfo{pages}{1205--1208}, \doiprefix\url{10.1126/science.1258122} (\bibinfo{year}{2014}).

\bibitem{Wang2015SHG}
\bibinfo{author}{Wang, G.} \emph{et~al.}
\newblock \bibinfo{journal}{\bibinfo{title}{Giant enhancement of the optical second-harmonic emission of {WSe}$_2$ monolayers by laser excitation at exciton resonances}}.
\newblock {\emph{\JournalTitle{Physical Review Letters}}} \textbf{\bibinfo{volume}{114}}, \bibinfo{pages}{097403}, \doiprefix\url{10.1103/PhysRevLett.114.097403} (\bibinfo{year}{2015}).

\bibitem{ma2023photocurrent}
\bibinfo{author}{Ma, Q.}, \bibinfo{author}{Krishna~Kumar, R.}, \bibinfo{author}{Xu, S.-Y.}, \bibinfo{author}{Koppens, F.~H.} \& \bibinfo{author}{Song, J.~C.}
\newblock \bibinfo{journal}{\bibinfo{title}{Photocurrent as a multiphysics diagnostic of quantum materials}}.
\newblock {\emph{\JournalTitle{Nature Reviews Physics}}} \textbf{\bibinfo{volume}{5}}, \bibinfo{pages}{170--184}, \doiprefix\url{10.1038/s42254-022-00551-2} (\bibinfo{year}{2023}).

\bibitem{yin2022tunable}
\bibinfo{author}{Yin, J.} \emph{et~al.}
\newblock \bibinfo{journal}{\bibinfo{title}{Tunable and giant valley-selective {H}all effect in gapped bilayer graphene}}.
\newblock {\emph{\JournalTitle{Science}}} \textbf{\bibinfo{volume}{375}}, \bibinfo{pages}{1398--1402}, \doiprefix\url{10.1126/science.abl4266} (\bibinfo{year}{2022}).

\bibitem{morimoto2016topological}
\bibinfo{author}{Morimoto, T.} \& \bibinfo{author}{Nagaosa, N.}
\newblock \bibinfo{journal}{\bibinfo{title}{Topological nature of nonlinear optical effects in solids}}.
\newblock {\emph{\JournalTitle{Science Advances}}} \textbf{\bibinfo{volume}{2}}, \bibinfo{pages}{e1501524}, \doiprefix\url{10.1126/sciadv.1501524} (\bibinfo{year}{2016}).

\bibitem{wen2020steering}
\bibinfo{author}{Wen, T.} \emph{et~al.}
\newblock \bibinfo{journal}{\bibinfo{title}{Steering valley-polarized emission of monolayer {MoS}$_2$ sandwiched in plasmonic antennas}}.
\newblock {\emph{\JournalTitle{Science Advances}}} \textbf{\bibinfo{volume}{6}}, \bibinfo{pages}{eaao0019}, \doiprefix\url{10.1126/sciadv.aao0019} (\bibinfo{year}{2020}).

\bibitem{zheng2023electron}
\bibinfo{author}{Zheng, L.} \emph{et~al.}
\newblock \bibinfo{journal}{\bibinfo{title}{Electron-induced chirality-selective routing of valley photons via metallic nanostructure}}.
\newblock {\emph{\JournalTitle{Advanced Materials}}} \textbf{\bibinfo{volume}{35}}, \bibinfo{pages}{2204908}, \doiprefix\url{10.1002/adma.202204908} (\bibinfo{year}{2023}).

\bibitem{li2018tailoring}
\bibinfo{author}{Li, Z.} \emph{et~al.}
\newblock \bibinfo{journal}{\bibinfo{title}{Tailoring {MoS}$_2$ valley-polarized photoluminescence with super chiral near-field}}.
\newblock {\emph{\JournalTitle{Advanced Materials}}} \textbf{\bibinfo{volume}{30}}, \bibinfo{pages}{1801908}, \doiprefix\url{10.1002/adma.201801908} (\bibinfo{year}{2018}).

\bibitem{liu2023controlling}
\bibinfo{author}{Liu, Y.} \emph{et~al.}
\newblock \bibinfo{journal}{\bibinfo{title}{Controlling valley-specific light emission from monolayer {MoS}$_2$ with achiral dielectric metasurfaces}}.
\newblock {\emph{\JournalTitle{Nano Letters}}} \textbf{\bibinfo{volume}{23}}, \bibinfo{pages}{6124--6131}, \doiprefix\url{10.1021/acs.nanolett.3c01630} (\bibinfo{year}{2023}).

\bibitem{wang2020routing}
\bibinfo{author}{Wang, J.} \emph{et~al.}
\newblock \bibinfo{journal}{\bibinfo{title}{Routing valley exciton emission of a {WS}$_2$ monolayer via delocalized {B}loch modes of in-plane inversion-symmetry-broken photonic crystal slabs}}.
\newblock {\emph{\JournalTitle{Light: Science and Applications}}} \textbf{\bibinfo{volume}{9}}, \bibinfo{pages}{148}, \doiprefix\url{10.1038/s41377-020-00387-4} (\bibinfo{year}{2020}).

\bibitem{sun2019separation}
\bibinfo{author}{Sun, L.} \emph{et~al.}
\newblock \bibinfo{journal}{\bibinfo{title}{Separation of valley excitons in a {MoS}$_2$ monolayer using a subwavelength asymmetric groove array}}.
\newblock {\emph{\JournalTitle{Nature Photonics}}} \textbf{\bibinfo{volume}{13}}, \bibinfo{pages}{180--184}, \doiprefix\url{10.1038/s41566-019-0348-z} (\bibinfo{year}{2019}).

\bibitem{Chervy2018}
\bibinfo{author}{Chervy, T.} \emph{et~al.}
\newblock \bibinfo{journal}{\bibinfo{title}{Room temperature chiral coupling of valley excitons with spin-momentum locked surface plasmons}}.
\newblock {\emph{\JournalTitle{ACS Photonics}}} \textbf{\bibinfo{volume}{5}}, \bibinfo{pages}{1281--1287}, \doiprefix\url{10.1021/acsphotonics.7b01032} (\bibinfo{year}{2018}).

\bibitem{Gong2018}
\bibinfo{author}{Gong, S.-H.}, \bibinfo{author}{Alpeggiani, F.}, \bibinfo{author}{Sciacca, B.}, \bibinfo{author}{Garnett, E.~C.} \& \bibinfo{author}{Kuipers, L.}
\newblock \bibinfo{journal}{\bibinfo{title}{Nanoscale chiral valley-photon interface through optical spin-orbit coupling}}.
\newblock {\emph{\JournalTitle{Science}}} \textbf{\bibinfo{volume}{359}}, \bibinfo{pages}{443--447}, \doiprefix\url{10.1126/science.aan8010} (\bibinfo{year}{2018}).

\bibitem{Liu2023}
\bibinfo{author}{Liu, B.} \emph{et~al.}
\newblock \bibinfo{journal}{\bibinfo{title}{Long-range propagation of exciton-polaritons in large-area {2D} semiconductor monolayers}}.
\newblock {\emph{\JournalTitle{ACS Nano}}} \textbf{\bibinfo{volume}{17}}, \bibinfo{pages}{14442–14448}, \doiprefix\url{10.1021/acsnano.3c03485} (\bibinfo{year}{2023}).

\bibitem{Raziman2019}
\bibinfo{author}{Raziman, T.~V.}, \bibinfo{author}{Godiksen, R.~H.}, \bibinfo{author}{M\"{u}ller, M.~A.} \& \bibinfo{author}{Curto, A.~G.}
\newblock \bibinfo{journal}{\bibinfo{title}{Conditions for enhancing chiral nanophotonics near achiral nanoparticles}}.
\newblock {\emph{\JournalTitle{ACS Photonics}}} \textbf{\bibinfo{volume}{6}}, \bibinfo{pages}{2583--2589}, \doiprefix\url{10.1021/acsphotonics.9b01200} (\bibinfo{year}{2019}).

\bibitem{bucher2024influence}
\bibinfo{author}{Bucher, T.} \emph{et~al.}
\newblock \bibinfo{journal}{\bibinfo{title}{Influence of resonant plasmonic nanoparticles on optically accessing the valley degree of freedom in {2D} semiconductors}}.
\newblock {\emph{\JournalTitle{Nature Communications}}} \textbf{\bibinfo{volume}{15}}, \bibinfo{pages}{10098}, \doiprefix\url{10.1038/s41467-024-54359-y} (\bibinfo{year}{2024}).

\bibitem{Raziman2022}
\bibinfo{author}{Raziman, T.~V.}, \bibinfo{author}{Visser, C.~P.}, \bibinfo{author}{Wang, S.}, \bibinfo{author}{G\'{o}mez~Rivas, J.} \& \bibinfo{author}{Curto, A.~G.}
\newblock \bibinfo{journal}{\bibinfo{title}{Exciton diffusion and annihilation in nanophotonic {P}urcell landscapes}}.
\newblock {\emph{\JournalTitle{Advanced Optical Materials}}} \textbf{\bibinfo{volume}{10}}, \bibinfo{pages}{2200103}, \doiprefix\url{10.1002/adom.202200103} (\bibinfo{year}{2022}).

\bibitem{duan2023valley}
\bibinfo{author}{Duan, X.} \emph{et~al.}
\newblock \bibinfo{journal}{\bibinfo{title}{Valley-addressable monolayer lasing through spin-controlled {B}erry phase photonic cavities}}.
\newblock {\emph{\JournalTitle{Science}}} \textbf{\bibinfo{volume}{381}}, \bibinfo{pages}{1429--1432}, \doiprefix\url{10.1126/science.adi7196} (\bibinfo{year}{2023}).

\bibitem{guddala2021all}
\bibinfo{author}{Guddala, S.} \emph{et~al.}
\newblock \bibinfo{journal}{\bibinfo{title}{All-optical nonreciprocity due to valley polarization pumping in transition metal dichalcogenides}}.
\newblock {\emph{\JournalTitle{Nature Communications}}} \textbf{\bibinfo{volume}{12}}, \bibinfo{pages}{3746}, \doiprefix\url{10.1038/s41467-021-24138-0} (\bibinfo{year}{2021}).

\bibitem{turunen2022quantum}
\bibinfo{author}{Turunen, M.} \emph{et~al.}
\newblock \bibinfo{journal}{\bibinfo{title}{Quantum photonics with layered {2D} materials}}.
\newblock {\emph{\JournalTitle{Nature Reviews Physics}}} \textbf{\bibinfo{volume}{4}}, \bibinfo{pages}{219--236}, \doiprefix\url{10.1038/s42254-021-00408-0} (\bibinfo{year}{2022}).

\bibitem{reserbat2021quantum}
\bibinfo{author}{Reserbat-Plantey, A.} \emph{et~al.}
\newblock \bibinfo{journal}{\bibinfo{title}{Quantum nanophotonics in two-dimensional materials}}.
\newblock {\emph{\JournalTitle{ACS Photonics}}} \textbf{\bibinfo{volume}{8}}, \bibinfo{pages}{85--101}, \doiprefix\url{10.1021/acsphotonics.0c01224} (\bibinfo{year}{2021}).

\bibitem{gonzalez2024light}
\bibinfo{author}{Gonz{\'a}lez-Tudela, A.}, \bibinfo{author}{Reiserer, A.}, \bibinfo{author}{Garc{\'\i}a-Ripoll, J.~J.} \& \bibinfo{author}{Garc{\'\i}a-Vidal, F.~J.}
\newblock \bibinfo{journal}{\bibinfo{title}{Light--matter interactions in quantum nanophotonic devices}}.
\newblock {\emph{\JournalTitle{Nature Reviews Physics}}} \textbf{\bibinfo{volume}{6}}, \bibinfo{pages}{166–179}, \doiprefix\url{10.1038/s42254-023-00681-1} (\bibinfo{year}{2024}).

\bibitem{Cui_NS_22}
\bibinfo{author}{Cui, X.} \emph{et~al.}
\newblock \bibinfo{journal}{\bibinfo{title}{On-chip photonics and optoelectronics with a van der {Waals} material dielectric platform}}.
\newblock {\emph{\JournalTitle{Nanoscale}}} \textbf{\bibinfo{volume}{14}}, \bibinfo{pages}{9459--9465}, \doiprefix\url{10.1039/D2NR01042A} (\bibinfo{year}{2022}).

\bibitem{Yang_AM_2025}
\bibinfo{author}{Luo, Y.}, \bibinfo{author}{Sun, Z.}, \bibinfo{author}{Sun, Z.} \& \bibinfo{author}{Dai, Q.}
\newblock \bibinfo{journal}{\bibinfo{title}{Ultrafast infrared plasmonics}}.
\newblock {\emph{\JournalTitle{Advanced Materials}}} \textbf{\bibinfo{volume}{37}}, \bibinfo{pages}{2413748}, \doiprefix\url{10.1002/adma.202413748} (\bibinfo{year}{2025}).

\bibitem{Zhang_SA_22}
\bibinfo{author}{Zhang, Y.} \emph{et~al.}
\newblock \bibinfo{journal}{\bibinfo{title}{Chirality logic gates}}.
\newblock {\emph{\JournalTitle{Science Advances}}} \textbf{\bibinfo{volume}{8}}, \bibinfo{pages}{eabq8246}, \doiprefix\url{10.1126/sciadv.abq8246} (\bibinfo{year}{2022}).

\bibitem{Zhang_APL_23}
\bibinfo{author}{Zhang, Y.}, \bibinfo{author}{Arias-Munoz, J.~C.}, \bibinfo{author}{Cui, X.} \& \bibinfo{author}{Sun, Z.}
\newblock \bibinfo{journal}{\bibinfo{title}{Prospect of optical chirality logic computing}}.
\newblock {\emph{\JournalTitle{Applied Physics Letters}}} \textbf{\bibinfo{volume}{123}}, \bibinfo{pages}{240501}, \doiprefix\url{10.1063/5.0178917} (\bibinfo{year}{2023}).

\bibitem{Cui_SA_2025}
\bibinfo{author}{Cui, X.} \emph{et~al.}
\newblock \bibinfo{journal}{\bibinfo{title}{Miniaturized spectral sensing with a tunable optoelectronic interface}}.
\newblock {\emph{\JournalTitle{Science Advances}}} \textbf{\bibinfo{volume}{11}}, \bibinfo{pages}{eado6886}, \doiprefix\url{10.1126/sciadv.ado6886} (\bibinfo{year}{2025}).

\bibitem{Das_LSA_2025}
\bibinfo{author}{Das, S.} \emph{et~al.}
\newblock \bibinfo{journal}{\bibinfo{title}{Nanoscale thickness octave-spanning coherent supercontinuum light generation}}.
\newblock {\emph{\JournalTitle{Light: Science \& Applications}}} \textbf{\bibinfo{volume}{14}}, \bibinfo{pages}{41}, \doiprefix\url{10.1038/s41377-024-01660-6} (\bibinfo{year}{2025}).

\bibitem{Wu_AM_22}
\bibinfo{author}{Wu, C.} \emph{et~al.}
\newblock \bibinfo{journal}{\bibinfo{title}{Ultrasensitive mid-infrared biosensing in aqueous solutions with graphene plasmons}}.
\newblock {\emph{\JournalTitle{Advanced Materials}}} \textbf{\bibinfo{volume}{34}}, \bibinfo{pages}{2110525}, \doiprefix\url{10.1002/adma.202110525} (\bibinfo{year}{2022}).

\bibitem{Hu_NC_19}
\bibinfo{author}{Hu, H.} \emph{et~al.}
\newblock \bibinfo{journal}{\bibinfo{title}{Gas identification with graphene plasmons}}.
\newblock {\emph{\JournalTitle{Nature Communications}}} \textbf{\bibinfo{volume}{10}}, \bibinfo{pages}{1131}, \doiprefix\url{10.1038/s41467-019-09008-0} (\bibinfo{year}{2019}).

\bibitem{zhong2020quantum}
\bibinfo{author}{Zhong, H.-S.} \emph{et~al.}
\newblock \bibinfo{journal}{\bibinfo{title}{Quantum computational advantage using photons}}.
\newblock {\emph{\JournalTitle{Science}}} \textbf{\bibinfo{volume}{370}}, \bibinfo{pages}{1460--1463}, \doiprefix\url{10.1126/science.abe8770} (\bibinfo{year}{2020}).

\bibitem{o2009photonic}
\bibinfo{author}{O'Brien, J.~L.}, \bibinfo{author}{Furusawa, A.} \& \bibinfo{author}{Vu{\v{c}}kovi{\'c}, J.}
\newblock \bibinfo{journal}{\bibinfo{title}{Photonic quantum technologies}}.
\newblock {\emph{\JournalTitle{Nature Photonics}}} \textbf{\bibinfo{volume}{3}}, \bibinfo{pages}{687--695}, \doiprefix\url{10.1038/nphoton.2009.229} (\bibinfo{year}{2009}).

\bibitem{tonndorf2015single}
\bibinfo{author}{Tonndorf, P.} \emph{et~al.}
\newblock \bibinfo{journal}{\bibinfo{title}{Single-photon emission from localized excitons in an atomically thin semiconductor}}.
\newblock {\emph{\JournalTitle{Optica}}} \textbf{\bibinfo{volume}{2}}, \bibinfo{pages}{347--352}, \doiprefix\url{10.1364/OPTICA.2.000347} (\bibinfo{year}{2015}).

\bibitem{montblanch2023layered}
\bibinfo{author}{Montblanch, A. R.-P.}, \bibinfo{author}{Barbone, M.}, \bibinfo{author}{Aharonovich, I.}, \bibinfo{author}{Atat{\"u}re, M.} \& \bibinfo{author}{Ferrari, A.~C.}
\newblock \bibinfo{journal}{\bibinfo{title}{Layered materials as a platform for quantum technologies}}.
\newblock {\emph{\JournalTitle{Nature Nanotechnology}}} \textbf{\bibinfo{volume}{18}}, \bibinfo{pages}{555--571}, \doiprefix\url{10.1038/s41565-023-01354-x} (\bibinfo{year}{2023}).

\bibitem{micevic2022demand}
\bibinfo{author}{Micevic, A.} \emph{et~al.}
\newblock \bibinfo{journal}{\bibinfo{title}{On-demand generation of optically active defects in monolayer {WS}$_2$ by a focused helium ion beam}}.
\newblock {\emph{\JournalTitle{Applied Physics Letters}}} \textbf{\bibinfo{volume}{121}}, \bibinfo{pages}{183101}, \doiprefix\url{10.1063/5.0118697} (\bibinfo{year}{2022}).

\bibitem{kumar2015strain}
\bibinfo{author}{Kumar, S.}, \bibinfo{author}{Kaczmarczyk, A.} \& \bibinfo{author}{Gerardot, B.~D.}
\newblock \bibinfo{journal}{\bibinfo{title}{Strain-induced spatial and spectral isolation of quantum emitters in mono-and bilayer {WSe\(_2\)}}}.
\newblock {\emph{\JournalTitle{Nano letters}}} \textbf{\bibinfo{volume}{15}}, \bibinfo{pages}{7567--7573} (\bibinfo{year}{2015}).

\bibitem{munkhbat2020electrical}
\bibinfo{author}{Munkhbat, B.} \emph{et~al.}
\newblock \bibinfo{journal}{\bibinfo{title}{Electrical control of hybrid monolayer tungsten disulfide--plasmonic nanoantenna light--matter states at cryogenic and room temperatures}}.
\newblock {\emph{\JournalTitle{ACS nano}}} \textbf{\bibinfo{volume}{14}}, \bibinfo{pages}{1196--1206}, \doiprefix\url{10.1021/acsnano.9b09684} (\bibinfo{year}{2020}).

\bibitem{klein2019site}
\bibinfo{author}{Klein, J.} \emph{et~al.}
\newblock \bibinfo{journal}{\bibinfo{title}{Site-selectively generated photon emitters in monolayer {MoS\(_2\)} via local helium ion irradiation}}.
\newblock {\emph{\JournalTitle{Nature Communications}}} \textbf{\bibinfo{volume}{10}}, \bibinfo{pages}{2755}, \doiprefix\url{10.1038/s41467-019-10632-z} (\bibinfo{year}{2019}).

\bibitem{tran2016quantum}
\bibinfo{author}{Tran, T.~T.}, \bibinfo{author}{Bray, K.}, \bibinfo{author}{Ford, M.~J.}, \bibinfo{author}{Toth, M.} \& \bibinfo{author}{Aharonovich, I.}
\newblock \bibinfo{journal}{\bibinfo{title}{Quantum emission from hexagonal boron nitride monolayers}}.
\newblock {\emph{\JournalTitle{Nature Nanotechnology}}} \textbf{\bibinfo{volume}{11}}, \bibinfo{pages}{37--41}, \doiprefix\url{10.1038/nnano.2015.242} (\bibinfo{year}{2016}).

\bibitem{palacios2017large}
\bibinfo{author}{Palacios-Berraquero, C.} \emph{et~al.}
\newblock \bibinfo{journal}{\bibinfo{title}{Large-scale quantum-emitter arrays in atomically thin semiconductors}}.
\newblock {\emph{\JournalTitle{Nature Communications}}} \textbf{\bibinfo{volume}{8}}, \bibinfo{pages}{15093}, \doiprefix\url{10.1038/ncomms15093} (\bibinfo{year}{2017}).

\bibitem{yu2021site}
\bibinfo{author}{Yu, L.} \emph{et~al.}
\newblock \bibinfo{journal}{\bibinfo{title}{Site-controlled quantum emitters in monolayer {MoSe}$_2$}}.
\newblock {\emph{\JournalTitle{Nano Letters}}} \textbf{\bibinfo{volume}{21}}, \bibinfo{pages}{2376--2381}, \doiprefix\url{10.1021/acs.nanolett.0c04282} (\bibinfo{year}{2021}).

\bibitem{errando-herranz_resonance_2021}
\bibinfo{author}{Errando-Herranz, C.} \emph{et~al.}
\newblock \bibinfo{journal}{\bibinfo{title}{Resonance {Fluorescence} from {Waveguide}-{Coupled}, {Strain}-{Localized}, {Two}-{Dimensional} {Quantum} {Emitters}}}.
\newblock {\emph{\JournalTitle{ACS Photonics}}} \textbf{\bibinfo{volume}{8}}, \bibinfo{pages}{1069--1076}, \doiprefix\url{10.1021/acsphotonics.0c01653} (\bibinfo{year}{2021}).

\bibitem{tonndorf2017chip}
\bibinfo{author}{Tonndorf, P.} \emph{et~al.}
\newblock \bibinfo{journal}{\bibinfo{title}{On-chip waveguide coupling of a layered semiconductor single-photon source}}.
\newblock {\emph{\JournalTitle{Nano Letters}}} \textbf{\bibinfo{volume}{17}}, \bibinfo{pages}{5446--5451}, \doiprefix\url{10.1021/acs.nanolett.7b02092} (\bibinfo{year}{2017}).

\bibitem{sortino2020dielectric}
\bibinfo{author}{Sortino, L.} \emph{et~al.}
\newblock \bibinfo{journal}{\bibinfo{title}{Dielectric nanoantennas for strain engineering in atomically thin two-dimensional semiconductors}}.
\newblock {\emph{\JournalTitle{ACS Photonics}}} \textbf{\bibinfo{volume}{7}}, \bibinfo{pages}{2413--2422}, \doiprefix\url{10.1021/acsphotonics.0c00294} (\bibinfo{year}{2020}).

\bibitem{drawer2023monolayer}
\bibinfo{author}{Drawer, J.-C.} \emph{et~al.}
\newblock \bibinfo{journal}{\bibinfo{title}{Monolayer-based single-photon source in a liquid-helium-free open cavity featuring 65\% brightness and quantum coherence}}.
\newblock {\emph{\JournalTitle{Nano Letters}}} \textbf{\bibinfo{volume}{23}}, \bibinfo{pages}{8683--8689}, \doiprefix\url{10.1021/acs.nanolett.3c02584} (\bibinfo{year}{2023}).

\bibitem{dastidar2022quantum}
\bibinfo{author}{Dastidar, M.~G.}, \bibinfo{author}{Thekkooden, I.}, \bibinfo{author}{Nayak, P.~K.} \& \bibinfo{author}{Bhallamudi, V.~P.}
\newblock \bibinfo{journal}{\bibinfo{title}{Quantum emitters and detectors based on 2{D} van der {W}aals materials}}.
\newblock {\emph{\JournalTitle{Nanoscale}}} \textbf{\bibinfo{volume}{14}}, \bibinfo{pages}{5289--5313}, \doiprefix\url{10.1039/D1NR08193D} (\bibinfo{year}{2022}).

\bibitem{So+2025}
\bibinfo{author}{So, J.-P.}
\newblock \bibinfo{journal}{\bibinfo{title}{Deterministic generation and nanophotonic integration of 2d quantum emitters for advanced quantum photonic functionalities}}.
\newblock {\emph{\JournalTitle{Nanophotonics}}} \doiprefix\url{10.1515/nanoph-2024-0629} (\bibinfo{year}{2025}).

\bibitem{senellart2017high}
\bibinfo{author}{Senellart, P.}, \bibinfo{author}{Solomon, G.} \& \bibinfo{author}{White, A.}
\newblock \bibinfo{journal}{\bibinfo{title}{High-performance semiconductor quantum-dot single-photon sources}}.
\newblock {\emph{\JournalTitle{Nature Nanotechnology}}} \textbf{\bibinfo{volume}{12}}, \bibinfo{pages}{1026--1039}, \doiprefix\url{10.1038/nnano.2017.218} (\bibinfo{year}{2017}).

\bibitem{paralikis2024tailoring}
\bibinfo{author}{Paralikis, A.} \emph{et~al.}
\newblock \bibinfo{journal}{\bibinfo{title}{Tailoring polarization in {WSe\(_2\)} quantum emitters through deterministic strain engineering}}.
\newblock {\emph{\JournalTitle{npj 2D Materials and Applications}}} \textbf{\bibinfo{volume}{8}}, \bibinfo{pages}{59}, \doiprefix\url{10.1038/s41699-024-00497-2} (\bibinfo{year}{2024}).

\bibitem{Piccinini2025}
\bibinfo{author}{Piccinini, C.} \emph{et~al.}
\newblock \bibinfo{journal}{\bibinfo{title}{High-purity and stable single-photon emission in bilayer {WS}e$_{2}$ via phonon-assisted excitation}}.
\newblock {\emph{\JournalTitle{Communications Physics}}} \textbf{\bibinfo{volume}{8}}, \doiprefix\url{10.1038/s42005-025-02080-7} (\bibinfo{year}{2025}).

\bibitem{Parto2021DefectK}
\bibinfo{author}{Parto, K.}, \bibinfo{author}{Azzam, S.~I.}, \bibinfo{author}{Banerjee, K.} \& \bibinfo{author}{Moody, G.}
\newblock \bibinfo{journal}{\bibinfo{title}{{Defect and strain engineering of monolayer {WSe\(_2\)} enables site-controlled single-photon emission up to 150 K}}}.
\newblock {\emph{\JournalTitle{Nature Communications}}} \textbf{\bibinfo{volume}{12}}, \bibinfo{pages}{1--9}, \doiprefix\url{10.1038/s41467-021-23709-5} (\bibinfo{year}{2021}).

\bibitem{branny2017deterministic}
\bibinfo{author}{Branny, A.}, \bibinfo{author}{Kumar, S.}, \bibinfo{author}{Proux, R.} \& \bibinfo{author}{Gerardot, B.~D.}
\newblock \bibinfo{journal}{\bibinfo{title}{Deterministic strain-induced arrays of quantum emitters in a two-dimensional semiconductor}}.
\newblock {\emph{\JournalTitle{Nature Communications}}} \textbf{\bibinfo{volume}{8}}, \bibinfo{pages}{15053}, \doiprefix\url{10.1038/ncomms15053} (\bibinfo{year}{2017}).

\bibitem{vannucci2024single}
\bibinfo{author}{Vannucci, L.} \emph{et~al.}
\newblock \bibinfo{journal}{\bibinfo{title}{Single-photon emitters in {WSe}$_2$: Critical role of phonons on excitation schemes and indistinguishability}}.
\newblock {\emph{\JournalTitle{Physical Review B}}} \textbf{\bibinfo{volume}{109}}, \bibinfo{pages}{245304}, \doiprefix\url{10.1103/PhysRevB.109.245304} (\bibinfo{year}{2024}).

\bibitem{chakraborty2019electrical}
\bibinfo{author}{Chakraborty, C.}, \bibinfo{author}{Jungwirth, N.~R.}, \bibinfo{author}{Fuchs, G.~D.} \& \bibinfo{author}{Vamivakas, A.~N.}
\newblock \bibinfo{journal}{\bibinfo{title}{Electrical manipulation of the fine-structure splitting of {WSe\(_2\)} quantum emitters}}.
\newblock {\emph{\JournalTitle{Physical Review B}}} \textbf{\bibinfo{volume}{99}}, \bibinfo{pages}{045308}, \doiprefix\url{10.1103/PhysRevB.99.045308} (\bibinfo{year}{2019}).

\bibitem{lenferink2022tunable}
\bibinfo{author}{Lenferink, E.~J.} \emph{et~al.}
\newblock \bibinfo{journal}{\bibinfo{title}{Tunable emission from localized excitons deterministically positioned in monolayer p-n junctions}}.
\newblock {\emph{\JournalTitle{ACS Photonics}}} \textbf{\bibinfo{volume}{9}}, \bibinfo{pages}{3067--3074}, \doiprefix\url{10.1021/acsphotonics.2c00811} (\bibinfo{year}{2022}).

\bibitem{howarth2024electroluminescent}
\bibinfo{author}{Howarth, J.} \emph{et~al.}
\newblock \bibinfo{journal}{\bibinfo{title}{Electroluminescent vertical tunneling junctions based on {WSe\(_2\)} monolayer quantum emitter arrays: Exploring tunability with electric and magnetic fields}}.
\newblock {\emph{\JournalTitle{Proceedings of the National Academy of Sciences}}} \textbf{\bibinfo{volume}{121}}, \bibinfo{pages}{e2401757121}, \doiprefix\url{10.1073/pnas.2401757121} (\bibinfo{year}{2024}).

\bibitem{kim2019position}
\bibinfo{author}{Kim, H.}, \bibinfo{author}{Moon, J.~S.}, \bibinfo{author}{Noh, G.}, \bibinfo{author}{Lee, J.} \& \bibinfo{author}{Kim, J.-H.}
\newblock \bibinfo{journal}{\bibinfo{title}{Position and frequency control of strain-induced quantum emitters in {WSe\(_2\)} monolayers}}.
\newblock {\emph{\JournalTitle{Nano Letters}}} \textbf{\bibinfo{volume}{19}}, \bibinfo{pages}{7534--7539}, \doiprefix\url{10.1021/acs.nanolett.9b03421} (\bibinfo{year}{2019}).

\bibitem{Autere_AM_18}
\bibinfo{author}{Autere, A.} \emph{et~al.}
\newblock \bibinfo{journal}{\bibinfo{title}{Nonlinear optics with {2D} layered materials}}.
\newblock {\emph{\JournalTitle{Advanced Materials}}} \textbf{\bibinfo{volume}{30}}, \bibinfo{pages}{1705963}, \doiprefix\url{10.1002/adma.201705963} (\bibinfo{year}{2018}).

\bibitem{Saynatjoki_NC_17}
\bibinfo{author}{Säynätjoki, A.} \emph{et~al.}
\newblock \bibinfo{journal}{\bibinfo{title}{Ultra-strong nonlinear optical processes and trigonal warping in {MoS}$_2$ layers}}.
\newblock {\emph{\JournalTitle{Nature Communications}}} \textbf{\bibinfo{volume}{8}}, \bibinfo{pages}{893}, \doiprefix\url{10.1038/s41467-017-00749-4} (\bibinfo{year}{2017}).

\bibitem{Kim_AM_17}
\bibinfo{author}{Lee, K.~F.} \emph{et~al.}
\newblock \bibinfo{journal}{\bibinfo{title}{Photon-pair generation with a 100 nm thick carbon nanotube film}}.
\newblock {\emph{\JournalTitle{Advanced Materials}}} \textbf{\bibinfo{volume}{29}}, \bibinfo{pages}{1605978}, \doiprefix\url{10.1002/adma.201605978} (\bibinfo{year}{2017}).

\bibitem{xu2022towards}
\bibinfo{author}{Xu, X.} \emph{et~al.}
\newblock \bibinfo{journal}{\bibinfo{title}{Towards compact phase-matched and waveguided nonlinear optics in atomically layered semiconductors}}.
\newblock {\emph{\JournalTitle{Nature Photonics}}} \textbf{\bibinfo{volume}{16}}, \bibinfo{pages}{698--706}, \doiprefix\url{10.1038/s41566-022-01053-4} (\bibinfo{year}{2022}).

\bibitem{Hong_PRL_23}
\bibinfo{author}{Hong, H.} \emph{et~al.}
\newblock \bibinfo{journal}{\bibinfo{title}{Twist phase matching in two-dimensional materials}}.
\newblock {\emph{\JournalTitle{Physical Review Letters}}} \textbf{\bibinfo{volume}{131}}, \bibinfo{pages}{233801}, \doiprefix\url{10.1103/PhysRevLett.131.233801} (\bibinfo{year}{2023}).

\bibitem{vermeulen2023post}
\bibinfo{author}{Vermeulen, N.} \emph{et~al.}
\newblock \bibinfo{journal}{\bibinfo{title}{Post-2000 nonlinear optical materials and measurements: data tables and best practices}}.
\newblock {\emph{\JournalTitle{Journal of Physics: Photonics}}} \textbf{\bibinfo{volume}{5}}, \bibinfo{pages}{035001}, \doiprefix\url{10.1088/2515-7647/ac9e2f} (\bibinfo{year}{2023}).

\bibitem{marini2018constraints}
\bibinfo{author}{Marini, L.}, \bibinfo{author}{Helt, L.}, \bibinfo{author}{Lu, Y.}, \bibinfo{author}{Eggleton, B.~J.} \& \bibinfo{author}{Palomba, S.}
\newblock \bibinfo{journal}{\bibinfo{title}{Constraints on downconversion in atomically thick films}}.
\newblock {\emph{\JournalTitle{Journal of the Optical Society of America B}}} \textbf{\bibinfo{volume}{35}}, \bibinfo{pages}{672}, \doiprefix\url{10.1364/JOSAB.35.000672} (\bibinfo{year}{2018}).

\bibitem{guo2023ultrathin}
\bibinfo{author}{Guo, Q.} \emph{et~al.}
\newblock \bibinfo{journal}{\bibinfo{title}{Ultrathin quantum light source with van der {Waals} {NbOCl}$_2$ crystal}}.
\newblock {\emph{\JournalTitle{Nature}}} \textbf{\bibinfo{volume}{613}}, \bibinfo{pages}{53--59}, \doiprefix\url{10.1038/s41586-022-05393-7} (\bibinfo{year}{2023}).

\bibitem{Du_M_20}
\bibinfo{author}{Du, L.}, \bibinfo{author}{Dai, Y.} \& \bibinfo{author}{Sun, Z.}
\newblock \bibinfo{journal}{\bibinfo{title}{Twisting for tunable nonlinear optics}}.
\newblock {\emph{\JournalTitle{Matter}}} \textbf{\bibinfo{volume}{3}}, \bibinfo{pages}{987--988}, \doiprefix\url{10.1016/j.matt.2020.09.013} (\bibinfo{year}{2020}).

\bibitem{weissflog2024tunable}
\bibinfo{author}{Weissflog, M.~A.} \emph{et~al.}
\newblock \bibinfo{journal}{\bibinfo{title}{A tunable transition metal dichalcogenide entangled photon-pair source}}.
\newblock {\emph{\JournalTitle{Nature Communications}}} \textbf{\bibinfo{volume}{15}}, \bibinfo{pages}{7600}, \doiprefix\url{10.1038/s41467-024-51843-3} (\bibinfo{year}{2024}).

\bibitem{feng2024polarization}
\bibinfo{author}{Feng, J.} \emph{et~al.}
\newblock \bibinfo{journal}{\bibinfo{title}{Polarization-entangled photon-pair source with van der {Waals} {3R}-{WS}$_2$ crystal}}.
\newblock {\emph{\JournalTitle{eLight}}} \textbf{\bibinfo{volume}{4}}, \bibinfo{pages}{16}, \doiprefix\url{10.1186/s43593-024-00074-6} (\bibinfo{year}{2024}).

\bibitem{Zuo_NN_20}
\bibinfo{author}{Zuo, Y.} \emph{et~al.}
\newblock \bibinfo{journal}{\bibinfo{title}{Optical fibres with embedded two-dimensional materials for ultrahigh nonlinearity}}.
\newblock {\emph{\JournalTitle{Nature Nanotechnology}}} \textbf{\bibinfo{volume}{15}}, \bibinfo{pages}{987--991}, \doiprefix\url{10.1038/s41565-020-0770-x} (\bibinfo{year}{2020}).

\bibitem{Chen_NP_19}
\bibinfo{author}{Chen, K.} \emph{et~al.}
\newblock \bibinfo{journal}{\bibinfo{title}{Graphene photonic crystal fibre with strong and tunable light–matter interaction}}.
\newblock {\emph{\JournalTitle{Nature Photonics}}} \textbf{\bibinfo{volume}{13}}, \bibinfo{pages}{754--759}, \doiprefix\url{10.1038/s41566-019-0492-5} (\bibinfo{year}{2019}).

\bibitem{Zhang_LSA_22}
\bibinfo{author}{Zhang, Y.} \emph{et~al.}
\newblock \bibinfo{journal}{\bibinfo{title}{Coherent modulation of chiral nonlinear optics with crystal symmetry}}.
\newblock {\emph{\JournalTitle{Light: Science \& Applications}}} \textbf{\bibinfo{volume}{11}}, \bibinfo{pages}{216}, \doiprefix\url{10.1038/s41377-022-00915-4} (\bibinfo{year}{2022}).

\bibitem{silberhorn2007detecting}
\bibinfo{author}{Silberhorn, C.}
\newblock \bibinfo{journal}{\bibinfo{title}{Detecting quantum light}}.
\newblock {\emph{\JournalTitle{Contemporary Physics}}} \textbf{\bibinfo{volume}{48}}, \bibinfo{pages}{143--156}, \doiprefix\url{10.1080/00107510701662538} (\bibinfo{year}{2007}).

\bibitem{abdullah2024recent}
\bibinfo{author}{Abdullah, M.} \emph{et~al.}
\newblock \bibinfo{journal}{\bibinfo{title}{Recent progress of {2D} materials-based photodetectors from {UV} to {THz} waves: Principles, materials, and applications}}.
\newblock {\emph{\JournalTitle{Small}}} \textbf{\bibinfo{volume}{20}}, \bibinfo{pages}{2402668}, \doiprefix\url{10.1002/smll.202402668} (\bibinfo{year}{2024}).

\bibitem{Koppens2014}
\bibinfo{author}{Koppens, F.~H.} \emph{et~al.}
\newblock \bibinfo{title}{Photodetectors based on graphene, other two-dimensional materials and hybrid systems}, \doiprefix\url{10.1038/nnano.2014.215} (\bibinfo{year}{2014}).

\bibitem{Roy_AM_2018}
\bibinfo{author}{Roy, K.} \emph{et~al.}
\newblock \bibinfo{journal}{\bibinfo{title}{Number-resolved single-photon detection with ultralow noise van der {Waals} hybrid}}.
\newblock {\emph{\JournalTitle{Advanced Materials}}} \textbf{\bibinfo{volume}{30}}, \bibinfo{pages}{1704412}, \doiprefix\url{10.1002/adma.201704412} (\bibinfo{year}{2018}).

\bibitem{Walsh_Science_2021}
\bibinfo{author}{Walsh, E.~D.} \emph{et~al.}
\newblock \bibinfo{journal}{\bibinfo{title}{Josephson junction infrared single-photon detector}}.
\newblock {\emph{\JournalTitle{Science}}} \textbf{\bibinfo{volume}{372}}, \bibinfo{pages}{409--412}, \doiprefix\url{10.1126/science.abf5539} (\bibinfo{year}{2021}).

\bibitem{metuh2025singlephotondetectionsuperconductingniobium}
\bibinfo{author}{Metuh, P.} \emph{et~al.}
\newblock \bibinfo{journal}{\bibinfo{title}{Toward single-photon detection with superconducting niobium diselenide nanowires}}.
\newblock {\emph{\JournalTitle{arXiv:2503.22670}}} \doiprefix\url{10.48550/arXiv.2503.22670} (\bibinfo{year}{2025}).

\bibitem{Quantum_sensing}
\bibinfo{author}{Degen, C.~L.}, \bibinfo{author}{Reinhard, F.} \& \bibinfo{author}{Cappellaro, P.}
\newblock \bibinfo{journal}{\bibinfo{title}{Quantum sensing}}.
\newblock {\emph{\JournalTitle{Reviews of Modern Physics}}} \textbf{\bibinfo{volume}{89}}, \bibinfo{pages}{035002}, \doiprefix\url{10.1103/RevModPhys.89.035002} (\bibinfo{year}{2017}).

\bibitem{Gottscholl_bn_2021}
\bibinfo{author}{Gottscholl, A.} \emph{et~al.}
\newblock \bibinfo{journal}{\bibinfo{title}{Spin defects in {hBN} as promising temperature, pressure and magnetic field quantum sensors}}.
\newblock {\emph{\JournalTitle{Nature Communications}}} \textbf{\bibinfo{volume}{12}}, \bibinfo{pages}{4480}, \doiprefix\url{10.1038/s41467-021-24725-1} (\bibinfo{year}{2021}).

\bibitem{Lyu_NL_2022}
\bibinfo{author}{Lyu, X.} \emph{et~al.}
\newblock \bibinfo{journal}{\bibinfo{title}{Strain quantum sensing with spin defects in hexagonal boron nitride}}.
\newblock {\emph{\JournalTitle{Nano Letters}}} \textbf{\bibinfo{volume}{22}}, \bibinfo{pages}{6553--6559}, \doiprefix\url{10.1021/acs.nanolett.2c01722} (\bibinfo{year}{2022}).

\bibitem{Rizzato_NC_2023}
\bibinfo{author}{Rizzato, R.} \emph{et~al.}
\newblock \bibinfo{journal}{\bibinfo{title}{Extending the coherence of spin defects in {hBN} enables advanced qubit control and quantum sensing}}.
\newblock {\emph{\JournalTitle{Nature Communications}}} \textbf{\bibinfo{volume}{12}}, \bibinfo{pages}{4480}, \doiprefix\url{10.1038/s41467-023-40473-w} (\bibinfo{year}{2021}).

\bibitem{Robertson_ACS_Nano_2023}
\bibinfo{author}{Robertson, I.~O.} \emph{et~al.}
\newblock \bibinfo{journal}{\bibinfo{title}{Detection of paramagnetic spins with an ultrathin van der {Waals} quantum sensor}}.
\newblock {\emph{\JournalTitle{ACS Nano}}} \textbf{\bibinfo{volume}{17}}, \bibinfo{pages}{13408--13417}, \doiprefix\url{10.1021/acsnano.3c01678} (\bibinfo{year}{2023}).

\bibitem{Gao_NL_2021}
\bibinfo{author}{Gao, X.} \emph{et~al.}
\newblock \bibinfo{journal}{\bibinfo{title}{High-contrast plasmonic-enhanced shallow spin defects in hexagonal boron nitride for quantum sensing}}.
\newblock {\emph{\JournalTitle{Nano Letters}}} \textbf{\bibinfo{volume}{21}}, \bibinfo{pages}{7708--7714}, \doiprefix\url{10.1021/acs.nanolett.1c02495} (\bibinfo{year}{2021}).

\bibitem{Sasaki_APL_2023}
\bibinfo{author}{Sasaki, K.} \emph{et~al.}
\newblock \bibinfo{journal}{\bibinfo{title}{Magnetic field imaging by h{BN} quantum sensor nanoarray}}.
\newblock {\emph{\JournalTitle{Applied Physics Letters}}} \textbf{\bibinfo{volume}{122}}, \bibinfo{pages}{244003}, \doiprefix\url{10.1063/5.0147072} (\bibinfo{year}{2023}).

\bibitem{Gao_ACS_Photonics_2023}
\bibinfo{author}{Gao, X.} \emph{et~al.}
\newblock \bibinfo{journal}{\bibinfo{title}{Quantum sensing of paramagnetic spins in liquids with spin qubits in hexagonal boron nitride}}.
\newblock {\emph{\JournalTitle{ACS Photonics}}} \textbf{\bibinfo{volume}{10}}, \bibinfo{pages}{2894--2900}, \doiprefix\url{10.1021/acsphotonics.3c00621} (\bibinfo{year}{2023}).

\bibitem{gilardoni2024singlespinhexagonalboron}
\bibinfo{author}{M.~Gilardoni, C.} \emph{et~al.}
\newblock \bibinfo{journal}{\bibinfo{title}{A single spin in hexagonal boron nitride for vectorial quantum magnetometry}}.
\newblock {\emph{\JournalTitle{Nature Communications}}} \textbf{\bibinfo{volume}{16}}, \doiprefix\url{10.1038/s41467-025-59642-0} (\bibinfo{year}{2025}).

\bibitem{Moon_AOM_2024}
\bibinfo{author}{Moon, J.~S.} \emph{et~al.}
\newblock \bibinfo{journal}{\bibinfo{title}{Fiber-integrated van der {Waals} quantum sensor with an optimal cavity interface}}.
\newblock {\emph{\JournalTitle{Advanced Optical Materials}}} \textbf{\bibinfo{volume}{12}}, \bibinfo{pages}{2401987}, \doiprefix\url{10.1002/adom.202401987} (\bibinfo{year}{2024}).

\bibitem{G14}
\bibinfo{author}{{Garc\'{\i}a de Abajo}, F.~J.}
\newblock \bibinfo{journal}{\bibinfo{title}{Graphene plasmonics: challenges and opportunities}}.
\newblock {\emph{\JournalTitle{ACS Photonics}}} \textbf{\bibinfo{volume}{1}}, \bibinfo{pages}{135--152}, \doiprefix\url{10.1021/ph400147y} (\bibinfo{year}{2014}).

\bibitem{G14_2}
\bibinfo{author}{{Garc\'{\i}a de Abajo}, F.~J.}
\newblock \bibinfo{journal}{\bibinfo{title}{Multiple excitation of confined graphene plasmons by single free electrons}}.
\newblock {\emph{\JournalTitle{ACS Nano}}} \textbf{\bibinfo{volume}{7}}, \bibinfo{pages}{11409--11419}, \doiprefix\url{10.1021/nn405367e} (\bibinfo{year}{2013}).

\bibitem{GP16}
\bibinfo{author}{Gon\c{c}alves, P. A.~D.} \& \bibinfo{author}{Peres, N. M.~R.}
\newblock \emph{\bibinfo{title}{An Introduction to Graphene Plasmonics}} (\bibinfo{publisher}{World Scientific}, \bibinfo{address}{Singapore}, \bibinfo{year}{2016}).

\bibitem{G25}
\bibinfo{author}{{{Garc\'{\i}a de Abajo}, F. J. {\it et al.}}}
\newblock \bibinfo{journal}{\bibinfo{title}{Roadmap for photonics with {2D} materials}}.
\newblock {\emph{\JournalTitle{ACS Photonics}}} \textbf{\bibinfo{volume}{12}}, \bibinfo{pages}{10.1021/acsphotonics.5c00353}, \doiprefix\url{10.1021/acsphotonics.5c00353} (\bibinfo{year}{2025}).

\bibitem{LCZ14}
\bibinfo{author}{Li, Y.} \emph{et~al.}
\newblock \bibinfo{journal}{\bibinfo{title}{Measurement of the optical dielectric function of monolayer transition-metal dichalcogenides: {MoS}$_2$, {MoS}e$_2$, {WS}$_2$, and {WS}e$_2$}}.
\newblock {\emph{\JournalTitle{Phys. Rev. B}}} \textbf{\bibinfo{volume}{90}}, \bibinfo{pages}{205422}, \doiprefix\url{10.1103/PhysRevB.90.205422} (\bibinfo{year}{2014}).

\bibitem{WSA24}
\bibinfo{author}{Woo, S.~Y.} \emph{et~al.}
\newblock \bibinfo{journal}{\bibinfo{title}{Engineering {2D} material exciton line shape with graphene/{hBN} encapsulation}}.
\newblock {\emph{\JournalTitle{Nano Lett.}}} \textbf{\bibinfo{volume}{24}}, \bibinfo{pages}{3678--3685}, \doiprefix\url{10.1021/acs.nanolett.3c05063} (\bibinfo{year}{2024}).

\bibitem{H65}
\bibinfo{author}{Hedin, L.}
\newblock \bibinfo{journal}{\bibinfo{title}{New method for calculating the one-particle {G}reen's function with application to the electron-gas problem}}.
\newblock {\emph{\JournalTitle{Phys. Rev.}}} \textbf{\bibinfo{volume}{139}}, \bibinfo{pages}{A796--A823}, \doiprefix\url{10.1103/PhysRev.139.A796} (\bibinfo{year}{1965}).

\bibitem{ORR02}
\bibinfo{author}{Onida, G.}, \bibinfo{author}{Reining, L.} \& \bibinfo{author}{Rubio, A.}
\newblock \bibinfo{journal}{\bibinfo{title}{Electronic excitations: density-functional versus many-body {G}reen's-function approaches}}.
\newblock {\emph{\JournalTitle{Rev. Mod. Phys.}}} \textbf{\bibinfo{volume}{74}}, \bibinfo{pages}{601--659}, \doiprefix\url{10.1103/RevModPhys.74.601} (\bibinfo{year}{2002}).

\bibitem{RL00}
\bibinfo{author}{Rohlfing, M.} \& \bibinfo{author}{Louie, S.~G.}
\newblock \bibinfo{journal}{\bibinfo{title}{Electron-hole excitations and optical spectra from first principles}}.
\newblock {\emph{\JournalTitle{Phys. Rev. B}}} \textbf{\bibinfo{volume}{62}}, \bibinfo{pages}{4927--4944}, \doiprefix\url{10.1103/PhysRevB.62.4927} (\bibinfo{year}{2000}).

\bibitem{PPD04}
\bibinfo{author}{Palummo, M.} \emph{et~al.}
\newblock \bibinfo{journal}{\bibinfo{title}{The {Bethe}-{Salpeter} equation: a first-principles approach for calculating surface optical spectra}}.
\newblock {\emph{\JournalTitle{J. Phys. Condens. Matter}}} \textbf{\bibinfo{volume}{16}}, \bibinfo{pages}{S4313}, \doiprefix\url{10.1088/0953-8984/16/39/006} (\bibinfo{year}{2004}).

\bibitem{WPI23}
\bibinfo{author}{Wang, L.} \emph{et~al.}
\newblock \bibinfo{journal}{\bibinfo{title}{Exciton-assisted electron tunnelling in {van} {der} {Waals} heterostructures}}.
\newblock {\emph{\JournalTitle{Nat. Mater.}}} \textbf{\bibinfo{volume}{22}}, \bibinfo{pages}{1094--1099}, \doiprefix\url{10.1038/s41563-023-01556-7} (\bibinfo{year}{2023}).

\bibitem{Li2018}
\bibinfo{author}{Li, Y.} \& \bibinfo{author}{Heinz, T.~F.}
\newblock \bibinfo{journal}{\bibinfo{title}{Two-dimensional models for the optical response of thin films}}.
\newblock {\emph{\JournalTitle{2D Materials}}} \textbf{\bibinfo{volume}{5}}, \bibinfo{pages}{025021}, \doiprefix\url{10.1088/2053-1583/aab0cf} (\bibinfo{year}{2018}).

\bibitem{SKR17}
\bibinfo{author}{S{\"a}yn{\"a}tjoki, A.} \emph{et~al.}
\newblock \bibinfo{journal}{\bibinfo{title}{Ultra-strong nonlinear optical processes and trigonal warping in {MoS}$_2$ layers}}.
\newblock {\emph{\JournalTitle{Nat. Commun.}}} \textbf{\bibinfo{volume}{8}}, \bibinfo{pages}{893}, \doiprefix\url{10.1038/s41467-017-00749-4} (\bibinfo{year}{2017}).

\bibitem{AJM18}
\bibinfo{author}{Autere, A.} \emph{et~al.}
\newblock \bibinfo{journal}{\bibinfo{title}{Optical harmonic generation in monolayer group-{VI} transition metal dichalcogenides}}.
\newblock {\emph{\JournalTitle{Phys. Rev. B}}} \textbf{\bibinfo{volume}{98}}, \bibinfo{pages}{115426}, \doiprefix\url{10.1103/PhysRevB.98.115426} (\bibinfo{year}{2018}).

\bibitem{WIR22}
\bibinfo{author}{Wang, Y.} \emph{et~al.}
\newblock \bibinfo{journal}{\bibinfo{title}{Probing electronic states in monolayer semiconductors through static and transient third-harmonic spectroscopies}}.
\newblock {\emph{\JournalTitle{Adv. Mater.}}} \textbf{\bibinfo{volume}{34}}, \bibinfo{pages}{2107104}, \doiprefix\url{10.1002/adma.202107104} (\bibinfo{year}{2022}).

\bibitem{WIB22}
\bibinfo{author}{Wang, Y.} \emph{et~al.}
\newblock \bibinfo{journal}{\bibinfo{title}{Optical control of high-harmonic generation at the atomic thickness}}.
\newblock {\emph{\JournalTitle{Nano Lett.}}} \textbf{\bibinfo{volume}{22}}, \bibinfo{pages}{8455--8462}, \doiprefix\url{10.1021/acs.nanolett.2c02711} (\bibinfo{year}{2022}).

\bibitem{Scuri2018}
\bibinfo{author}{Scuri, G.} \emph{et~al.}
\newblock \bibinfo{journal}{\bibinfo{title}{Large excitonic reflectivity of monolayer {MoS}$_2$ encapsulated in hexagonal boron nitride}}.
\newblock {\emph{\JournalTitle{Physical Review Letters}}} \textbf{\bibinfo{volume}{120}}, \bibinfo{pages}{37402}, \doiprefix\url{10.1103/PhysRevLett.120.037402} (\bibinfo{year}{2018}).

\bibitem{Epstein2020}
\bibinfo{author}{Epstein, I.} \emph{et~al.}
\newblock \bibinfo{journal}{\bibinfo{title}{Near-unity light absorption in a monolayer {WS}$_2$ van der {Waals} heterostructure cavity}}.
\newblock {\emph{\JournalTitle{Nano Letters}}} \textbf{\bibinfo{volume}{20}}, \bibinfo{pages}{3545--3552}, \doiprefix\url{10.1021/acs.nanolett.0c00492} (\bibinfo{year}{2020}).

\bibitem{Lindberg1988}
\bibinfo{author}{Lindberg, M.} \& \bibinfo{author}{Koch, S.~W.}
\newblock \bibinfo{journal}{\bibinfo{title}{Effective {B}loch equations for semiconductors}}.
\newblock {\emph{\JournalTitle{Physical Review B}}} \textbf{\bibinfo{volume}{38}}, \bibinfo{pages}{3342--3350}, \doiprefix\url{10.1103/PhysRevB.38.3342} (\bibinfo{year}{1988}).

\bibitem{Stroucken2013}
\bibinfo{author}{Stroucken, T.}, \bibinfo{author}{Grönqvist, J.~H.} \& \bibinfo{author}{Koch, S.~W.}
\newblock \bibinfo{journal}{\bibinfo{title}{Screening and gap generation in bilayer graphene}}.
\newblock {\emph{\JournalTitle{Physical Review B}}} \textbf{\bibinfo{volume}{87}}, \bibinfo{pages}{245428}, \doiprefix\url{10.1103/PhysRevB.87.245428} (\bibinfo{year}{2013}).

\bibitem{Selig2016}
\bibinfo{author}{Selig, M.} \emph{et~al.}
\newblock \bibinfo{journal}{\bibinfo{title}{Excitonic linewidth and coherence lifetime in monolayer transition metal dichalcogenides}}.
\newblock {\emph{\JournalTitle{Nature Communications}}} \textbf{\bibinfo{volume}{7}}, \bibinfo{pages}{13279}, \doiprefix\url{10.1038/ncomms13279} (\bibinfo{year}{2016}).

\bibitem{Selig2018}
\bibinfo{author}{Selig, M.} \emph{et~al.}
\newblock \bibinfo{journal}{\bibinfo{title}{Dark and bright exciton formation, thermalization, and photoluminescence in monolayer transition metal dichalcogenides}}.
\newblock {\emph{\JournalTitle{2D Materials}}} \textbf{\bibinfo{volume}{5}}, \bibinfo{pages}{035017}, \doiprefix\url{10.1088/2053-1583/aabea3} (\bibinfo{year}{2018}).

\bibitem{Maduro2022}
\bibinfo{author}{Maduro, L.}, \bibinfo{author}{Noordam, M.}, \bibinfo{author}{Bolhuis, M.}, \bibinfo{author}{Kuipers, L.} \& \bibinfo{author}{Conesa-Boj, S.}
\newblock \bibinfo{journal}{\bibinfo{title}{Position-controlled fabrication of vertically aligned {Mo}/{MoS}$_2$ core--shell nanopillar arrays}}.
\newblock {\emph{\JournalTitle{Advanced Functional Materials}}} \textbf{\bibinfo{volume}{32}}, \bibinfo{pages}{2107880}, \doiprefix\url{10.1002/adfm.202107880} (\bibinfo{year}{2022}).

\bibitem{Bolhuis2020}
\bibinfo{author}{Bolhuis, M.} \emph{et~al.}
\newblock \bibinfo{journal}{\bibinfo{title}{Vertically-oriented {MoS}$_2$ nanosheets for nonlinear optical devices}}.
\newblock {\emph{\JournalTitle{Nanoscale}}} \textbf{\bibinfo{volume}{12}}, \bibinfo{pages}{10491--10497}, \doiprefix\url{10.1039/D0NR00755B} (\bibinfo{year}{2020}).

\bibitem{Fitzgerald:24}
\bibinfo{author}{Fitzgerald, J.~M.} \emph{et~al.}
\newblock \bibinfo{journal}{\bibinfo{title}{Circumventing the polariton bottleneck via dark excitons in {2D} semiconductors}}.
\newblock {\emph{\JournalTitle{Optica}}} \textbf{\bibinfo{volume}{11}}, \bibinfo{pages}{1346--1351}, \doiprefix\url{10.1364/OPTICA.528699} (\bibinfo{year}{2024}).

\bibitem{Kennes2021Moire}
\bibinfo{author}{Kennes, D.~M.} \emph{et~al.}
\newblock \bibinfo{journal}{\bibinfo{title}{Moir\'{e} heterostructures as a condensed-matter quantum simulator}}.
\newblock {\emph{\JournalTitle{Nature Physics}}} \textbf{\bibinfo{volume}{17}}, \bibinfo{pages}{155--163}, \doiprefix\url{10.1038/s41567-020-01154-3} (\bibinfo{year}{2021}).

\bibitem{Bhimanapati2015Recent}
\bibinfo{author}{Bhimanapati, G.~R.} \emph{et~al.}
\newblock \bibinfo{journal}{\bibinfo{title}{Recent {Advances} in {Two}-{Dimensional} {Materials} beyond {Graphene}}}.
\newblock {\emph{\JournalTitle{ACS Nano}}} \textbf{\bibinfo{volume}{9}}, \bibinfo{pages}{11509--11539}, \doiprefix\url{10.1021/acsnano.5b05556} (\bibinfo{year}{2015}).

\bibitem{Greten2024Dipolar}
\bibinfo{author}{Greten, L.} \emph{et~al.}
\newblock \bibinfo{journal}{\bibinfo{title}{Dipolar {Coupling} at {Interfaces} of {Ultrathin} {Semiconductors}, {Semimetals}, {Plasmonic} {Nanoparticles}, and {Molecules}}}.
\newblock {\emph{\JournalTitle{Physica Status Solidi (a)}}} \textbf{\bibinfo{volume}{221}}, \bibinfo{pages}{2300102}, \doiprefix\url{10.1002/pssa.202300102} (\bibinfo{year}{2024}).

\bibitem{Sajid2020Single}
\bibinfo{author}{Sajid, A.}, \bibinfo{author}{Ford, M.~J.} \& \bibinfo{author}{Reimers, J.~R.}
\newblock \bibinfo{journal}{\bibinfo{title}{Single-photon emitters in hexagonal boron nitride: a review of progress}}.
\newblock {\emph{\JournalTitle{Reports on Progress in Physics}}} \textbf{\bibinfo{volume}{83}}, \bibinfo{pages}{044501}, \doiprefix\url{10.1088/1361-6633/ab6310} (\bibinfo{year}{2020}).

\bibitem{Yazyev2010Emergence}
\bibinfo{author}{Yazyev, O.~V.}
\newblock \bibinfo{journal}{\bibinfo{title}{Emergence of magnetism in graphene materials and nanostructures}}.
\newblock {\emph{\JournalTitle{Reports on Progress in Physics}}} \textbf{\bibinfo{volume}{73}}, \bibinfo{pages}{056501}, \doiprefix\url{10.1088/0034-4885/73/5/056501} (\bibinfo{year}{2010}).

\bibitem{Ghorashi2024Highly}
\bibinfo{author}{Ghorashi, A.} \emph{et~al.}
\newblock \bibinfo{journal}{\bibinfo{title}{Highly confined, low-loss plasmonics based on two-dimensional solid-state defect lattices}}.
\newblock {\emph{\JournalTitle{Physical Review Materials}}} \textbf{\bibinfo{volume}{8}}, \bibinfo{pages}{L011001}, \doiprefix\url{10.1103/PhysRevMaterials.8.L011001} (\bibinfo{year}{2024}).

\bibitem{Huang2022Carbon}
\bibinfo{author}{Huang, P.} \emph{et~al.}
\newblock \bibinfo{journal}{\bibinfo{title}{Carbon and vacancy centers in hexagonal boron nitride}}.
\newblock {\emph{\JournalTitle{Physical Review B}}} \textbf{\bibinfo{volume}{106}}, \bibinfo{pages}{014107}, \doiprefix\url{10.1103/PhysRevB.106.014107} (\bibinfo{year}{2022}).

\bibitem{Vinichenko2017Accurate}
\bibinfo{author}{Vinichenko, D.}, \bibinfo{author}{Sensoy, M.~G.}, \bibinfo{author}{Friend, C.~M.} \& \bibinfo{author}{Kaxiras, E.}
\newblock \bibinfo{journal}{\bibinfo{title}{Accurate formation energies of charged defects in solids: A systematic approach}}.
\newblock {\emph{\JournalTitle{Physical Review B}}} \textbf{\bibinfo{volume}{95}}, \bibinfo{pages}{235310}, \doiprefix\url{10.1103/PhysRevB.95.235310} (\bibinfo{year}{2017}).

\bibitem{Sajid2020V}
\bibinfo{author}{Sajid, A.} \& \bibinfo{author}{Thygesen, K.~S.}
\newblock \bibinfo{journal}{\bibinfo{title}{V$_{{N}}${C}$_{{B}}$ defect as source of single photon emission from hexagonal boron nitride}}.
\newblock {\emph{\JournalTitle{2D Materials}}} \textbf{\bibinfo{volume}{7}}, \bibinfo{pages}{031007}, \doiprefix\url{10.1088/2053-1583/ab8f61} (\bibinfo{year}{2020}).

\bibitem{Kurman2020Tunable}
\bibinfo{author}{Kurman, Y.} \& \bibinfo{author}{Kaminer, I.}
\newblock \bibinfo{journal}{\bibinfo{title}{Tunable bandgap renormalization by nonlocal ultra-strong coupling in nanophotonics}}.
\newblock {\emph{\JournalTitle{Nature Physics}}} \textbf{\bibinfo{volume}{16}}, \bibinfo{pages}{868--874}, \doiprefix\url{10.1038/s41567-020-0890-0} (\bibinfo{year}{2020}).

\bibitem{Yan20202d}
\bibinfo{author}{Yan, S.}, \bibinfo{author}{Zhu, X.}, \bibinfo{author}{Dong, J.}, \bibinfo{author}{Ding, Y.} \& \bibinfo{author}{Xiao, S.}
\newblock \bibinfo{journal}{\bibinfo{title}{{2D} materials integrated with metallic nanostructures: fundamentals and optoelectronic applications}}.
\newblock {\emph{\JournalTitle{Nanophotonics}}} \textbf{\bibinfo{volume}{9}}, \bibinfo{pages}{1877--1900}, \doiprefix\url{10.1515/nanoph-2020-0074} (\bibinfo{year}{2020}).

\bibitem{Alpeggiani2014Semiclassical}
\bibinfo{author}{Alpeggiani, F.} \& \bibinfo{author}{Andreani, L.~C.}
\newblock \bibinfo{journal}{\bibinfo{title}{Semiclassical theory of multisubband plasmons: {Nonlocal} electrodynamics and radiative effects}}.
\newblock {\emph{\JournalTitle{Physical Review B}}} \textbf{\bibinfo{volume}{90}}, \bibinfo{pages}{115311}, \doiprefix\url{10.1103/PhysRevB.90.115311} (\bibinfo{year}{2014}).

\bibitem{Lundeberg2017Tuning}
\bibinfo{author}{Lundeberg, M.~B.} \emph{et~al.}
\newblock \bibinfo{journal}{\bibinfo{title}{Tuning quantum nonlocal effects in graphene plasmonics}}.
\newblock {\emph{\JournalTitle{Science}}} \textbf{\bibinfo{volume}{357}}, \bibinfo{pages}{187--191}, \doiprefix\url{10.1126/science.aan2735} (\bibinfo{year}{2017}).

\bibitem{Rosner2016Two-dimensional}
\bibinfo{author}{Rösner, M.} \emph{et~al.}
\newblock \bibinfo{journal}{\bibinfo{title}{Two-{Dimensional} {Heterojunctions} from {Nonlocal} {Manipulations} of the {Interactions}}}.
\newblock {\emph{\JournalTitle{Nano Letters}}} \textbf{\bibinfo{volume}{16}}, \bibinfo{pages}{2322--2327}, \doiprefix\url{10.1021/acs.nanolett.5b05009} (\bibinfo{year}{2016}).

\bibitem{Yu2024_2}
\bibinfo{author}{Yu, H.} \emph{et~al.}
\newblock \bibinfo{journal}{\bibinfo{title}{Eight in. wafer-scale epitaxial monolayer {MoS}$_2$}}.
\newblock {\emph{\JournalTitle{Advanced Materials}}} \textbf{\bibinfo{volume}{36}}, \bibinfo{pages}{2402855}, \doiprefix\url{10.1002/adma.202402855} (\bibinfo{year}{2024}).

\bibitem{Lee2020}
\bibinfo{author}{Lee, D.~H.}, \bibinfo{author}{Sim, Y.}, \bibinfo{author}{Wang, J.} \& \bibinfo{author}{Kwon, S.~Y.}
\newblock \bibinfo{journal}{\bibinfo{title}{Metal-organic chemical vapor deposition of {2D} van der {Waals} materials - the challenges and the extensive future opportunities}}.
\newblock {\emph{\JournalTitle{APL Materials}}} \textbf{\bibinfo{volume}{8}}, \bibinfo{pages}{030901}, \doiprefix\url{10.1063/1.5142601} (\bibinfo{year}{2020}).

\bibitem{Han2015}
\bibinfo{author}{Han, G.~H.} \emph{et~al.}
\newblock \bibinfo{journal}{\bibinfo{title}{Seeded growth of highly crystalline molybdenum disulphide monolayers at controlled locations}}.
\newblock {\emph{\JournalTitle{Nature Communications}}} \textbf{\bibinfo{volume}{6}}, \bibinfo{pages}{6128}, \doiprefix\url{10.1038/ncomms7128} (\bibinfo{year}{2015}).

\bibitem{Ferrando2023}
\bibinfo{author}{Ferrando, G.} \emph{et~al.}
\newblock \bibinfo{journal}{\bibinfo{title}{{Flat-optics hybrid {MoS}$_2$/polymer films for photochemical conversion}}}.
\newblock {\emph{\JournalTitle{Nanoscale}}} \textbf{\bibinfo{volume}{15}}, \bibinfo{pages}{1953--1961}, \doiprefix\url{10.1039/d2nr05004h} (\bibinfo{year}{2022}).

\bibitem{Martella2017}
\bibinfo{author}{Martella, C.} \emph{et~al.}
\newblock \bibinfo{journal}{\bibinfo{title}{Anisotropic {MoS}$_2$ nanosheets grown on self‐organized nanopatterned substrates}}.
\newblock {\emph{\JournalTitle{Advanced Materials}}} \textbf{\bibinfo{volume}{29}}, \bibinfo{pages}{1605785}, \doiprefix\url{10.1002/adma.201605785} (\bibinfo{year}{2017}).

\bibitem{Gardella2024}
\bibinfo{author}{Gardella, M.} \emph{et~al.}
\newblock \bibinfo{journal}{\bibinfo{title}{Large area van der waals {MoS}$_2$–{WS}$_2$ heterostructures for visible-light energy conversion}}.
\newblock {\emph{\JournalTitle{RSC Applied Interfaces}}} \textbf{\bibinfo{volume}{1}}, \bibinfo{pages}{1001--1011}, \doiprefix\url{10.1039/D3LF00220A} (\bibinfo{year}{2024}).

\bibitem{Yan2018}
\bibinfo{author}{Yan, P.} \emph{et~al.}
\newblock \bibinfo{journal}{\bibinfo{title}{Chemical vapor deposition of monolayer {MoS}$_2$ on sapphire, {Si} and {GaN} substrates}}.
\newblock {\emph{\JournalTitle{Superlattices and Microstructures}}} \textbf{\bibinfo{volume}{120}}, \bibinfo{pages}{235--240}, \doiprefix\url{10.1016/j.spmi.2018.05.049} (\bibinfo{year}{2018}).

\bibitem{Xia2023}
\bibinfo{author}{Xia, Y.} \emph{et~al.}
\newblock \bibinfo{journal}{\bibinfo{title}{12-inch growth of uniform {MoS}$_2$ monolayer for integrated circuit manufacture}}.
\newblock {\emph{\JournalTitle{Nature Materials}}} \textbf{\bibinfo{volume}{22}}, \bibinfo{pages}{1324--1331}, \doiprefix\url{10.1038/s41563-023-01671-5} (\bibinfo{year}{2023}).

\bibitem{Song2021}
\bibinfo{author}{Song, Y.}, \bibinfo{author}{Zou, W.}, \bibinfo{author}{Lu, Q.}, \bibinfo{author}{Lin, L.} \& \bibinfo{author}{Liu, Z.}
\newblock \bibinfo{journal}{\bibinfo{title}{Graphene transfer: Paving the road for applications of chemical vapor deposition graphene}}.
\newblock {\emph{\JournalTitle{Small}}} \textbf{\bibinfo{volume}{17}}, \bibinfo{pages}{2007600}, \doiprefix\url{10.1002/smll.202007600} (\bibinfo{year}{2021}).

\bibitem{Mannix2022}
\bibinfo{author}{Mannix, A.~J.} \emph{et~al.}
\newblock \bibinfo{journal}{\bibinfo{title}{{Robotic four-dimensional pixel assembly of van der {Waals} solids}}}.
\newblock {\emph{\JournalTitle{Nature Nanotechnology}}} \textbf{\bibinfo{volume}{17}}, \bibinfo{pages}{361--366}, \doiprefix\url{10.1038/s41565-021-01061-5} (\bibinfo{year}{2022}).

\bibitem{Liu2022}
\bibinfo{author}{Liu, G.} \emph{et~al.}
\newblock \bibinfo{journal}{\bibinfo{title}{{Graphene-assisted metal transfer printing for wafer-scale integration of metal electrodes and two-dimensional materials}}}.
\newblock {\emph{\JournalTitle{Nature Electronics}}} \textbf{\bibinfo{volume}{5}}, \bibinfo{pages}{275--280}, \doiprefix\url{10.1038/s41928-022-00764-4} (\bibinfo{year}{2022}).

\bibitem{Yang2023}
\bibinfo{author}{Yang, X.} \emph{et~al.}
\newblock \bibinfo{journal}{\bibinfo{title}{{Highly reproducible van der {Waals} integration of two-dimensional electronics on the wafer scale}}}.
\newblock {\emph{\JournalTitle{Nature Nanotechnology}}} \textbf{\bibinfo{volume}{18}}, \bibinfo{pages}{471–478}, \doiprefix\url{10.1038/s41565-023-01342-1} (\bibinfo{year}{2023}).

\bibitem{Huang2022}
\bibinfo{author}{Huang, J.~K.} \emph{et~al.}
\newblock \bibinfo{journal}{\bibinfo{title}{{High-$\kappa$ perovskite membranes as insulators for two-dimensional transistors}}}.
\newblock {\emph{\JournalTitle{Nature}}} \textbf{\bibinfo{volume}{605}}, \bibinfo{pages}{262--267}, \doiprefix\url{10.1038/s41586-022-04588-2} (\bibinfo{year}{2022}).

\bibitem{Mootheri2021}
\bibinfo{author}{Mootheri, V.} \emph{et~al.}
\newblock \bibinfo{journal}{\bibinfo{title}{{Graphene based Van der {Waals} contacts on {MoS}$_2$ field effect transistors}}}.
\newblock {\emph{\JournalTitle{2D Materials}}} \textbf{\bibinfo{volume}{8}}, \bibinfo{pages}{015003}, \doiprefix\url{10.1088/2053-1583/abb959} (\bibinfo{year}{2021}).

\bibitem{Nguyen2023}
\bibinfo{author}{Nguyen, V.~L.} \emph{et~al.}
\newblock \bibinfo{journal}{\bibinfo{title}{{Wafer-scale integration of transition metal dichalcogenide field-effect transistors using adhesion lithography}}}.
\newblock {\emph{\JournalTitle{Nature Electronics}}} \textbf{\bibinfo{volume}{6}}, \bibinfo{pages}{146--153}, \doiprefix\url{10.1038/s41928-022-00890-z} (\bibinfo{year}{2023}).

\bibitem{jung2019transferred}
\bibinfo{author}{Jung, Y.} \emph{et~al.}
\newblock \bibinfo{journal}{\bibinfo{title}{Transferred via contacts as a platform for ideal two-dimensional transistors}}.
\newblock {\emph{\JournalTitle{Nature Electronics}}} \textbf{\bibinfo{volume}{2}}, \bibinfo{pages}{187--194}, \doiprefix\url{10.1038/s41928-019-0245-y} (\bibinfo{year}{2019}).

\bibitem{Giordano2023}
\bibinfo{author}{Giordano, M.~C.}, \bibinfo{author}{Zambito, G.}, \bibinfo{author}{Gardella, M.} \& \bibinfo{author}{{Buatier de Mongeot}, F.}
\newblock \bibinfo{journal}{\bibinfo{title}{{Deterministic Thermal Sculpting of Large-Scale {2D} Semiconductor Nanocircuits}}}.
\newblock {\emph{\JournalTitle{Advanced Materials Interfaces}}} \textbf{\bibinfo{volume}{10}}, \bibinfo{pages}{1--7}, \doiprefix\url{10.1002/admi.202201408} (\bibinfo{year}{2023}).

\bibitem{Bhatnagar2021}
\bibinfo{author}{Bhatnagar, M.} \emph{et~al.}
\newblock \bibinfo{journal}{\bibinfo{title}{{Broadband and Tunable Light Harvesting in Nanorippled {MoS}$_2$ Ultrathin Films}}}.
\newblock {\emph{\JournalTitle{ACS Applied Materials and Interfaces}}} \textbf{\bibinfo{volume}{13}}, \bibinfo{pages}{13508--13516}, \doiprefix\url{10.1021/acsami.0c20387} (\bibinfo{year}{2021}).

\bibitem{quellmalz2021large}
\bibinfo{author}{Quellmalz, A.} \emph{et~al.}
\newblock \bibinfo{journal}{\bibinfo{title}{Large-area integration of two-dimensional materials and their heterostructures by wafer bonding}}.
\newblock {\emph{\JournalTitle{Nature Communications}}} \textbf{\bibinfo{volume}{12}}, \bibinfo{pages}{917}, \doiprefix\url{10.1038/s41467-021-21136-0} (\bibinfo{year}{2021}).

\bibitem{lanza2020yield}
\bibinfo{author}{Lanza, M.}, \bibinfo{author}{Smets, Q.}, \bibinfo{author}{Huyghebaert, C.} \& \bibinfo{author}{Li, L.-J.}
\newblock \bibinfo{journal}{\bibinfo{title}{Yield, variability, reliability, and stability of two-dimensional materials based solid-state electronic devices}}.
\newblock {\emph{\JournalTitle{Nature Communications}}} \textbf{\bibinfo{volume}{11}}, \bibinfo{pages}{5689}, \doiprefix\url{10.1038/s41467-020-19053-9} (\bibinfo{year}{2020}).

\bibitem{eizagirre2019}
\bibinfo{author}{Eizagirre~Barker, S.} \emph{et~al.}
\newblock \bibinfo{journal}{\bibinfo{title}{Preserving the emission lifetime and efficiency of a monolayer semiconductor upon transfer}}.
\newblock {\emph{\JournalTitle{Advanced Optical Materials}}} \textbf{\bibinfo{volume}{7}}, \bibinfo{pages}{1900351}, \doiprefix\url{10.1002/adom.201900351} (\bibinfo{year}{2019}).

\bibitem{Godiksen2020}
\bibinfo{author}{Godiksen, R.~H.} \emph{et~al.}
\newblock \bibinfo{journal}{\bibinfo{title}{Correlated exciton fluctuations in a two-dimensional semiconductor on a metal}}.
\newblock {\emph{\JournalTitle{Nano Letters}}} \textbf{\bibinfo{volume}{20}}, \bibinfo{pages}{4829--4836}, \doiprefix\url{10.1021/acs.nanolett.0c00756} (\bibinfo{year}{2020}).

\end{thebibliography}

\end{document}